\documentclass[fleqn,usenatbib]{mnras}

\usepackage{newtxtext,newtxmath}

\usepackage[T1]{fontenc}
\usepackage{standalone}
\usepackage{multirow}
\usepackage{longtable}
\DeclareRobustCommand{\VAN}[3]{#2}
\let\VANthebibliography\thebibliography
\def\thebibliography{\DeclareRobustCommand{\VAN}[3]{##3}\VANthebibliography}
\usepackage{graphicx}

\usepackage{float}
\floatstyle{boxed}
\restylefloat{figure}

\usepackage{amsmath}	
\usepackage{pdflscape}
\usepackage{deluxetable}
\usepackage{lscape}

\def\rasec {\hbox{$\,$\raise 0.6 ex \hbox{\rm s}\kern-.35em
                  \lower 0.0 ex \hbox{.}$\,$}}        
\def\decsec{\hbox{$\,$\raise 0.5 ex \hbox{$\prime\prime$}\kern-.45em
                  \lower 0.0 ex \hbox{.}$\,$}}         
\def\decmin{\hbox{$\,$\raise 0.5 ex \hbox{$\prime$}\kern-.45em
    \lower 0.0 ex \hbox{}$\,$}}

\title[CHANG-ES XXVII]{CHANG-ES XXVII: A Radio/X-ray Catalogue of Compact Sources in and around Edge-on Galaxies}

\author[J. A. Irwin et al.]{Judith Irwin$^1$\thanks{E-mail: irwinja@queensu.ca}, Jacqueline Dyer$^1$, Leonardo Drake$^2$, Q. Daniel Wang$^2$,   
\newauthor{Jeroen Stil$^3$, Yelena Stein$^{4,5}$, Jayanne English$^6$, and Theresa Wiegert$^1$}
\\
$^1$ Department of Physics, Engineering Physics \& Astronomy, Queens University, Kingston, Ontario, K7L 3N6, Canada\\ $^2$ Department of Astronomy, University of Massachusetts, North Pleasant Street, Amherst, MA 01003-9305, USA,LGRT-B 619E, 710\\
$^3$ Dept. of Physics \& Astronomy, 834 Campus Place N. W., University of Calgary, 2500 University Drive N W, Calgary, AB, Canada, T2N 1N4\\
$^4$ Ruhr University Bochum, Faculty of Physics and Astronomy, AIRUB, Germany\\
$^5$ Observatoire astronomique de Strasbourg, Universit\'e de Strasbourg, CNRS, UMR 7550, 11 rue de l'Universit\'e, 67000 Strasbourg, France\\
$^6$ Dept. of Physics and Astronomy, University of Manitoba, Winnipeg, Manitoba, Canada, R3T 2N2\\
}

\date{Accepted XXX. Received YYY; in original form ZZZ}

\pubyear{2020}

\begin{document}
\label{firstpage}
\pagerange{\pageref{firstpage}--\pageref{lastpage}}
\maketitle

\begin{abstract}
We present catalogues of discrete, compact radio sources in and around the discs of 35 edge-on galaxies in the Continuum Halos in Nearby Galaxies -- an EVLA Survey (CHANG-ES).  The sources were extracted using the PyBDSF program at both 1.6 GHz (L-band) and 6.0 GHz (C-band) from matching resolution ($\approx$ 3 arcsec) data.  We also present catalogues of X-ray sources from Chandra data sets for 27 of the galaxies.  The sources at the two radio frequency bands were positionally cross-correlated with each other, and the result cross-correlated with the X-ray sources.  All catalogues  are included for download with this paper. We detect a total of 2507 sources at L-band and 1413 sources at C-band.  Seventy-five sources have been successfully cross-correlated in both radio bands plus X-ray.  
Three new nuclear sources are  candidates for Low Luminosity Active Galactic Nuclei in NGC~3877, NGC~4192, and NGC~5792; the one in NGC~3877 also appears to be variable. 
 We also find new nuclear sources in two companion galaxies: NGC~4435 (companion to NGC~4438) and NGC~4298 (companion to NGC~4302).  
We have also discovered what appears to be a foreground double-star; each star has X-ray emission and there is radio emission at both L-band and C-band in between them.  This could be a colliding wind binary system.  Suggestions for follow-up studies are offered.
\end{abstract}

\begin{keywords}
catalogues -- radio continuum: galaxies -- X-rays: galaxies -- galaxies: nuclei 
\end{keywords}



\section{Introduction}\label{s:Intro}

We present a radio continuum and X-ray catalogue of discrete sources in and around a sample of nearby edge-on galaxies (Table~\ref{t:g_param}). All galaxies in this sample are from the CHANG-ES (Continuum Halos in Nearby Galaxies -- an EVLA\footnote{The Expanded Very Large Array (EVLA) is now known as the Karl G. Jansky Very Large Array.} Survey) program. CHANG-ES data \citep{Irwin_2012a} were obtained with the Karl G. Jansky Very Large Array (hereafter, the VLA). {A summary of CHANG-ES goals and selected results can be found in \cite{irw19a}.} The X-ray data were obtained using the Chandra X-ray satellite.  This new catalogue represents the first systematic investigation of discrete radio/X-ray sources in such galaxies. In this paper, we present the catalogues, point out some intriguing new results, and suggest future research projects that could make use of the data.

{Radio continuum emission consists of thermal and non-thermal contributions and these components are often mixed together. However, when discrete sources are isolated spatially, which can be accomplished with high resolution data,  one or the other of these emission processes may {dominate}.  A discrete thermal source is typically an HII region which has the well-known $I_\nu\propto\,\nu^{-0.1}$ flat thermal Bremsstrahlung spectrum, where $I_\nu$ is the specific intensity and $\nu$ is the frequency. A non-thermal discrete source will display a synchrotron spectrum,  $I_\nu\propto\,\nu^\alpha$, with $\alpha$ values that are typically much steeper than $-0.1$.  For example, a sample of Galactic and extragalactic supernova remnants (SNRs) exhibit a range of $\alpha$ from $-0.4$ to $-1.1$ \citep{bel11} and even so-called `flat-spectrum SNRs' show spectra with $-0.5\,<\,\alpha\,<\,-0.2$, significantly steeper than HII regions \citep{uro14}. 

Another important category of non-thermal sources is the energetic active galactic nuclei (AGNs), or galactic cores, for which the spectral index could take on a range of values.  For example, a recent study \citep{lah16} lists $-0.8$ to $+0.2$, depending on source type and frequency.
The identifiable location of a core at the centre of a galaxy is key.
An earlier detailed study of CHANG-ES galaxies using high resolution L-band (1.6 GHz) data alone, along with information on whether or not the galaxies harbour AGNs, can be found in
\cite{irw19b}.  With the addition of the C-band (6 GHz) data  and the source finding algorithm used here (Sect.~\ref{ss:ana_radio}) we now identify new nuclear sources} to add to this lexicon.

{It is clear from the above discussion that both high spatial resolution as well as measurements at two different radio frequencies (which permits spectral index measurements) provide key constraints on the type of object being identified. CHANG-ES has provided the requisite high resolution ($\approx\,3\,$arcsec) radio data at two different frequencies: L-band in the B-configuration of the VLA,  and C-band in C-configuration.  As can be seen in Table~\ref{t:g_param}, of the entire 35 galaxy sample, the corresponding linear resolutions range from 55 to 646 pc over galaxy distances that range from 4.4 to 42 Mpc.  HII region sizes span many orders of magnitude, from less than $\approx\,0.01$ pc to $\approx\,1$ kpc \citep[][their Fig. 2]{hun09}. As for SNRs, in very close galaxies such as the Large and Small Magellanic Clouds, M~31, M~33, and M~82, their sizes range from about a parsec to a few hundred pc \citep{lon17}, with tens of pc being typical. Consequently, CHANG-ES linear resolutions span the range from unresolved to resolved discrete sources in the discs of galaxies. AGN, on the other hand, will likely be unresolved.}

In addition to discrete sources in galaxy discs, there are also background sources (typically radio galaxies and quasars), which are visible in the field, some of which could be just projected in the fields of the foreground galaxies under consideration. 
The identification of such sources can be interesting in their own rights, but can also sometimes be very useful for the study of foreground galaxies. For example, a background source that is seen through the disc or halo of a foreground galaxy experiences Faraday rotation of its polarized emission. A good example, also from CHANG-ES, is given in \cite{irw13}. Although we do not discuss polarization in this paper, our catalogues may be useful for future follow-up studies.

A final interesting category is foreground sources, i.e. stars.  While there should be very few such cases, we will provide an interesting example in Sect.~\ref{s:discussion}.

{Important additional constraints are provided by X-ray data. For AGN, for example, observations in radio and X-ray can provide sensitive probes of the accretion state of AGN, as well as their jet ejection and interaction with the ambient medium \citep[e.g.,][]{Perlman2021}.  Compact X-ray sources in the disc typically represent very young SNRs, including plerions or pulsar wind nebulae, plus some giant HII regions or superbubbles produced by fast stellar winds and supernovae (SNe) collectively \citep[e.g.,][]{Yew2018}.  They should also include luminous X-ray binaries -- black holes or neutron stars accreting from their companions; some of such accreting binaries could produce radio jets. The observations of such off-nucleus sources in radio and X-ray together can provide information about the end-products of massive stars (e.g., their stellar origins, populations, emission mechanisms, circumstellar environments, etc. \citep[e.g.,][]{Maddox2006,Pannuti2007,Wang2012}. 
}

Our X-ray source detections are based on Chandra observations that were available for 27 of the CHANG-ES galaxies by June, 2020. Sixteen of them have been used in the survey of galactic gaseous coronae of edge-on galaxies \citep{Li2013}. Some of these galaxies now have more available observations. The Chandra spatial resolution is $\sim 0.4^{\prime\prime}$ \citep{Weisskopf2000}.

Existing radio and X-ray studies of high-energy sources in nearby galaxies are typically on face-on ones \citep[e.g.,][]{Maddox2006,Pannuti2007}. In face-on galaxies, it is relatively easy to examine the association of sources with galactic disc structures such as spiral arms.  However, because of the larger projected areas on the sky, it is often statistically difficult to resolve the confusion with background sources (e.g., distant AGNs), as well as the galactic halo emission of the host galaxies. The advantages of studying edge-on disc galaxies include relatively cleaner separation of the disc and halo emission and smaller confusion with background sources because of their discs' smaller footprints in the sky. 
{In addition, the edge-on orientation of the CHANG-ES galaxies provides a unique opportunity to search for discrete sources that are associated with a galaxy, but may not be embedded in the disc.  For example, 
one HII region that is far outside of the disc of the CHANG-ES galaxy, NGC~4157, has already been identified \citep{var19}.  
Thus, our catalogues will lay the groundwork for  potential future follow-up studies of discrete in-disc or halo objects. }



{\begin{table*}
\fontsize{7}{11}\selectfont
\begin{center}
\caption{Image Parameters of CHANGES galaxies}\label{t:g_param}
    \begin{tabular}{lcccccccccc}
    \hline
Galaxy & 
RA$^{\rm a}$ & Dec$^{\rm a}$ 
& $D^{\rm b}$ &$d_{25}$, $r_{25}$, PA$^{\rm c}$ & \multicolumn{3}{c}{L-band$^{\rm d}$}  & \multicolumn{3}{c}{C-band$^{\rm d}$}\\ 
& & & & & Ang. Res. & Lin. Res. & rms & Ang. Res. & Lin. Res. & rms\\
Name   & $^{\rm h}$ $^{\rm m}$ $^{\rm s}$  & $^\circ$ $^\prime$ $^{\prime\prime}$& Mpc &$^{\prime}$, $^{\prime}$, $^\circ$&
$^{\prime\prime}$, $^{\prime\prime}$, $^\circ$ & pc&$\mu$Jy beam$^{-1}$& $^{\prime\prime}$, $^{\prime\prime}$, $^\circ$ & pc & $\mu$Jy beam$^{-1}$
\\
\hline
N 660   & 01 43 02.40   & +13 38 42.2   & 12.3 & 3.16, 0.98, 41 $^{\rm e}$ &3.39, 3.27, 44.4 & 199
&24& 3.15, 2.64, -55.7& 173& 3.8\\
N 891   & 02 22 33.41   & +42 20 56.9   & 9.1  & 13.5, 2.51, 24 & 3.15, 2.90, 54.2 & 133
&16.0& 2.77, 2.61, 85.9 & 119 & 3.1\\
N 2613  & 08 33 22.84   & -22 58 25.2   &23.4 & 7.24, 1.78, 114 & 5.18, 3.02, -179.7 & 465
&19.6& 4.95, 2.37, -6.2 & 415& 3.3\\
N 2683  & 08 52 41.35   & +33 25 18.5   &6.27& 9.33, 2.19, 41& 3.06, 2.98, 57.8& 92
&14.5& 2.71, 2.61, -31.5& 81& 5.0\\
N 2820  & 09 21 45.58   & +64 15 28.6   &26.5& 4.07, 0.49, 60 $^{\rm f}$& 3.23, 3.17, 52.8& 411
& 16.3& 2.64, 2.62, 49.0 & 338& 3.2\\
N 2992  & 09 45 42.00   & -14 19 35.0   &34& 3.55, 1.10, 22& 4.87, 3.57, 16.4 & 696
& 16.5& 4.12, 2.48, -8.0 &544& 3.2\\
N 3003  & 09 48 36.05   & +33 25 17.4   &25.4 &5.75, 1.35, 78 & 3.11, 3.00, 70.1 & 376
&14.0& 2.65, 2.56, -72.7& 321& 3.0\\
N 3044  & 09 53 40.88   & +01 34 46.7   &20.3& 4.90, 0.71, 115 & 3.67, 3.39, 68.8 & 347
&15.0& 3.34, 2.82,-34.2& 303& 3.6\\
N 3079  & 10 01 57.80   & +55 40 47.3   &20.6& 7.94, 1.45, 179 & 3.14, 3.00, 58.4 & 307
& 18.0& 2.68, 2.58, -89.0& 263& 3.4\\
N 3432  & 10 52 31.13   & +36 37 07.6   &9.42 & 6.76, 1.48, 42.0 & 3.20, 3.12, 82.8 & 144
& 21.0& 2.75, 2.52, -79.0& 120& 3.5\\
N 3448  & 10 54 39.24   & +54 18 18.8   &24.5 & 5.62, 1.78, 65 & 3.16, 2.98, 63.9  & 365
& 17.0& 2.57, 2.55, 59.0& 304& 3.2\\
N 3556  & 11 11 30.97   & +55 40 26.8   &14.09& 8.71, 2.24, 81& 3.12, 2.98, 58.2 & 207
& 16.0& 2.58, 2.55, 38.9 & 174& 3.1\\
N 3628  & 11 20 17.01   & +13 35 22.9   &8.5& 14.8, 2.95, 104& 3.21, 3.13,  3.7 & 131
&14.5& 3.68, 2.74, -79.0 & 132& 4.6\\
N 3735  & 11 35 57.30   & +70 32 08.1   &42 & 4.17, 0.83, 130& 3.24, 3.11, 33.8  & 646
&16.0& 2.64, 2.60, 24.4 & 533& 4.1\\
N 3877  & 11 46 07.80   & +47 29 41.2   &17.7 & 5.50, 1.29, 34& 3.01, 2.87, 22.4 & 252
&11.5& 2.56, 2.50, -75.5& 217& 3.4\\
N 4013  & 11 58 31.38   & +43 56 47.7   &16& 5.25, 1.02, 66& 3.01, 2.90, -84.2 & 229
&14.0 & 2.65, 2.50, -87.9& 200& 2.9\\
N 4096  & 12 06 01.13   & +47 28 42.4   &10.32 & 6.61, 1.78, 17 & 3.06, 2.94, -84.9 & 150
&14.5& 2.55, 2.49, -75.5& 126& 3.4\\
N 4157  & 12 11 04.37   & +50 29 04.8   &15.6 & 6.76, 1.35, 63 & 3.02, 2.84, 29.5 & 222
&11.7& 2.60, 2.49, -83.2& 192& 3.3 \\
N 4192  & 12 13 48.29   & +14 54 01.2   &13.55& 9.77, 2.75, 152& 3.21, 3.07, -7.5 & 206
&14.5 & 2.78, 2.65, -67.0 & 178& 3.4\\
N 4217  & 12 15 50.90   & +47 05 30.4   &20.6 & 5.25, 1.55, 49 & 3.07, 2.94, -85.5 & 300
&14.5 & 2.58, 2.50, -78.7& 254& 3.1\\
N 4244  & 12 17 29.66   & +37 48 25.6   &4.4 & 16.6, 1.91, 45& 3.09, 3.00, 45.0 & 65
&14.4& 2.64, 2.54, -81.0& 55& 3.0\\
N 4302  & 12 21 42.48   & +14 35 53.9   &19.41 & 5.50, 1.00, 178& 3.50, 3.13, -8.4 & 312
&13.5& 2.78, 2.66, -67.0& 256& 3.4\\
N 4388  & 12 25 46.75   & +12 39 43.5   &16.6 & 5.62, 1.29, 89& 3.57, 3.22, -2.1 & 273
&16.0 & 2.76, 2.67, -10.4& 219& 3.2\\
N 4438  & 12 27 45.59   & +13 00 31.8   &10.39& 8.51, 3.16, 20& 3.32, 2.91, -6.3 & 156
&25.0 & 2.75, 2.66, -2.86& 135& 3.2\\
N 4565  & 12 36 20.78   & +25 59 15.6   &11.9&15.9, 2.14, 134 & 3.31, 3.01, 45.5 & 182
&15.0& 2.63, 2.59, -61.7 & 151& 3.0\\
N 4594  & 12 39 59.43   & -11 37 23.0   &12.7& 8.71, 3.55, 87&4.36, 3.25, -14.0 & 234
&17.5& 3.90, 2.58, -3.3& 199& 2.9\\
N 4631  & 12 42 08.01   & +32 32 29.4   &7.4 & 15.5, 2.69, 84 & 3.40, 3.05, 63.4 & 116
&16.0& 2.61, 2.50, -76.8& 92& 3.1\\
N 4666  & 12 45 08.59   & -00 27 42.8   &27.5&4.57, 1.29, 40 & 3.80, 3.48, 39.6 & 485
&15.0& 3.14, 2.73, -25.1& 391& 3.8\\
N 4845  & 12 58 01.19   & +01 34 33.0   &16.98& 5.01, 1.32, 75& 3.51, 3.33, 22.7& 280
&18.0& 3.05, 2.75, -11.7& 238& 3.9\\
N 5084  & 13 20 16.92   & -21 49 39.3   &23.4& 9.33, 1.74, 80& 5.63, 2.95, -11.8 & 487
&17.0& 4.81, 2.38, -5.3& 408& 3.0\\
N 5297  & 13 46 23.68   & +43 52 20.5   &40.4 &5.62, 1.26, 149 & 3.13, 2.99, 52.8& 599
&13.6 & 2.63, 2.50, -72.6& 502& 2.9\\
N 5775  & 14 53 57.60   & +03 32 40.0   &28.9&4.17, 1.00, 146 &3.65, 3.44, 64.3 & 497
&14.0& 3.06, 2.69, -5.28& 403& 3.0\\
N 5792  & 14 58 22.71   & -01 05 27.9   &31.7& 6.92, 1.74, 82& 3.89, 3.42, 48.6 & 562
&15.0 & 3.20, 2.78, -6.93& 460& 3.0\\
N 5907  & 15 15 53.77   & +56 19 43.6   &16.8 & 12.6, 1.38, 155& 3.35, 2.79, -4.6& 250
&13.5& 2.66, 2.60, 82.5 & 214& 3.2\\
U10288  & 16 14 24.80   & -00 12 27.1 &34.1 & 4.79, 0.58, 91& 3.80, 3.58, 66.2& 610
&14.0& 3.14, 2.87, -11.0& 497& 3.1\\
\hline
    \end{tabular}
\end{center}
$^{\rm a}$ Pointing and image centre.\\
$^{\rm b}$ Distance, from \cite{irw19b}.\\
$^{\rm c}$ Galaxy major and minor axis diameters at the 25 magnitude per square arcsec level from the Third Reference Catalogue of Bright Galaxies \citep[RC3,][]{dev91}, unless otherwise indicated.  The 3rd value gives the galaxy position angle (PA) (counterclockwise from north) from the K$_s$ passband as given in the Nasa Extragalactic Database (NED), with minor adjustments ($<$ 5 deg), where necessary, to match the C-configuration, C-band angle, as specified in \cite{kra20}.\\
$^{\rm d}$ L-band and C-band image parameters are listed in 3 columns each. First column: Angular size of the synthesized beam -- Major axis FWHM, minor axis FWHM, and position angle, respectively. Second column: Linear resolution corresponding to the average of the major and minor axes FWHM. Third column: average rms noise in the map measured over multiple boxes at positions away from the galaxy or background sources. \\
$^{\rm e}$ The RC3 value for the major and minor axes is grossly overestimated because of the extended tidal features in the north-south direction.  For this galaxy, we adopt the K-band major and minor axes from The Two Micron All Sky Survey (2MASS) \citep{2mass}.\\
$^{\rm f}$ The K$_s$ PA is 65$^\circ$.  This galaxy's major axis differs between the innermost and outermost regions due to an interaction with a companion.\\
\end{table*}
}

 {This new catalogue of discrete sources includes all 35 CHANG-ES galaxies  in the radio, 27 of which  have X-ray.}
The rest of the paper is organized as follows. In Sect.~\ref{s:Obs}, after a brief review of the observations and data reduction, we describe the algorithms that were used to detect radio and X-ray sources. We present the results in Sect.~\ref{s:Res}, including the cross-matching of the two radio bands and the additional cross-matching with X-ray sources. In Sect.~\ref{s:quality}, we present some statistics and quality checks of the data. Sect.~\ref{s:discussion} provides a general discussion, and points out several intriguing sources for follow-up. Sect.~\ref{s:sum} gives a summary.

\section{Observations and data analyses}\label{s:Obs}


\subsection{Radio Observations}\label{ss:Obs_radio}

The images relevant to this study consist of B-configuration, L-band (hereafter B/L) observations centred at 1.6 GHz using a 500 MHz bandwidth, and approximately matching resolution C-configuration, C-band (hereafter C/C) observations at 6.0 GHz with a 2 GHz bandwidth.

Data reduction was carried out using the Common Astronomy Software Application (CASA)\footnote{See casa.nrao.edu.}. 
Reduction of the VLA data has been fully described in previous CHANG-ES papers, for example,  \cite{Irwin_2012b}, \cite{Irwin_2013},  and \cite{irw19b}. 

As indicated in \cite{wie15}, maps were made with two different uv weightings, corresponding to two different spatial resolutions.  Here we use only the highest resolution maps which used robust = 0 uv weighting, as implemented in CASA, resulting in a resolution of approximately 3 arcsec.  The rms noise, however, was significantly higher in L-band than in C-band (Table~\ref{t:g_param}).

Very wide fields were made for almost all galaxies with pixel sizes of $0.5$ arcsec in both bands.  We then extracted field sizes of $2000\,\times\,2000$ (16.7 arcmin $\times$ 16.7 arcmin) at L-band and $1441\,\times\,1441$ pixels (12.0 arcmin $\times$ 12.0 arcmin) at C-band for analysis, as summarized in Table~\ref{t:fieldsize}.  These field sizes cover 0.65 and 1.6 times the primary beam (PB) full-width at half-maximum (FWHM) at L-band and C-band, respectively.  However, the galaxies that are largest in angular size (NGC~891, NGC~3628, NGC~4244, NGC~4565, NGC~4594, NGC~4631, NGC~5084, and NGC~5907), were observed with two pointings at C-band, so the effective PB for those cases does not decline as rapidly from the field centre, compared to  single-pointing galaxies. See \cite{wie15} for details.

For three galaxies, NGC~3735, NGC~4096, and NGC~4666, the original C/C map size was slightly less than the 1441 X 1441 field size specified in Table~\ref{t:fieldsize}.  We therefore remade these maps with a larger field size for consistency with the other galaxies. The rms values of Table~\ref{t:g_param} correspond to these remade maps.

The largest angular scale visible at L-band is 2 arcmin and at C-band it is 4 arcmin. These limits are because of missing short spacings in the uv plane.
All of our galaxies are larger than these limits so these high resolution observations do not recover all of the flux of any of the galaxies.  They will, however, recover all of the flux of any source within the galaxy of smaller angular size.  Consequently, we do not expect missing spacings to affect any results in this paper.

Maps were also made with and without a primary beam (PB) correction.  With no PB correction, the noise is uniform across the map (except for residual imperfectly cleaned sidelobes) but the fluxes are incorrect.  With a PB correction, the noise increases with distance from the centre of the map but the fluxes are correct.  In order to identify and extract discrete sources, it was necessary to run the source-extraction algorithm (next section) on maps that had uniform noise. Corrections to the flux densities and related spectral indices were made after the extraction process. Formulae for the PB correction as a function of distance from the map centre are well known\footnote{ https://library.nrao.edu/public/memos/evla/EVLAM\_{195}} for an assumption of circular symmetry.

The B/L data have been described in \cite{irw19b} and can be downloaded from
https://www.queensu.ca/changes/.  Details of the C/C data will be described in \cite{wal21} with images to be downloaded from the same website.

\subsection{Radio Source Extraction (PyBDSF)}\label{ss:ana_radio}

\begin{table*}
\begin{center}
\caption{Image size in L-band and C-band}\label{t:fieldsize}
\begin{tabular}{lccccc}
\hline
Frequency & Field Size (pix) &Field Size (arcmin)& Cell Size (") &  Res. (arcsec)$^{\rm a}$ &PB FWHM (arcmin)$^{\rm b}$\\ 
\hline
L-band (1.575 GHz$^{\rm c}$) & $2000\times2000$ & 16.7& $0.5$ & 3 &25.4   \\ 
\hline
C-band (6.000 GHz) & $1441\times1441$ & 12.0& $0.5$ & 3 &7.02    \\ 
\hline
\end{tabular}
\end{center}
$^{\rm a}$ Approximate spatial resolution.  Accurate values are given in Table~\ref{t:g_param}. \\
$^{\rm b}$ Primary beam full-width at half-maximum.\\
$^{\rm c}$ For NGC~4438, $\nu\,=\,1.77$ GHz and PB FWHM = 23.6 arcmin.
\end{table*}

We use Python Blob Detector and Source Finder \cite[PyBDSF,][]{moh15} software for our radio source detection. 
The algorithm first computes an rms (root-mean-square, or $\sigma$) and mean level as a function of position for a given map by using a sliding box whose parameters can be specified.  It then identifies `islands' which are regions of {\it contiguous} emission that are above a user-specified threshold.  For this work, we accept an island if all contiguous points are more than 3 times the local rms, above the local mean value. PyBDSF then fits Gaussians to sources within the island that meet another user-specified threshold.  In this work, a source is fit with a Gaussian if it is more than 5 times the rms value above the mean.  

\begin{figure}
   \centering
  \includegraphics[width=0.4\textwidth]{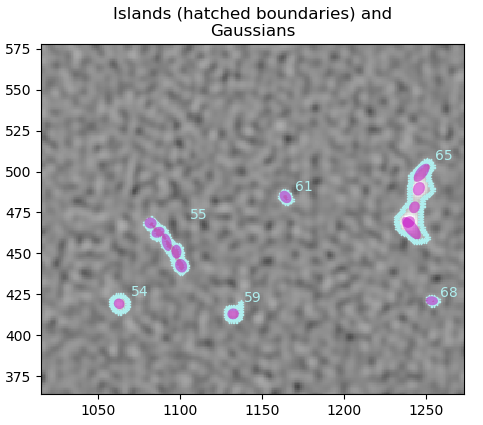}  
   \caption{Qualitative view of the islands and Gaussians showing background sources in the field of NGC~4565 as calculated by PyBDSF. The axis labels are in pixels. The hatched blue boundaries outline the locations of individual islands, while the magenta ellipses represent the Gaussians which have been fit to the individual detected sources. Islands 55 and 65 likely represent background extra-galactic radio sources.  Island 65 is located at RA = 12$^{\rm h}$ 36$^{\rm m}$ 11$\rasec$77, Dec =  25$^\circ$ 54$^\prime$ ~54$\decsec$94, and Island 55 is located at RA = 
    12$^{\rm h}$ 36$^{\rm m}$ 17${\rasec}$34, Dec=  25$^\circ$ 54$^{\prime}$~42${\decsec}$92. A quantitative view of the NGC~4565 field can be seen in Appendix~\ref{a:param}. }
  \label{f:N4565-regions}
\end{figure}

{A visual representation of islands and Gaussians can be seen for some background off-galaxy sources in Fig.~\ref{f:N4565-regions}. The Gaussian sources can be seen in magenta with islands shown as blue boundaries. This introductory image is meant to be qualitative only. }

The software also checks whether sources are significantly overlapped or can be considered separate sources. PyBDSF will group overlapping Gaussian sources within an island into a larger single source with a reconstructed Gaussian size, provided all emission on a line between them exceeds a specified value.  This is important because, often, the entire disc of of a galaxy has contiguous emission and is therefore identified as a single island. The emission throughout the island has variable intensity but we do not necessarily expect such variability to belong to discrete sources that we wish to highlight. Such a multi-Gaussian source is easily identifiable by the larger FWHM of its reconstructed Gaussian and by its label ($M$, next paragraph) in Table~\ref{a:c-radio}.
PyBDSF flags questionable Gaussians (e.g. if the centre is outside of the island) and we do not include flagged Gaussians in this analysis.

PyBDSF then provides tabular output (our catalogue) which includes the sizes of sources, their fluxes, and positions, including errors.   A source size that is deconvolved from the known synthesized beam (Ang. Res. in Table~\ref{t:g_param}) is also provided, enabling an estimate of the true source size. A value of zero means that the source is unresolved. This source list is given in Appendix~\ref{a:c-radio}. An important identifier is a label that is given to each source so that it is obvious whether or not the source is: {\it S:} a single-Gaussian source that is unique on the island, {\it C:} a single-Gaussian source that is on an island with other sources, or {\it M:} a multi-Gaussian source.

It is technically possible to convert all emission into a series of Gaussians.  However, the fact that PyBDSF allows us to identify sources that exceed a threshold above the background contiguous mean makes it ideal in identifying {\it discrete} sources. The default threshold of the 3-$\sigma$ island and 5-$\sigma$ Gaussian source  is a good compromise between having so many `detections' as to be meaningless and so few detections that potentially real discrete sources are missed.

We provide further information about the PyBDSF inputs and tests in
 Appendix~\ref{a:param}, and also show typical images that are generated during the fitting process. { Note that PyBDSF computes uncertainties that take into account both the rms map noise (which includes random errors) as well as errors in the ellipse fitting, as given in the output of Appendix~\ref{a:c-radio}. When we compute the primary-beam-corrected flux densities and spectral index, we also include uncertainties in the primary beam and a 1\% calibration error on the primary flux calibrator, added in quadrature (Appendix~\ref{a:cross_L_C})}. Previous use of this program by others can be found in, for example, \cite{van14}, \cite{stw} and \cite{hale}.  The code and documentation are available at
https://www.astron.nl/citt/pybdsf/.


\subsection{X-ray Data and Analysis}\label{ss:Obs_xray}
The 27 galaxies in the sample are covered by 55 Chandra imaging observations. All but three of the observations were performed with the Advanced CCD Imaging Spectrometer S Array (ACIS-S), while the others were obtained by the ACIS-I Array
\citep{garmire03}. The observation log is reported in Table~\ref{t:obslog_xray}. 
The ACIS detector was operated in the standard timed exposure mode, with FAINT or VERY FAINT telemetry format. In five observations, the subarray mode was used. 
We re-processed the data with the Chandra Interactive Analysis of Observations
software package (\textsc{ciao}) version 4.12
and Calibration Database (\textsc{caldb}) version 4.9.1\footnote{http://cxc.harvard.edu/ciao/} \citep{Fruscione06}.
We followed standard data reduction procedures: after downloading the data from the public archive, we ran the \textsc{chandra\_repro} tool to perform all the recommended data processing steps. We used the \textsc{merge\_obs} tool to merge the event files of all of the observations for the same target with more than one exposure, create point spread function (PSF) maps and exposure maps for each observation and for the merged observations, and then created exposure-corrected images of the single and merged observations limited to the 0.3-7.0 keV energy range. The exposure-corrected images are restricted to the S2, S3, and S4 chips (CCD 6, 7, 8), with the exception of the three observations using ACIS-I, which were produced using all four ACIS-I chips along with one of the three ACIS-S chips listed above. For ObsID 3956 (NGC~2992) we removed a very bright readout streak using the \textsc{acisreadcorr} tool.

\begin{table*}
\caption{{\sl Chandra} Observation Log}
\label{t:obslog_xray}
\begin{tabular}{llccccc}
\hline
\hline 
Object & OBSID & RA (J2000)\tablenotemark{a} & Dec (J2000)\tablenotemark{a} & Roll Angle & Exposure\tablenotemark{b} & OBS Date \\
 &  & (h~~m~~s) & ($^\circ~~~^{\prime}~~~^{\prime\prime}$) & ($^\circ$) & (s) & (yyyy-mm-dd) \\
\hline \\
NGC 660 & 15333 & 1 43 03.1 & 13 38 40 & 287 & 22763 & 2012-12-18 \\
 & 15587 & 1 43 03.1 & 13 38 40 & 287 & 27698 & 2012-11-20 \\
 & 1633 & 1 43 03.8 & 13 36 44 & 291 & 1916 & 2001-01-28 \\
 & 18352 & 1 43 01.2 & 13 38 56 & 106 & 9994 & 2015-08-26 \\
 & 4010 & 1 43 02.2 & 13 38 00 & 292 & 5068 & 2003-02-12 \\
NGC 891 & 14376 & 2 22 34.2 & 42 20 19 & 263 & 1823 & 2011-12-20 \\
 & 19297 & 2 22 31.3 & 42 19 41 & 217 & 39540 & 2016-11-14 \\
 & 4613 & 2 22 29.9 & 42 19 26 & 253 & 118878 & 2003-12-10 \\
 & 794 & 2 22 22.5 & 42 20 48 & 187 & 50905 & 2000-11-01 \\
NGC 2683 & 11311 & 8 52 26.0 & 33 28 41 & 107 & 38529 & 2011-01-05 \\
 & 1636 & 8 52 44.2 & 33 26 48 & 72 & 1737 & 2000-10-26 \\
NGC 2992 & 3956 & 9 45 55.2 & -14 20 05 & 352 & 49548 & 2003-02-16 \\
NGC 3079 & 19307 & 10 01 56.2 & 55 40 43 & 154 & 53158 & 2018-01-30 \\
 & 2038 & 10 01 53.7 & 55 40 41 & 200 & 26578 & 2001-03-07 \\
 & 20947 & 10 01 56.4 & 55 40 43 & 154 & 44478 & 2018-02-01 \\
NGC 3432 & 7091 & 10 52 29.0 & 36 34 56 & 261 & 1929 & 2006-06-25 \\
NGC 3448 & 19360 & 10 54 37.5 & 54 18 02 & 181 & 9972 & 2018-03-04 \\
NGC 3556 & 2025 & 11 11 45.7 & 55 40 19 & 359 & 59365 & 2001-09-08 \\
NGC 3628 & 2039 & 11 20 18.2 & 13 35 52 & 71 & 57960 & 2000-12-02 \\
 & 395 & 11 20 20.3 & 13 36 52 & 60 & 1753 & 1999-11-03 \\
NGC 3877 & 1971 & 11 46 06.3 & 47 29 31 & 97 & 29180 & 2001-01-14 \\
 & 1972 & 11 46 09.4 & 47 29 06 & 27 & 28722 & 2001-10-17 \\
 & 767 & 11 46 06.6 & 47 29 32 & 92 & 18914 & 2000-01-10 \\
 & 768 & 11 46 02.7 & 47 29 06 & 174 & 23455 & 2000-03-07 \\
 & 952 & 11 46 06.9 & 47 28 21 & 293 & 19772 & 2000-08-01 \\
NGC 4013 & 4013 & 11 58 28.2 & 43 56 58 & 174 & 4899 & 2003-03-16 \\
 & 4739 & 11 58 26.5 & 43 59 06 & 113 & 79100 & 2004-02-03 \\
NGC 4096 & 19345 & 12 06 01.4 & 47 29 04 & 36 & 7963 & 2017-11-01 \\
 & 7103 & 12 05 52.0 & 47 30 21 & 135 & 1734 & 2006-02-18 \\
NGC 4157 & 11310 & 12 11 08.6 & 50 25 09 & 314 & 59255 & 2010-08-21 \\
NGC 4192 & 19390 & 12 13 48.1 & 14 54 27 & 51 & 14864 & 2017-11-04 \\
NGC 4217 & 4738 & 12 15 43.2 & 47 07 09 & 132 & 72728 & 2004-02-19 \\
NGC 4244 & 942 & 12 17 22.3 & 37 46 51 & 230 & 49215 & 2000-05-20 \\
NGC 4302 & 19392 & 12 21 42.0 & 14 35 28 & 209 & 14222 & 2018-04-09 \\
 & 19397 & 12 21 32.3 & 14 35 57 & 210 & 7808 & 2018-04-09 \\
NGC 4388 & 12291 & 12 25 46.4 & 12 39 53 & 61 & 27606 & 2011-12-07 \\
 & 1619 & 12 25 45.2 & 12 39 16 & 242 & 19977 & 2001-06-08 \\
NGC 4438 & 2883 & 12 27 46.4 & 13 01 05 & 79 & 25072 & 2002-01-29 \\
 & 8042 & 12 27 40.0 & 13 05 00 & 83 & 4895 & 2008-02-11 \\
NGC 4565 & 3950 & 12 36 20.5 & 26 01 19 & 94 & 59202 & 2003-02-08 \\
 & 404 & 12 36 17.5 & 25 57 39 & 249 & 2829 & 2000-06-30 \\
NGC 4594 & 1586 & 12 39 57.0 & -11 38 37 & 251 & 18518 & 2001-05-31 \\
 & 407 & 12 40 01.9 & -11 35 51 & 68 & 1766 & 1999-12-20 \\
 & 9532* & 12 39 57.9 & -11 39 07 & 257 & 84914 & 2008-04-29 \\
 & 9533* & 12 40 02.8 & -11 35 17 & 71 & 88975 & 2008-12-02 \\
NGC 4631 & 797 & 12 41 57.1 & 32 31 49 & 200 & 59215 & 2000-04-16 \\
NGC 4666 & 4018 & 12 45 09.8 & -0 27 11 & 71 & 4937 & 2003-02-14 \\
NGC 5084 & 12173* & 13 20 16.5 & -21 50 07 & 240 & 9923 & 2011-08-22 \\
NGC 5297 & 19370 & 13 46 25.1 & 43 52 37 & 4 & 9944 & 2017-10-25 \\
NGC 5775 & 2940 & 14 53 57.1 & 3 33 17 & 109 & 58213 & 2002-04-05 \\
NGC 5907 & 12987 & 15 15 57.6 & 56 18 17 & 86 & 15977 & 2012-02-11 \\
 & 14391 & 15 15 57.6 & 56 18 18 & 86 & 13086 & 2012-02-11 \\
 & 20830 & 15 16 00.6 & 56 18 24 & 353 & 51270 & 2017-11-07 \\
 & 20994 & 15 15 56.2 & 56 18 22 & 104 & 32621 & 2018-02-27 \\
 & 20995 & 15 15 55.9 & 56 18 24 & 104 & 16033 & 2018-03-01 \\
\hline
\tablecomments{All observations were obtained using the ACIS-S detector with the exception of the observations marked with (*), which were obtained by the ACIS-I detector. \tablenotemark{a} The coordinates correspond to the time-averaged location of the optical axis (the on-axis position). \tablenotemark{b} The exposure
represents the live (dead-time-corrected) time of each observation.}
\end{tabular}
\end{table*}

Source detection was performed with the \textsc{wavedetect} tool on a merged image of the data generated by \textsc{merge\_obs}. In order to optimize the output of each \textsc{wavedetect} run, we calculated the significance threshold for source detection as the inverse of the total number of pixels so that the expected number of false sources per field is one. We also increased the maximum number of source-cleansing iterations as well as the number of wavelet scales for a better estimate of the background map and to distinguish between sources varying in size, respectively. For galaxies NGC~4013 and NGC~4594, we also lowered the minimum relative exposure needed in a pixel in order to analyze it because the images had large differences in exposure time, leading to one observation dominating the others. 

We estimate the position error of each detected source by including both statistical and systematic contributions, which are added in quadrature. The statistical contribution is obtained directly from the \textsc{wavedetect} detection output, while the systematic one from $0.2\arcsec+1.4\arcsec(r/8^\prime)^2$, which is an analytical approximation of the result presented in \citet[Fig.~4 of][]{fei02}. Here $r$ is the off-axis angle. For a source detected in a merged image of involved observations,  $r$ is their exposure-weighted mean.

\section{Results}\label{s:Res}

We present the catalogue of our detected radio sources in  Appendix~\ref{a:c-radio}.  Appendix~\ref{a:cross_L_C} gives the results of the cross-matching of the radio sources at L-band and C-band.  The catalogue of X-ray sources is in Appendix~\ref{a:c-xray}. The catalogue of sources that have been cross-matched in L-band, C-band and X-rays is presented in Table~\ref{t:cross_LCX}.  H$\alpha$ figures of all galaxies with detected sources overlaid are in Appendix~\ref{a:figs_r}.  Tables and figures are available for download as described in these appendices.


\subsection{Radio Source Detections}

\begin{table*}
\begin{center}
\caption{Number and Fluxes of Detected Radio Sources}\label{t:number_L_C}
    \begin{tabular}{| l | c | c| c| c| c| c|c c c c}
    \hline
         Galaxy & \multicolumn{5}{c}{\bf L-band}       &\multicolumn{5}{c}{\bf C-band}  \\
                      & Number\tablenotemark{a} & Number\tablenotemark{b} & Deproj.\tablenotemark{c}&
                      Flux Density\tablenotemark{d} & F\_Fract.\tablenotemark{e} &
                      Number\tablenotemark{a} & Number\tablenotemark{b} & Deproj.\tablenotemark{c} &
                      Flux Density\tablenotemark{d} & F\_Fract.\tablenotemark{e}\\
                      &     Total&On-disc &On-disc &On-disc 
                      & On-disc & Total& On-disc&On-disc &On-disc & On-disc \\
                      &       & & (kpc$^{-2}$)& (mJy)& (\%)& & (kpc$^{-2}$)&(mJy)& (mJy) &(\%)\\
         \hline
         NGC~660 & 51  & 1 & 0.00996 & 399.2 &76 &20 & 4 & 0.03984&608.0 &92\\
         NGC~891 & 69  & 9 &0.00897& 102.2&14 &43 & 15 & 0.01496&34.46&17\\
         NGC~2613 & 55  & 7&0.00367 &3.472& 5.8& 25& 10 & 0.00524&2.896&19\\
         NGC~2683 & 76 & 9 & 0.03957& 3.72& 5.6&20 & 9 & 0.03957&3.173&16\\
         NGC~2820 & 65 &2 &0.00259 & 9.521& 15&25 & 3& 0.00388&4.36&23\\
         NGC~2992 & 61 & 3 &0.00310 & 192.0 &94 &25 & 2 & 0.00207&77.63&97\\
         NGC~3003 & 89 & 4& 0.00282 & 3.024& 8.7&41 & 16 & 0.01129&3.190&30\\
         NGC~3044 & 65 & 2 &0.00304 & 44.080&42 &13 &2 &0.00304&14.34&38\\
         NGC~3079 & 34 & 1 &0.00056 &496.9& 61& 27& 7 & 0.00394&305.8&84\\
         NGC~3432 & 59 & 3& 0.01113&8.220& 9.9 &37&10 &0.03711&7.381&28\\
         NGC~3448 & 76 & 5& 0.00397& 20.56&45 & 44 & 9 & 0.00714&10.10&49\\
         NGC~3556 & 67 & 14 & 0.01399 &36.00 & 12&54 &30&0.02997&27.44&35\\
         NGC~3628 & 58 & 10 &0.00951 & 233.9 &44 &56 & 32 & 0.03043&113.0&61\\
         NGC~3735 & 75 & 2& 0.00098& 13.54 &17 &24 & 4 & 0.00196&9.962&40\\
         NGC~3877 & 128 & 13& 0.02064& 7.373 & 17& 50&20&0.03176&4.985&39\\
         NGC~4013 & 78 &1 & 0.00213& 10.11&27 & 24 & 2 & 0.00427&8.142&65\\ 
         NGC~4096 & 70 & 9& 0.02910& 1.598&2.8 & 47&27&0.08731&3.684&23\\ 
         NGC~4157 & 92 &4  & 0.00541& 5.516 & 3.0&30 &8&0.01082&2.595&4.7\\
         NGC~4192 & 78 & 9& 0.00773&15.15 & 19& 46&23&0.01975&8.441&35\\
         NGC~4217 & 90 &2  & 0.00257& 12.17 & 11&30 & 3 & 0.00386&6.327&18\\
         NGC~4244 & 105 & 17 & 0.04795& 5.077 & 28& 100&48 & 0.13539&4.616&51\\
         NGC~4302 & 79 & 4 &0.00528 &4.672 & 10&32&9&0.01188&6.262&52\\
         NGC~4388 &  36  & 2&0.00346 &77.41 & 59& 27&4&0.00692&37.61&61\\ 
         NGC~4438 & 30 & 4& 0.00770& 75.39 & 57& 32&10&0.01925&30.75&56\\ 
         NGC~4565 & 85  &12 & 0.00504& 6.540 & 4.3& 72&33&0.01387&8.529&20.2\\
         NGC~4594 & 70 & 6& 0.00738& 73.58& 79&74 & 28 & 0.03443&129.6&100\\
         NGC~4631 & 70 & 18& 0.02059& 115.8 & 11& 90 & 62 & 0.07091&80.69&28\\ 
         NGC~4666 &  66 &3 &0.00286& 41.59&10 & 25&3& 0.00286&30.47&24\\
         NGC~4845 & 54 &  4&0.00832&237.3  & 100&27 &4&0.00832 &361.7&84\\
         NGC~5084 & 62 & 2 & 0.00063 &34.07 &84 &35 & 7 & 0.00221&33.73&93\\
         NGC~5297 & 84 &  6&0.00175 & 0.9376 & 3.8&39& 17 & 0.00496&1.732&26\\
         NGC~5775 & 90 & 1& 0.00104& 33.59& 13& 27& 1&0.00104&24.63&33\\
         NGC~5792 & 65 & 4 & 0.00125& 28.27& 49& 46&15&0.00469&13.61&67\\
         NGC~5907 & 99 & 12 & 0.00403 &68.09 & 38&69&27&0.00907&19.84&39\\
         UGC~10288\tablenotemark{f}  & 76 & 0& 0.000  & 0.000& ---& 37&4&0.00226&12.13&---\\
         \hline
    \end{tabular}
\end{center}
{\tablenotetext{\rm a}{~Total number of sources detected per galaxy over the entire field (field sizes are given in Table~\ref{t:fieldsize}).}}
{\tablenotetext{\rm b}{~Total number of sources on the projected optical disc whose dimensions are given in Table~\ref{t:g_param}, and shown by the green ellipses in the figures of Appendix~\ref{a:figs_r}.}}
{\tablenotetext{\rm c}{~Total number of disc sources per square kpc, for a deprojected disc size.  The deprojected disc area is taken to be $\pi\,a^2$, where $a$ is the galaxy's semi-major axis, in kpc.}}
{\tablenotetext{\rm d}{~Total flux density of all sources on the projected disc. The flux density of each source (as in Column 7 of Table~\ref{t:s_n4565_lb}) was corrected for the primary beam (Column 35) and then the corrected flux densities of all sources on the disc were summed. Uncertainties can be read from Table~\ref{t:s_n4565_lb} { and the reader can adjust significant figures accordingly}.}}
{\tablenotetext{\rm e}{~Fraction of the galaxy's on-disc flux that is contained in compact sources, calculated from the on-disc flux fensity divided by the total flux of the galaxy, as given in \cite{wie15}. }}
{\tablenotetext{\rm f}{~The radio emission from this galaxy is dominated by a strong background source \citep{irw13} whose centre is just at the boundary of the disc. The centre of the background source at L-band is just outside of the disc and the centring at C-band is just inside of it. We therefore exclude this galaxy when calculating F\_Fract.}}
\end{table*}

We detect a total of 2507 and 1413 sources at L-band and C-band, respectively (Table~\ref{t:number_L_C}).
The number of detected sources depends on a number of factors as we now discuss.

The galaxy that shows the lowest number of sources at L-band is NGC~4438, with only 30 sources.
The main reason for this low number is the higher rms of this field compared to other galaxies, at 25 $\mu$Jy beam$^{-1}$ (Table~\ref{t:g_param}).  Consequently, there are fewer off-galaxy background sources showing up.  There is some variation in the number of background sources from field to field, in any event.  For example, NGC~4157 and NGC~3877 have similar rms values, but NGC~3877 shows more off-galaxy sources. Such variation is to be expected, given the different background environments from galaxy to galaxy.  

Galaxies of smaller angular size and/or smoother emission in which individual sources are less likely to be distinguished compared to the background disc emission will also show fewer sources. For example, NGC~4565 is a galaxy that has a large angular size and `patchier' emission than NGC~3044 (both having the same rms).  Few discrete sources on the disc of NGC~3044 have been detected in comparison to NGC~4565.  This is apparent in the figures of these two galaxies in Appendix~\ref{a:figs_r}.  It is important to keep in mind that the radio continuum intensity distribution does not necessarily follow the appearance of the H$\alpha$ emission that is shown as greyscale in the figures.

At C-band, the arguments are similar. In this band, the rms noise is lower than at L-band, favouring the detection of more sources. However, the primary beam is only 26\% as large as at L-band, so the adopted field size at C-band is 52\% smaller than at L-band (Table~\ref{t:fieldsize}). Consequently, 
the total number of sources detected at C-band is only 56\% of the total number of sources detected at L-band.  

Additional differences depend on the spectral indices of the sources, which favour L-band detections for steep spectrum sources (see also Sect.~\ref{ss:cross_L_C} in which we indicate the spectral index at which the sensitivity in L-band and C-band are equal).

\begin{figure}
   \centering
   \includegraphics[width=0.5\textwidth]{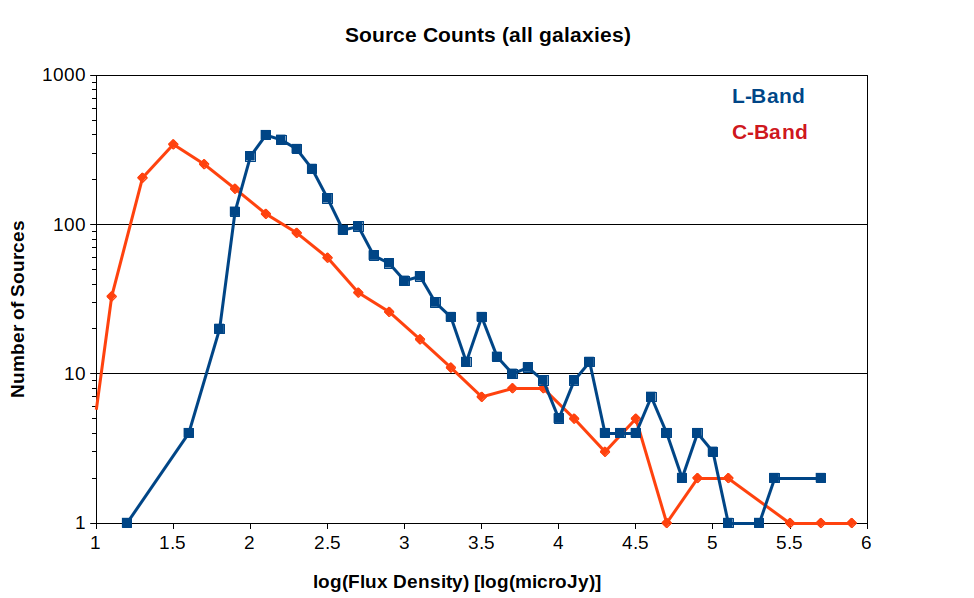}
   \caption{Number of detected radio sources at L-band and C-band.  All galaxies are included.  At L-band and C-band, the flux density divisions are separated by 0.1 dex and 0.2 dex, respectively. The median flux density at L-band is 163 $\mu$Jy (2.2 in the log), and at C-band, it is 39.2 $\mu$Jy (1.6 in the log). All values are primary-beam-corrected.}
  \label{f:sourcecounts}
\end{figure}

Figure~\ref{f:sourcecounts} shows the source counts for all sources detected. The median flux density of sources at L-band is 163 $\mu$Jy and the median at C-band is 39.2 $\mu$Jy.  The higher flux density, on average, for L-band sources compared to C-band sources is consistent with a steep non-thermal spectral index in most cases (Sect.~\ref{ss:cross_L_C}).

In Table~\ref{t:number_L_C}, we also list the number of sources within the optical disc of the galaxy in projection (`On-disc') at both bands, as determined by the galaxy disc size given in Table~\ref{t:g_param}.  Because the radio sources do not suffer from extinction, we also compute the number of sources per kpc$^2$ for each deprojected galaxy, 
i.e. as if the galaxy were face-on (`Deproj.'). The total flux density on-disc is also given by summing the flux density of each source over the disc region.  

{The column labelled, F\_Fract., lists the fraction of the flux that is contained in compact sources compared to the galaxy's total flux, the latter from \cite{wie15}.  
Excluding UGC~10288 (see table notes), the median fraction at L-band is 17\% and the median fraction at C-band is 38\%.  In general, therefore, the compact sources constitute a small fraction of the galaxy's total flux\footnote{For our largest galaxies, in addition, the total flux may be underestimated, especially at C-band, because of possible missing zero-spacing flux \citep[see][]{wie15}.}.  Thus the fraction could be slightly lower in a few cases.  

There are, however, some exceptions.  For example, those galaxies for which compact sources account for more than 75\% of the total flux at L-band are: 
NGC~660, NGC~2992, NGC~4845, and NGC~5084. At C-band, they are: NGC~660, NGC~2992, NGC~3079, NGC~4594, NGC~4845, and NGC~5084.  Each of these galaxies are well-known AGN galaxies in which the nuclear source dominates the flux density of the compact sources. }

Over all galaxies, we detect 205 on-disc sources at L-band and about 2.5 $\times$ more at C-band, i.e. 508. Since on-disc sources are generally centred in the field of view (not as down-weighted by the primary beam), and since most sources show a steep spectral index with lower flux at C-band than L-band, this suggests that it is the lower noise level in C-band that results in the higher number of sources detected.

\begin{figure}
    \centering
    \includegraphics[width=0.4\textwidth]{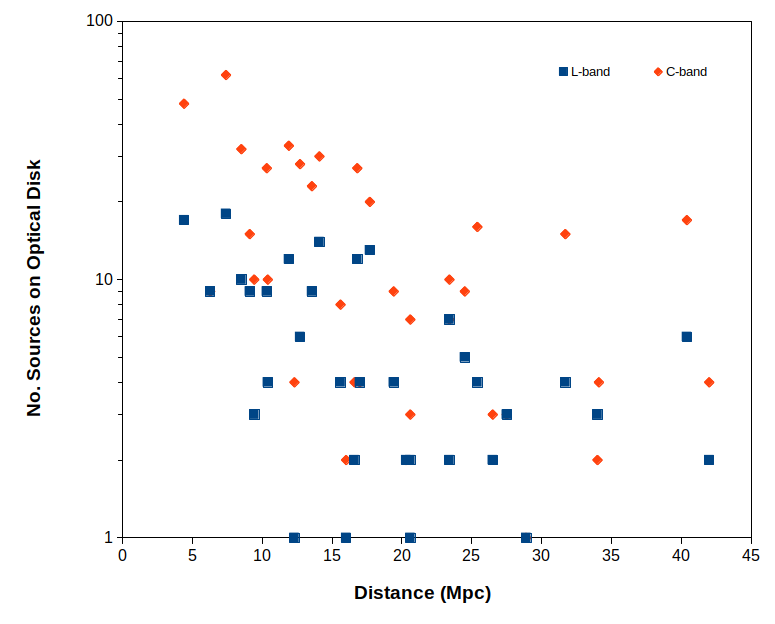}
    \caption{Number of detected radio sources on the galaxy disc as a function of galaxy distance. }
    \label{f:rad-dist}
\end{figure}
\begin{figure}
    \centering
    \includegraphics[width=0.48\textwidth]{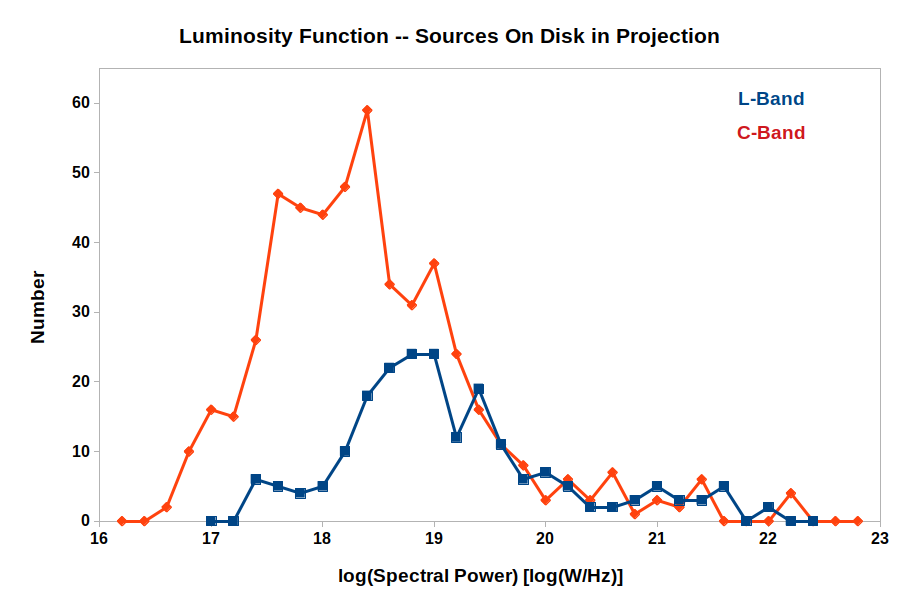}
    \caption{Number of detected radio sources on the galaxy disc in projection, as a function of spectral power at L-band (blue) and C-band (red). The x-axis values were computed via $L_\nu\,=\,4\,\pi\,D^2\,F_\nu$, where $F_\nu$ corresponds to Total\_Flux (Column 7), corrected for the primary beam (Column 35) in Table~\ref{t:s_n4565_lb}, and $D$ is the distance. The median uncertainty in power is 19\% at L-band and 15\% at C-band. The bin width is 0.2 dex. }
    \label{f:rad-lum}
\end{figure}

Figure~\ref{f:rad-dist} shows the number of sources detected on the galaxy disc as a function of galaxy distance. A decline in the number of sources would suggest that PyBDSF was detecting more discrete sources in the nearby galaxies because of the higher resolution.  We see that this is hinted at, but it appears to be a minor effect. There are still galaxies at larger distances with many sources. Note that the ordinate axis is represented as a logarithmic plot because the purpose of the figure is to check for any distance dependence; the differences between L-band and C-band numbers are therefore not as obvious as in Figure~\ref{f:rad-lum} (below).

We finally plot the luminosity function for all sources that are on the galaxy disc (in projection -- {see Sect.~\ref{sec:ondisk_background}}) in Figure~\ref{f:rad-lum}. Consistent with Table~\ref{t:number_L_C}, the figures of Appendix~\ref{a:figs_r}, and comments above, there are more on-disc sources at C-band than at L-band. {Of the reasons offered for differences given above, the most likely explanation for this is the lower rms at C-band because the on-disc sources are clustered around the centres of the fields where primary beam effects are minimal.}

The median spectral power is $L_{L-Band}\,=\,7.56\,\times\,10^{18}$ W/Hz
(log$L_{L-band}\,=\,18.9$), and
$L_{C-Band}\,=\,1.61\,\times\,10^{18}$ W/Hz
(log$L_{C-band}\,=\,18.2$). Again, this is consistent with steep spectral indices, on average. By comparison, the median L-band flux density of a collection of SNRs in M~82 is $L_{L-Band}\,=\,3\,\times\,10^{18}$ W/Hz \citep{irw19b} and 
$L_{C-Band}\,=\,1.6\,\times\,10^{18}$ W/Hz \citep{mux94}.  These values are quite comparable and suggest that most detected sources represent SNRs, although nuclear sources are also included in our sample (Sect.~\ref{sec:nuclear}).

\cite{mascoop2021} derived the 1.4 GHz luminosity function of 797 HII regions in the first Galactic quadrant, with the most luminous HII regions around $L_{L-Band}\,=\,2\,\times\,10^{16}$ W Hz$^{-1}$, corresponding with the ionizing flux of an early O star. Clearly, single star HII regions are well below the sensitivity of CHANG-ES in either L or C band. The thermal free-free luminosity of the 13 largest star formation regions in the Milky Way outside the Galactic centre, measured at 90 GHz and reduced to 1.5 GHz using a spectral index $\alpha_{ff} = -0.1$, is in the range $L_{L-Band}\,=5\,\times\,10^{18}$ W Hz$^{-1}$ to $L_{L-Band}\,=1\,\times\,10^{19}$ W Hz$^{-1}$ with the corresponding C-band luminosity $15\%$ lower \citep[using data listed in][]{rahman2010}. The 30 Dor region in the Large Magellanic Cloud with a thermal radio luminosity $L_{L-Band}\,=1\,\times\,10^{19}$ W Hz$^{-1}$, derived from the 22 GHz flux density in \citet{sabalisck1991} for a distance of 49.6 kpc, is on the extreme luminosity side of Local Group HII regions. These values are near the peak of the distributions in Figure~\ref{f:rad-lum}.

Resolution is a factor in determining the luminosity of the brightest HII regions. 
Substructure may lead to subdivision of larger structures in nearby galaxies. 
\citet{ken89} found that the HII region luminosity function is fairly independent of resolution up to a resolution of 200 pc, while the effects of blending became apparent at resolutions of more than 300 to 500 pc. The linear resolution of the C-band data exceeds 300 pc in 13 galaxies, but it exceeds 500 pc in only 3 galaxies. At this scale, it is likely that SNRs and HII regions are blended as well. Arguably, blending is more significant in an extinction-free tracer applied to edge-on galaxies. Also, the radio flux density may be confused with brighter diffuse non-thermal emission from the spiral arms. If this is a factor, the most luminous sources should have a steeper spectrum than the lower luminosity sources. Resolution also affects the contribution of diffuse ionized gas, but this is effectively subtracted by the source finder. {We have suggested above, however, that most sources have fluxes that are typical of SNRs and will also see in the next section that the sample is dominated by steep spectrum sources. Therefore, the sources found in this paper are generally not dominated by HII regions.}

{In summary, the observed peak of the distribution of Figure~\ref{f:rad-lum} is consistent with either SNRs or HII regions. However, we show in the next section, that most spectral indices are consistent with non-thermal emission. The high luminosity tail, which is two to three orders of magnitude higher than the luminosity at the peak of the distribution, is influenced by some relatively strong AGNs.} 


\subsection{Cross-matching of Radio Sources between L-band and C-band}\label{ss:cross_L_C}

In Appendix~\ref{a:cross_L_C}, we provide a description of the catalogue of all radio sources that have been positionally matched between L-band and C-band. The catalogue is available for download, as described in the appendix.

The cross-matching was carried out by Python programming language scripts developed by the authors which looked for sources in both bands that were within a radius of $1$ arcsec of each other.   The choice of $1$ arcsec was based on the typical positional error in RA and Dec for each of the detected sources. 

A total of 723 sources were found to be positionally matched at both bands over all 35 galaxies.  Of these, 684 cross-matched sources had good PB-corrected flux densities. The reason for unreliable flux densities was if the source was located far from the map centre where the PB fell below 11\%. In such cases,  we did not accept the resulting PB-corrected flux.  However, the positions are still acceptable so the cross-matched sources are still included in the catalogue without flux values.

For each of the matched sources, we further calculate the radio spectral index ($\alpha$) between L-band and C-band, according to
\begin{equation}
    \frac{S_L}{S_C}=\left(\frac{\nu_L}{\nu_C}\right)^\alpha \label{eq:alpha}
\end{equation}
where $S_L$ and $S_C$ are the primary-beam-corrected source flux densities in L-band and C-band, respectively, and $\nu_L$ and $\nu_C$ are their centre frequencies, respectively. Standard error propagation was carried out in quadrature { including uncertainties in the primary beam and a 1\% calibration uncertainty (see Appendix~\ref{a:cross_L_C}).} 


The average spectral index and average spectral index error are $\alpha_a\,=\,-0.58$ and $\pm\,0.18$, and the median values are $\alpha_m\,=\,-0.67$ and $\pm\,0.18$. {These spectral indices have not been corrected for any differences in size between L-band and C-band. There could be minor modifications to the spectral index values if, for example, the deconvolved source sizes (Columns 21 - 26 in Table~\ref{t:s_n4565_lb}) differ between L-band and C-band.}  However,  the average and median indicate that
the sources are clearly dominated by non-thermal emission, as suggested previously in Figs.~\ref{f:sourcecounts} and \ref{f:rad-dist}. 
{Note that, for median rms values of 15 $\mu$Jy beam$^{-1}$ at L-band and
3.2 $\mu$Jy beam$^{-1}$ at C-band, the $3\sigma$ sensitivity in both bands corresponds to a spectral index of $\alpha_{3\sigma}\,=\,-1.17$.  That is, the two bands have equal sensitivity at a spectral index of $\approx\,-1$. }

{The maximum spectral index in the entire sample is 
$\alpha_{max}\,=\,+0.91\,\pm\,0.02$ in the NGC~2683 field, and the minimum spectral index in the entire sample is
$\alpha_{min}\,=\,-1.93\,\pm\,0.10$ in the NGC~5775 field.  Both of these sources are not on the discs of the galaxies, in projection, and are likely background sources.}

We further separately identify sources that are within 3 arcsec ($\approx$ the spatial resolution)
of the nucleus of the galaxy, as listed in Table~\ref{t:g_param}. {We label these `nuclear sources', by virtue of their position only.}
Each of these sources was then compared with the CHANG-ES galaxies for which there was evidence of an AGN, as given in Table~10 of \cite{irw19b}. That paper identified 18 AGN candidates using a variety of criteria and using B/L images only, plus the B/L to C/C spectral index at the galaxy core.    Using PyBDSF extraction, we now find three new galaxies for which there are coincident L-band and C-band sources right at the nucleus {and whose deconvolved major axis size is less than 15 arcsec (5 $\times$ the spatial resolution). The size limit was imposed in order to exclude any galaxy whose emission simply peaks towards the nucleus in a more gradual, smooth fashion\footnote{An example is NGC~3044.}. These new nuclear sources are listed in Table~\ref{t:new_AGNs}.}

The new, tabulated nuclear source galaxies have not previously been identified as AGN candidates, to our knowledge.  However, recently,
\cite{yan21} have independently argued that the galaxy, NGC~5792, harbours an AGN. {Moreover, NGC~3877 appears to be variable, which is a strong argument for the presence of an AGN.  \citet{ulv02}, in a search for weak AGNs in galaxies at C-band, did {\it not} detect a nuclear source in NGC~3877 to an rms limit of 70 $\mu$Jy beam$^{-1}$ with a resolution of approximately one arcsec.  We measure 2 mJy for the core of this galaxy (Table~\ref{t:new_AGNs}) which is $\approx$ 30$\sigma$ of the Ulvestad \& Ho noise.  If the flux of the nuclear source had been constant, \cite{ulv02} would have easily detected this source. Therefore, between the time of the Ulvestad \& Ho observations (Oct. 28, 1999) and the CHANG-ES observations (April 4th 2012), the core of NGC~3877 has brightened at C-band by at least a factor of 6, if we take 5$\sigma$ as a detection limit. The CHANG-ES core spectral power is $L_{C-band}\,=\,7.5\,\times\,10^{19}$ W/Hz. {The spectral index of the NGC~3877 core is also very flat ($\alpha_{LC}\,=\,-0.092$, Table~\ref{t:cross_LCX}) which is typical of an AGN. Although the CHANG-ES L-band and C-band observations are not at the same time, only  74 days separate the two observations, making it unlikely that the spectral index has changed significantly. }
}

{These new nuclear sources warrant further investigation to confirm their nature. 
Note that the sources are resolved, suggesting that if AGNs are indeed embedded in these galaxy nuclei, other more extended emission is also present. 
 In Sect.~\ref{sec:nuclear}, we indicate which nuclear sources also show X-ray emission. These may be 
 `low-luminosity AGNs' (LLAGNs). For more discussion of LLAGNs, see \cite{irw19b}. }

\begin{table*}
\caption{New Nuclear Sources -- Radio}\label{t:new_AGNs}
    \begin{tabular}[b]{ c  c  c  c  c  c  c c}
    \hline\hline
         Galaxy & LC\_id\tablenotemark{a} & Source\_id\tablenotemark{b} & Flux$_{\rm cor}$\tablenotemark{c} & $\alpha$\tablenotemark{d}& DC$_{\rm Maj}$\tablenotemark{e} & DC$_{\rm Min}$\tablenotemark{e} & DC$_{\rm PA}$\tablenotemark{e}
\\
          & & &  ($\mu$Jy)  & & (arcsec)&(arcsec) & (deg) \\
         \hline\hline
         NGC~3877& 278&  & &  $-0.092\,\pm\,0.037$& 
\\
        ~~~~L-band& & 73  & $2367\,\pm\,68$   & & $4.06\,\pm\,0.09$ & $1.53\,\pm\,0.04$ & $32.3\,\pm\,1.9$
        \\
        ~~~~C-band& & 23  & $2093\,\pm\,81$   & & $4.49\,\pm\,0.06$ & $2.58\,\pm\,0.04$ & $29.1\,\pm\,1.7$
        \\
 \hline
           NGC~4192& 364&  & &  $-0.672\,\pm\,0.019$& 
\\
        ~~~~L-band& & 33  & $13370\,\pm\,275$   & & $9.61\,\pm\,0.06$ & $4.26\,\pm\,0.03$ & $155.6\,\pm\,0.5$
        \\
        ~~~~C-band& & 22  & $5434\,\pm\,73$   & & $9.39\,\pm\,0.04$ & $3.68\,\pm\,0.02$ & $153.4\,\pm\,0.3$
        \\ 
    \hline
           NGC~5792& 663&  & &  $-0.619\,\pm\,0.015$& 
\\
        ~~~~L-band& & 30  & $27102\,\pm\,389$   & & $12.57\,\pm\,0.10$ & $3.61\,\pm\,0.03$ & $79.0\,\pm\,0.5$
        \\
        ~~~~C-band& & 23  & $11828\,\pm\,153$   & & $12.10\,\pm\,0.06$ & $3.72\,\pm\,0.02$ & $80.0\,\pm\,0.4$
        \\ 
   \hline
%
    \end{tabular}
    \tablecomments{Galaxy information has been extracted from the tables of Appendices~\ref{a:c-radio} and \ref{a:cross_L_C}. \\
    \tablenotemark{a} {~Unique L-band to C-band cross-identification number.} \\
    \tablenotemark{b} {~Unique source identification number for the specified galaxy and band.}\\
    \tablenotemark{c} {~Source flux, corrected for the primary beam, and its error.}\\
    \tablenotemark{d} {~Spectral index of the source, and its error.}\\
    \tablenotemark{e} {~Major axis, minor axis, and position angle (North through East) of the source, after deconvolution from the beam. }
    }
    \end{table*}

\subsection{X-ray sources}\label{ss:Res_xray}

We present a catalogue of our detected X-ray sources in Appendix~\ref{a:c-xray}. The count rates are calculated in the 0.3-1.2~keV (Soft, Sft), 1.2-2~keV (Medium, Med), 2-7~keV (Hard, Hrd) and 0.3-7~keV (Broad, Brd) bands, while the hardness ratios, defined as HR1 = (Med-Sft)/(Med+Sft) and HR2 = (Hrd-Med)/(Hrd+Med), are listed for sources with broad counts greater than 10.
X-ray sources are marked with blue crosses in the Figures of Appendix~\ref{a:figs_r}.



\subsection{Cross-matching of Radio and X-ray Sources}\label{ss:Obs_radio_xray}


We cross-match the positions of the radio sources {that have already been cross-matched in the two radio bands,} with the  X-ray sources. L-band and C-band cross-matches are within 1 arcsec, as described in Sect.~\ref{ss:cross_L_C}. For the X-ray sources, the cross-match uses a matching radius of 3 $\times$ the uncertainty radius of the X-ray sources (3 times delta\_x, as given in the table of Appendix~\ref{a:c-xray}). A typical value of delta\_x is $\approx$ 0.5 arcsec. 

We present the results in their entirety in Table~\ref{t:cross_LCX} and also provide this table in  downloadable csv format associated with this paper, called {Table\_6\_LCX\_cross\_matched.csv}.

This table gives the identification numbers of the L-band, C-band and X-ray sources, as well as assigns a unique number for all sources for which cross-matches occurred.  The average RA and Dec of the sources, and the spectral index between L-band and C-band are specified.  The distance
 from the map centre and angle from x=0 (due west, counter-clockwise) are also listed to aid in identifying the sources on the maps of Appendix~\ref{a:figs_r}. Note that NGC~4666 is missing because there were no successful cross-matches for that galaxy.



 \subsubsection{Nuclear Sources -- All Bands}\label{sec:nuclear}

Sources in Table~\ref{t:cross_LCX} were flagged as `nuclear' if the average distance of the cross-matched L-band, C-band and X-ray source is within 3.0 arcsec of the galaxy centre (approximately the radio synthesized beam's full-width at half-maximum). This tolerance is slightly larger than the 1.5 arcsec radius used in Sect.~\ref{ss:cross_L_C} for the nuclear sources based on radio data only. When the X-ray sources are included, somewhat higher positional uncertainties are tolerated.

With only two exceptions, each of the nuclear sources has been listed previously as an AGN candidate in 
\cite{irw19b}.
The two nuclear sources that were not previously identified as AGN candidates are
NGC~4192 and NGC~3877, both of which were identified as `nuclear' in Table~\ref{t:new_AGNs} from radio data alone\footnote{The third source in that table, NGC~5792, had no X-ray data and so does not appear in Table~\ref{t:cross_LCX}.}.   As indicated in Sect~\ref{ss:cross_L_C}, this is the first evidence for a compact nuclear source in these two galaxies.




\subsubsection{On-disc versus Background Sources}
\label{sec:ondisk_background}
In Table~\ref{t:cross_LCX} we also list whether or not the source falls on the disc.  As before, the interpretation of a source being `on-disc' simply means that the source is within the green ellipse as displayed in the figures of Appendix~\ref{a:figs_r} and listed in Table~\ref{t:g_param}.  That is, the source is `in the disc' in projection.

It is clear that most cross-matched sources are likely background sources in the field. Of all 75 sources, only 19 fall on the discs of the galaxies.

\clearpage
\onecolumn
\begin{center}
{\setlength\tabcolsep{2pt}
  \begin{longtable}{lcccccccccc}
\caption{
Cross-matched Radio and X-ray Sources}\label{t:cross_LCX}\\
\hline
\hline
 Galaxy& ID$\_$LCX$^{\rm a}$& ID$\_$L$^{\rm b}$&ID$\_$C$^{\rm b}$&ID$\_$X$^{\rm c}$&RA\_avg$^{\rm d}$&Dec\_avg$^{\rm d}$&alpha\_{LC}$^{\rm e}$  & Dist$\_$to$\_$center$^{\rm f}$&Angle$\_$from$\_$x=0$^{\rm f}$& Position$^{\rm g}$\\
&&&&&(deg) & (deg) & & (arcmin)&(deg)&\\
\hline
\hline
\endfirsthead
\hline
\multicolumn{3}{l} {\bf Continued from previous page}\\
\hline
\hline
 Galaxy& ID$\_$LCX$^{\rm a} $&ID$\_$L$^{\rm b}$&ID$\_$C$^{\rm b} $&ID$\_$X$^{\rm c}$&RA\_avg$^{\rm d}$&Dec\_avg$^{\rm d}$&alpha\_{LC}$^{\rm e}$ &Dist$\_$to$\_$center$^{\rm f}$&Angle$\_$from$\_$x=0$^{\rm f}$& Position$^{\rm g}$\\
&&&&&(deg) & (deg) & & (arcmin)&(deg)\\
 \hline 
 \hline
\endhead
         \hline
NGC~660&0&11&2&36&25.8189&13.6334&-0.738&3.5&191.5& \\
NGC~660&1&20&13&19&25.7597&13.6458&0.313&0.05&68.7& nuclear\\
NGC~660&2&27&15&41&25.7449&13.6824&0.572&2.41&68.5& \\
\hline
NGC~891&0&18&4&92&35.707&42.3332&-0.050&3.15&197.7& \\
NGC~891&1&22&10&97&35.6908&42.3803&-1.196&2.95&140.7& \\
NGC~891&2&30&23&137&35.6623&42.3983&-0.835&3.13&109& on-disc \\
NGC~891&3&35&26&87&35.6358&42.4389&0.382&5.39&88.4& \\
NGC~891&4&39&30&82&35.6129&42.3934&-0.960&2.91&66.2&  \\
NGC~891&5&43&33&50&35.6018&42.3606&-0.821&1.8&22.6& \\
NGC~891&6&52&36&69&35.565&42.323&-0.816&3.64&334.6& \\
NGC~891&7&57&39&133&35.5519&42.3246&-1.156&4.14&339.2& \\
\hline
NGC~2683&0&41&11&1&133.1721&32.42178&-0.058&0.01&22.1&nuclear\\
\hline
NGC~2992&0&18&2&8&146.4831&-14.3539&-0.680&3.76&206& \\
NGC~2992&1&22&4&4&146.4515&-14.3682&-0.679&2.94&238.6& \\
NGC~2992&2&32&12&13&146.4246&-14.3266&-0.685&0.02&332.8&nuclear\\
NGC~2992&3&32&12&14&146.4247&-14.3264&-0.685&0.02&332.8&nuclear\\
NGC~2992&4&40&19&40&146.3932&-14.2766&0.511&3.51&58.3& \\
NGC~2992&5&41&22&35&146.3905&-14.2574&-0.673&4.6&64.1& \\
\hline
NGC~3079&0&9&0&108&150.6108&55.7318&0.116&5.11&142.4& \\
NGC~3079&1&21&22&94&150.4293&55.7419&0.441&4.26&60.8& \\
\hline
NGC~3432&0&20&11&3&163.1411&36.6043&-0.846&1.03&237.7& \\
\hline
NGC~3448&0&31&16&23&163.7081&54.3669&-0.954&4.01&112.8& \\
\hline
NGC~3556&0&26&17&22&167.9123&55.684&-0.557&1.27&152.3& on-disc\\
NGC~3556&1&28&20&25&167.9089&55.7101&-0.703&2.38&115& \\
NGC~3556&2&31&24&24&167.893&55.6969&-0.883&1.45&109& \\
NGC~3556&3&36&33&62&167.8531&55.6782&-0.189&0.91&15.5& on-disc\\
NGC~3556&4&43&40&27&167.8242&55.6712&-0.384&1.86&354.6& on-disc\\
NGC~3556&5&45&42&41&167.819&55.6932&0.156&2.33&29.5& \\
\hline
NGC~3628&0&17&20&25&170.1042&13.6078&-0.692&2.22&150.8& \\
NGC~3628&1&37&47&49&170.0118&13.5978&-0.275&3.48&8.00& on-disc\\
\hline
NGC~3877&0&54&8&75&176.568&47.4783&-0.734&1.75&214.5& \\
NGC~3877&1&71&21&66&176.5362&47.5792&-0.844&5.07&91.7& \\
NGC~3877&2&73&23&25&176.5322&47.4945&-0.092&0.02&310.7&nuclear\\
NGC~3877&3&81&32&1&176.5187&47.4447&-0.657&3.05&280.6& \\
\hline
NGC~4013&0&25&2&95&179.7041&43.9498&-0.773&3.18&176.4& \\
NGC~4013&1&32&6&1&179.6607&43.9183&0.192&2.13&232.7& \\
NGC~4013&2&62&22&91&179.5331&43.997&-0.924&5.19&35.7& \\
\hline
NGC~4096&0&17&2&15&181.5938&47.4947&-0.049&3.74&164.9& \\
\hline
NGC~4157&0&34&6&7&182.8259&50.4478&-0.995&3.12&225.2& \\
NGC~4157&1&58&23&17&182.7107&50.4625&-1.048&2.57&328.8& on-disc\\
\hline
NGC~4192&0&24&8&18&183.4844&14.9387&-0.827&3.01&130& \\
NGC~4192&1&33&22&3&183.4512&14.9005&-0.672&0.01&96.8&nuclear\\
NGC~4192&2&41&36&13&183.4192&14.8915&0.065&1.93&344.1& \\
\hline
NGC~4217&0&64&26&53&183.8763&47.1028&-0.819&3.56&10.6& \\
\hline
NGC~4244&0&34&25&9&184.4174&37.7971&-0.723&2.16&196.2& \\
NGC~4244&1&54&52&3&184.3648&37.7369&-0.413&4.24&275.6& \\
\hline
NGC~4302&0&40&4&8&185.4427&14.6159&0.505&1.39&130.9& \\
NGC~4302&1&56&25&3&185.3865&14.6062&-0.383&2.4&11.2&  nucleus of N~4298\\
NGC~4302&2&58&26&24&185.3823&14.6351&-0.855&3.41&40.3&\\
\hline
NGC~4388&0&5&0&50&186.5368&12.7262&&6.62&144.5& \\
NGC~4388&1&12&6&36&186.4859&12.6053&-1.132&4.17&234.8& \\
NGC~4388&2&20&15&34&186.4529&12.7403&-0.827&4.72&95.8& \\
NGC~4388&3&21&17&3&186.4472&12.6228&-1.041&2.36&266.5& \\
\hline
NGC~4438&0&8&2&12&186.9883&13.0425&-0.522&3.47&144.5& \\
NGC~4438&1&9&5&49&186.9768&13.0216&-0.299&2.28&160.3& \\
NGC~4438&2&15&18&7&186.9401&13.0089&-0.756&0.01&162.2&nuclear\\
NGC~4438&3&16&19&21&186.934&12.976&-0.313&2.00&280& on-disc\\
NGC~4438&4&18&24&29&186.9188&13.079&-0.522&4.39&73.6& nucleus of N~4435\\
\hline
NGC~4565&0&66&55&32&189.0453&26.0133&-0.334&2.71&34.7& on-disc\\
\hline
NGC~4594&0&22&18&202&190.0309&-11.6672&-0.703&3.28&233.8& \\
NGC~4594&1&22&18&307&190.0303&-11.6672&-0.703&3.28&233.8& \\
NGC~4594&2&27&25&34&190.0151&-11.6275&-1.073&1.06&194.5& on-disc\\
NGC~4594&3&30&29&52&189.9976&-11.6231&0.427&0.00&234.9&nuclear\\
NGC~4594&4&46&61&233&189.930&-11.6124&-0.140&4.03&9.1& \\
\hline
NGC~4631&0&7&0&59&190.6467&32.5612&-0.946&5.85&168.4& on-disc\\
NGC~4631&1&35&45&20&190.5218&32.6097&-0.687&4.13&81.9& \\
NGC~4631&2&49&76&6&190.4615&32.5313&-0.819&3.69&350.4& on-disc\\
\hline
NGC~5084&0&33&15&8&200.0702&-21.8273&-0.147&0.03&38.3& nuclear\\
\hline
NGC~5297&0&57&32&0&206.5634&43.8389&-0.865&2.52&307.3& \\
\hline
NGC~5775&0&55&12&50&223.4737&3.6065&0.411&3.85&75.3& \\
NGC~5775&1&56&13&18&223.4714&3.5486&-0.991&1.14&12.5& \\
NGC~5775&2&60&16&66&223.4622&3.5234&-0.627&2.09&322.7& \\
\hline
NGC~5907&0&9&0&69&229.1461&56.3139&-0.183&5.79&188.7& \\
NGC~5907&1&21&5&8&229.067&56.2838&0.656&4.11&221& \\
NGC~5907&2&82&67&70&228.8484&56.3201&-0.652&4.22&352.9& \\
\hline
\multicolumn{11}{l}{This table is available in the file, Table\_6\_LCX\_cross\_matched.csv.}\\
\multicolumn{11}{l}{$^{\rm a}$ Unique identifer for the cross-matched L-band, C-band, and X-ray sources, starting at zero. }\\
\multicolumn{11}{l}{$^{\rm b}$ Unique identifers for the L-band and C-band sources, equivalent to Source$\_$id for the respective radio bands in Appendices~\ref{a:c-radio} and \ref{a:cross_L_C}.} \\
\multicolumn{11}{l}{$^{\rm c}$ Unique identifer for the X-ray sources, equivalent to ID in Appendix~\ref{a:c-xray}.} \\
\multicolumn{11}{l}{$^{\rm d}$ Average of the positions in the 3 bands. Radio positions match to within 1 arcsec and X-ray positions match to within 3$\,$delta\_x.} \\
\multicolumn{11}{l}{ where delta\_x is given in the tables of Appendix~\ref{a:c-xray} and is typically $\approx$ 0.5 arcsec.} \\
\multicolumn{11}{l}{$^{\rm e}$ Radio spectral index between L and C-bands, as described in Appendix~\ref{a:cross_L_C} and given in Column 9 of Table~\ref{t:cross_L_C}. }\\
\multicolumn{11}{l}{$^{\rm f}$ Average distance to map centre and angle, CCW, from the direction, $x=0$ (due west) as described in Appendix~\ref{a:cross_L_C}. }\\
\multicolumn{11}{l}{$^{\rm g}$ If Dist$\_$to$\_$center is less than 3.0 arcsec, the source is flagged as `nuclear'. If the source falls within the optical disc, in projection,} \\
\multicolumn{11}{l}{as delineated by the green ellipse in Fig.~\ref{a:figs_r}, then it is denoted, `on-disc'. }\\
    \end{longtable}
}
\end{center}

\twocolumn
\clearpage
\section{Statistics and Quality Checks}\label{s:quality}

\begin{figure}
    \centering
    \includegraphics[width=0.5\textwidth]{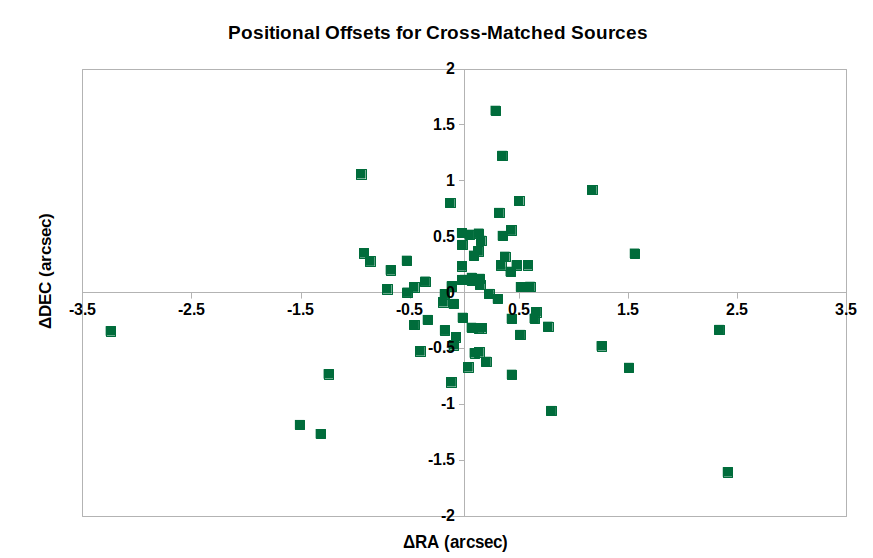}
    \caption{Plot of the offsets in RA and Dec between the average radio position and the X-ray position for all cross-matched sources from Table~\ref{t:cross_LCX}.}
    \label{f:offsets}
\end{figure}

As indicated in Sect.~\ref{ss:cross_L_C}, the radio sources at the two bands are taken to be cross-matched if their positions agree to within 1 arcsec in RA and Dec.  However, agreement between radio and the X-ray sources depends on the X-ray error, $\delta\,x$, whose value is given in Table~\ref{t:660_xray} and related downloadable files. { The median $\delta\,x$ for all X-ray sources is 0.48 arcsec and the average is 0.75 arcsec.} By design, all cross-matched sources have positional agreement within 3$\delta \,x$. { Consequently, positive matches are typically within $\sim$1.5 arcsec, although the value varies with individual $\delta \,x$. } The offsets are visualized in Fig.~\ref{f:offsets}, where we have plotted the offsets between the average radio position and the X-ray position. The
rms values are $\Delta$RA = 0.78 arcsec and
$\Delta$Dec = 0.56 arcsec. 
{The median |$\Delta$RA| and |$\Delta$Dec| are 0.35 arcsec  and 0.33 arcsec, respectively.}

The number of successfully cross-matched radio/X-ray sources is not large for any given galaxy (Table~\ref{t:cross_LCX}).  However, there is some possibility that some matches could be chance coincidences rather than physical associations.  

To test this, we created fields of artificial X-ray sources that are randomly scattered over the cross-match field for each galaxy.
We first count the number of detected X-ray sources for each galaxy from the catalogue of X-ray sources in Appendix~\ref{a:c-xray}. The  number of detected X-ray sources are listed in Table~\ref{t:X_false}, first the total number of sources, then the number of sources in the cross-match field (i.e. the 12 arcmin square field, Table~\ref{t:fieldsize}) and finally, the number of detected sources that are on the galaxy disc in projection, i.e. within the green ellipse as drawn in Appendix~\ref{a:figs_r}. 
 The locations of the detected sources were determined by calculation using spherical coordinates as outlined in Appendix~\ref{a:c-radio}.

  \begin{table*}
\caption{X-ray Statistics and Possible False Matches\label{t:X_false}}
\vspace{0.1truein}
\begin{center}
\begin{tabular}{lccccc}
 Galaxy& \multicolumn{3}{c}{No. Detected X-ray Sources$^{\rm a}$}&   \multicolumn{2}{c}{No. Possible False Matches$^{\rm b}$} \\
 & Total & In Cross-match field & On-disc&In Cross-match field&On-disc\\
\hline
\hline
NGC~660& 72&57&11& 0&0\\
NGC~891&160&118& 61&0&0\\
NGC~2683&53&44&15&0&0\\
NGC~2992&56&42& 6&0&0\\
NGC~3079&123&78&46&0&0\\
NGC~3432&7&7&3& 0&0\\
NGC~3448&31&22&13&0&0\\
NGC~3556&97&68&39&0&0\\
NGC~3628&87&74& 41&0&0\\
NGC~3877&129&78&23&0&0\\
NGC~4013&101&66&21&0&0\\
NGC~4096&36&29&19&0& 0\\
NGC~4157&69&45&18& 0&0\\
NGC~4192&31&26& 13&0&0\\
NGC~4217&78&48&19&0&0\\
NGC~4244&57&42&9&0&0\\
NGC~4302&29&22& 4&0&0\\
NGC~4388&50&44& 13&0&0\\
NGC~4438&58&47&22&0&0\\
NGC~4565&120&101&65&0&0\\
NGC~4594&388&301& 187&0&2\\
NGC~4631&72&52&31&0&0\\
NGC~4666 & 34& 29& 17&0&0 \\
NGC~5084&43&36&15&0&0\\
NGC~5297&20&17& 5& 0&0\\
NGC~5775&100&86&26&0&0\\
NGC~5907&97&66& 44&1&0\\
\hline
\end{tabular}
\end{center}
{$^{\rm a}${~Number of X-ray sources detected, as specified in Appendix~\ref{a:c-xray}. `Total' is all sources in the catalogue. `In Cross-match field' specifies the number in the overlap region for cross-matching, which is the C-band field (12 arcmin $\times$ 12 arcmin) as given in Table~\ref{t:fieldsize}. `On-disc' is the number of sources within the optical disc as delineated by the green ellipses in Appendix~\ref{a:figs_r}.}}\\
{$^{\rm b}${~Number of possible false-matches when artificial X-ray sources are randomly inserted `In Cross-match field' and `On-disc' as defined above. }}\\
    \end{table*}

We next generated the same number of artificial X-ray sources as actually occur in the cross-match region (e.g. 57 for NGC~660) and placed them randomly within this region. We then search for cross-matches between the radio sources and the artificial X-ray sources within a cross-match radius of 3$\times$ the average of RA and Dec rms offsets shown in Figure~\ref{f:offsets}, i.e. 2.0 arcsec. Because all radii are the same for all artificial sources, we are able to use the Aladin Sky Atlas\footnote{https://aladin.u-strasbg.fr/} for this cross-matching.  Any matches are labelled as `False Matches' in Table~\ref{t:X_false}.
The result for the cross-match fields show that only one chance alignment was found (for NGC~5907) when the artificial X-ray sources are sprinkled randomly over the field.  

The above analysis assumes that all sources are randomly placed over the cross-match field area.  However, the { radio} sources cluster more centrally because of the primary beam pattern weighting.  
 We therefore repeated the above analysis for the on-disc regions by randomly distributing artificial X-ray sources over only on the disc, as outlined above. We ensure that the number of artificial X-ray sources is the same as the number of actual X-ray sources detected. The results are listed in the last column of Table~\ref{t:X_false}.  

It is clear that any possible false detections in the sample are very small and therefore the cross-matches that are listed in Table~\ref{t:cross_LCX} have a high probability of being physically associated with each other. {The above randomization exercises are meant to be a useful guide and result from a single randomization exercise for the two cases denoted above.} Therefore, follow-up observations on cross-matched sources may be required to ensure that chance matches have not occurred for any specific source.
When examining the figures of Appendix~\ref{a:figs_r}, is helpful to remember the the symbol sizes are larger than the error bars on the source positions.

\section{Discussion and Suggestions for Follow-up Studies}
\label{s:discussion}

A detailed analysis of the sources in our catalogues remains for follow-up studies.  However, a brief preliminary analysis was carried out to better understand the sources in our sample.

\begin{figure}
    \centering
    \includegraphics[width=0.5\textwidth]{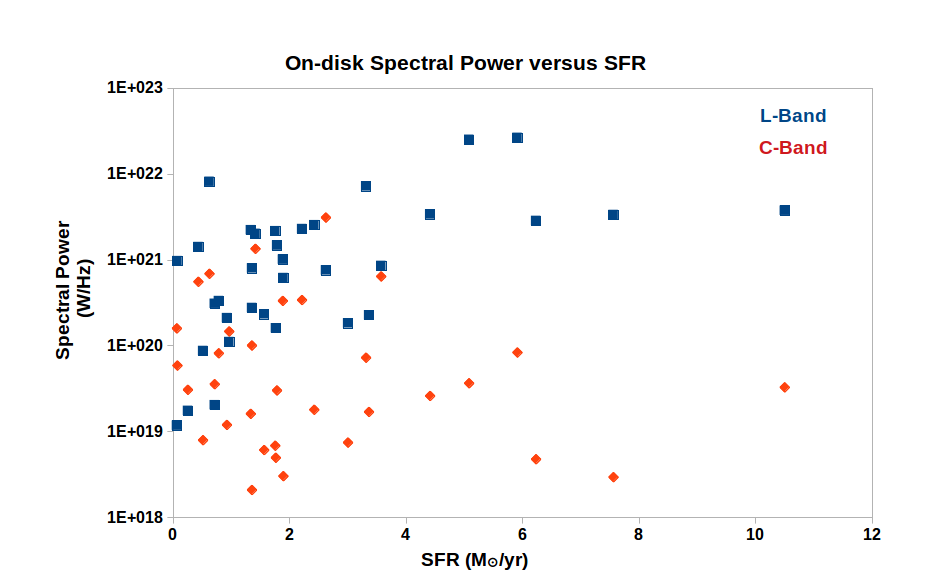}
    \caption{Spectral power of all sources on the disc, in projection, as a function of SFR. The median uncertainty at L-band is 19\% and, at C-band, it is 15\% (assuming accurate distances).}
    \label{f:powerSFR}
\end{figure}

Fig.~\ref{f:powerSFR} shows a plot of the spectral power of the sum of all sources on the disc (in projection) as a function of star formation rate (SFR), as given in \cite{var19}.   It is apparent that there is no convincing trend from these data.  Similarly there is no convincing trend for the number of sources as a function of SFR (not shown). {Given that the total flux density of all compact sources in any given galaxy (except for AGNs) represents only a minor fraction of the total galaxy flux (Table~\ref{t:number_L_C}), this lack of correlation is perhaps not surprising.  The total flux reflects an integration of previous star formation activity over time, whereas the compact sources are more likely to represent current SF. The current work, then, may be a starting point for follow-up studies of active star forming regions in these galaxies. A study of compact source flux over the disc (i.e. per kpc$^2$ of disc area) in comparison with various measures of SFR over equivalent disc area could help in investigating the scale over which these SFR-measures apply. }


We then focussed only on the radio+X-ray cross-correlated sources of Table~\ref{t:cross_LCX}.  Each of these sources were loaded into the Aladin Sky Atlas over an optical representation of each galaxy.  All objects from the Set of Identifications, Measurements, and Bibliography for Astronomical Data \cite[Simbad,][]{simbad} were then loaded for comparison.  In addition, sources from the Gaia satellite \citep{gaia1} were also loaded, in the event that stars might also be represented in the catalogues. The sources in the fields of each galaxy were then visually inspected.

\begin{figure}
    \centering
    \includegraphics[width=0.45\textwidth]{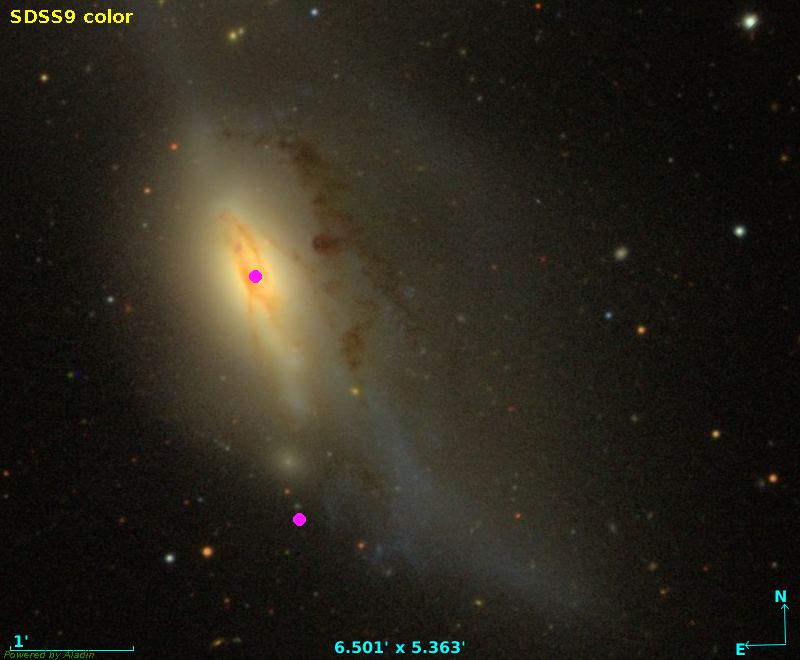}
    \caption{SDSS9 image of the galaxy, NGC~4438, showing two sources, which are cross-matched in X-ray and both radio bands, from Table~\ref{t:cross_LCX} in magenta.  The source near the centre of the image is the nuclear source (ID\_LCX = 2) and the source approximately 2 armin to the south is
    an X-ray source with a previously unknown corresponding radio source (ID\_LCX = 3). }
    \label{f:N4438_source}
\end{figure}

Some sources in our galaxies are previously known compact X-ray sources \citep[e.g.][]{wan03,col04,mac04,li08}.  However, they have not previously been identified with compact radio sources.  An example is NGC~4438 (Fig.~\ref{f:N4438_source}), where we show two of the cross-matched radio/X-ray sources in solid magenta circles superimposed on the Sloan Digital Sky Survey Data Release 9 (SDSS9, \cite{sdssdr9}) image in colour. The nuclear source (ID\_LCX = 2 in Table~\ref{t:cross_LCX}), indicative of an AGN, has been reported earlier \citep[Table 10 of ][]{irw19b}. The off-nuclear source, ID\_LCX = 3 for this galaxy (Table~\ref{t:cross_LCX}), 2 arcmin to the SSW of the nucleus, has previously been identified as an X-ray source in \cite{mac04}, but never before identified as a radio source. There is no Simbad source at this position.
These X-ray/radio detections are particularly interesting because they suggest the possibility of energetic activity, such as jets, microquasars, or colliding winds.  In this particular case, the source is located in the outer part of the disc galaxy, which is undergoing strong ram-pressure stripping in the Virgo cluster. This location and the lack of an optical counterpart, together with the apparent X-ray/radio association, indicate that the source {may be} a background blazar, although a deep optical confirmation is needed.

\begin{figure}
    \centering
    \hspace*{-1.6cm}{\includegraphics[width=0.65\textwidth,height=0.4\textwidth]{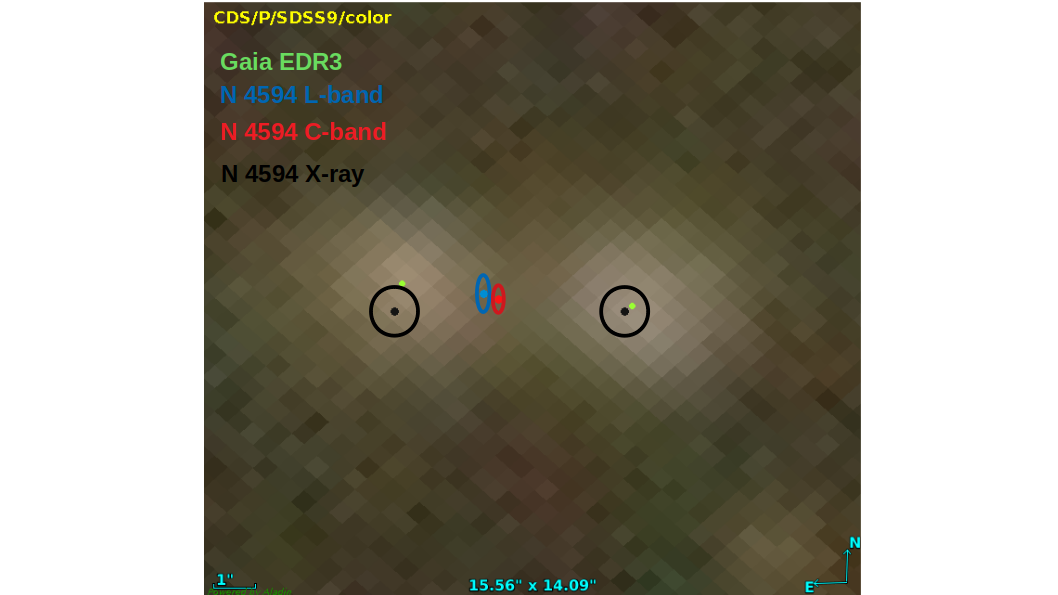}}
    \caption{The binary star system near the galaxy, NGC~4594 (data in Table~\ref{t:doublestar}), is shown as the background image in colour 
    using the Aladin Sky Atlas$^{\rm 8}$.
    The stellar identifications from Gaia EDR3 are 3530134264552397440 (right star, our S1 in Table~\ref{t:doublestar}) and 
   3530134264552424960 (left star, our S2); the Gaia centres are denoted by green dots.  The Gaia error is too small to show (less than 20 mas). The small blue and red dots with corresponding blue and red error ellipses (1$\sigma$), located between the stars, denote the positions of the L-band and C-band sources, respectively.  The black dots with surrounding black circles denote the positions and error circles of the two X-ray sources. These are the original positions, rather than the average of the cross-matched source positions given in Table~\ref{t:cross_LCX}. The scale showing 1 arcsec is displayed at lower left, the field of view is marked at the bottom, and the orientation is marked at lower right.}
    \label{f:doublestar}
\end{figure}

An unexpected result is what is likely a foreground double-star near NGC~4594.  There are no Simbad sources at this location. In Fig.~\ref{f:doublestar}, we plot the positions of the various sources with their error ellipses, taken from the original data of Appendices~\ref{a:c-radio} and \ref{a:c-xray}, rather than the average of the cross-correlated positions of Table~\ref{t:cross_LCX}. The original error ellipses are also shown, rather than the 3$\sigma$ tolerance that we used to do the radio/X-ray cross-correlations. Relevant data for the double-star are given in Table~\ref{t:doublestar}.

The background colour image of Fig.~\ref{f:doublestar}, from the SDSS9, shows two stars separated by 5.4 arcsec. The green dots represent the centres from the Gaia satellite Early Data Release 3 \cite[EDR3,][]{gaiaedr3} and are clearly located right at the centres of stars that are apparent in the SDSS9 image. X-ray emission is seen from the locations of both stars (black dots with circles).  Radio emission is seen at L-band (blue dot with 1$\sigma$ error ellipse) and C-band (red dot with 1$\sigma$ error ellipse) almost midway between the stars.  All sources satisfy the cross-correlation tolerances described in Sect.~\ref{ss:Obs_radio_xray}. 

At the time of writing, parallax measurements are not available for the EDR3 Gaia objects.
The G-band mean magnitude for S1 (right star in figure) is G = 20.965$\,\pm\,0.015$ and for S2, (left star in figure) it is G = 21.632$\,\pm\,0.046$.
The listed colour value for S1 is G$_{\rm BP}$ - G$_{\rm RP}$ = 1.075, and for S2 it is
G$_{\rm BP}$ - G$_{\rm RP}$ = 0.932.  These colours would imply stellar effective temperatures between 5000 and 6000 K \citep{and18}. However, because these stars are so faint, the G$_{\rm BP}$ - G$_{\rm RP}$ colour index cannot be used with confidence (Gaia General Enquiry, Private Communication).  The spectral index of the radio source is $\alpha\,=\,-0.703\,\pm\,0.083$. 

{The two X-ray detections shown in black are ID$\_$X: 203 and 308 (Table 6) which are separated by 5.4$^{\prime\prime}$. Unfortunately, with only about 19 and 13 net counts in the 0.3-7~keV band (Table~D1), the detections give little X-ray spectral information about these two X-ray sources. Nevertheless, from the count rates, we estimate their energy fluxes, which are included in Table 8.
They have similar luminosities of $\sim 1 \times 10^{28} {\rm~ergs~s^{-1}} (D/10^{2}{\rm~pc})^{2}$, where $D$ is the distance to the source.}

Follow-up data are clearly required.  However, we note that radio emission and X-ray emission have been observed from colliding wind binary systems \citep[CWBs,][]{dou03,rom19,cal19}.  {Further study is needed to confirm whether this system is indeed a CWB {or whether there is some other explanation for this activity.} It is also possible that there are other such systems in our data, which could be revealed with more thorough scrutiny.}

 \begin{table*}  
  \caption{Double-Star in NGC~4594 Field\label{t:doublestar}}
\vspace{0.1truein}
\begin{center}
\begin{tabular}{lcccccccc}
 Source or Band & ID$^{\rm a}$ &  RA$^{\rm b}$ & E\_RA$^{\rm b}$ & Dec$^{\rm b}$ & E\_Dec$^{\rm b}$& Flux$^{\rm c}$&E\_flux$^{\rm c}$&Comments\\
  & & (deg) & (arcsec) & (deg) & (arcsec)\\ 
\hline
\hline
Gaia EDR3& S1 &  190.02969
&2.94&  -11.66722& 1.57& --- & ---\\
Gaia EDR3& S2  & 190.03124& 19.95
& -11.66707& 8.86 & --- & ---\\
This work L-band & 22 &190.03069
& 0.15& -11.66713 &0.43
& 592.3 &	57.3 & unresolved
\\ 
This work C-band & 17 & 190.03059
& 0.092
& -11.66717
& 0.27& 231.1 &	11.7& unresolved\\
This work X-ray & 307 & 190.02974 &0.591	& -11.66725	&0.591 & 0.4 & 0.1
\\
This work X-ray&202& 190.03128 & 0.534
 & -11.66725 &
0.534 & 0.3 & 0.1
\\
\hline
\end{tabular}
\end{center}
{$^{\rm a}${~ Source identification number as in Table~\ref{t:cross_LCX} (ID\_L, ID\_C, or ID\_X).  For Gaia, we give our new ID number, where S1 is the Gaia identification number, 3530134264552397440, and 
S2 is the Gaia identification number, 3530134264552424960.}}\\
{$^{\rm b}${~ Positions and errors as given in the tables of Appendices~\ref{a:c-radio} and \ref{a:c-xray}, or the EDR3 for Gaia. }}\\
{$^{\rm c}${~ Flux and the associated errors, where available.  For the radio sources, these are flux densities and errors that are primary beam corrected, as given in Columns 5 - 8 of Appendix~\ref{a:cross_L_C} in
in units of $\mu$Jy. A dash means that the data are not available. For the X-ray sources, the unabsorbed energy fluxes in units of $10^{-14} {\rm~ergs~s^{-1}~cm^{-2}}$ are crudely estimated from the count rates (Appendix D), assuming a power law spectrum with a photon index of 1.7 and a foreground absorption of $N_H = 1\times 10^{21} {\rm~cm^{-2}}$.}}\\
    \end{table*}

We finally report on the discovery of new nuclear sources in two companion galaxies.  
NGC~4298, which is the companion galaxy to NGC~4302 has a nuclear L-band/C-band/X-ray source right at its nucleus.  This source is ID\_LCX =1 in Table~\ref{t:cross_LCX} in the NGC~4302 field.  To our knowledge, there appears to have been no previously known nuclear source in this galaxy.  
NGC~4435, which is the companion galaxy to NGC~4438, also has a new nuclear source, namely ID\_LCX = 4 in Table~\ref{t:cross_LCX} in the NGC~4438 field. There appear to have been X-ray detections in the nucleus of this galaxy \citep[e.g.][]{wan16} but no radio source has previously been detected. 

{A NED\footnote{https://ned.ipac.caltech.edu.} search for NGC~4435 resulted in little new information about the X-ray source in the literature, including  \cite{wan16}, except for the X-ray detection as 2CXO J122740.7+130447  with the net counts of $\sim 65.6$ and a count rate of 2.5 counts/ks. We estimate the energy flux of the source as $\sim 5 \times 10^{-14} {\rm~ergs~s^{-1}~cm^{-2}}$ or the luminosity as $\sim 4.0 \times 10^{38} {\rm~ergs~s^{-1}}$ at the distance of the galaxy, assuming the same spectrum for the conversion as described in the caption to Table 8.}

The catalogues that have been produced represent a wealth of data that can be probed in much more detail. We have looked briefly only at the sources that are observed in L-band, C-band, and the X-ray as listed in Table~\ref{t:cross_LCX}. Many more of these listed sources could be investigated in their own rights.  Moreover, there are other sources that might have radio observations at only one frequency (due, for example, to signal-to-noise issues), yet a cross-correlation with the X-ray could still be very fruitful. 

It could be useful to examine the number and fluxes of sources as a function of height above the plane in the galaxies.  Such studies may point {either to compact halo objects, or background sources which could be used to probe the foreground halo, as described in  Sect.~\ref{s:Intro}.  To our knowledge, such a study has not been carried out before}. An investigation of radio source number and spectral power as a function of SFR (both per unit area) could provide information on the relation between in-disc sources and SFR, as suggested earlier in this section. 

In the introduction, we discussed the issue of distinguishing between HII regions and SNRs. Although most sources appear to represent SNRs, the degree to which some sources could represent HII regions has yet to be explored.   As an aid, we
have presented H$\alpha$ maps in Appendix~\ref{a:figs_r} which show only HII regions that are unobscured by dust. Thus the distribution of H$\alpha$ emission is not the same as seen in the radio band, {but these data still provide clues as to which sources could be dominated by (or have a significant fraction of) thermal emission.}
 Because the deconvolved sizes of the radio sources are provided in the catalogues, the size and luminosity distributions of compact sources can also be compared with known distributions for SNRs and HII regions {\cite[e.g.][]{loc96,tre14,gre19}}.


Most sources in this work are actually background sources {that are in the field of view but not near the galaxy disc, likely AGNs, quasars, or extra-galactic radio sources. Those background sources that are close to the foreground galaxy in projection, however, could provide opportunities to explore the foreground galaxy disc or halo via Faraday Rotation of the polarized emission. }A simple example of such a study is given in \cite{irw13}. Such studies will become more feasible with upcoming S-band (4 GHz) observations of the galaxies.

\section{Summary}\label{s:sum}

We have systematically identified compact radio and X-ray  sources in the CHANG-ES sample of nearby edge-on galaxies. The radio data used are from VLA observations in the B-array/L-band and C-array/C-band, covering all 35 CHANGES galaxies, while the X-ray data are from archival Chandra observations available for 27 galaxies in this sample. The radio detections utilized the PyBDSF algorithm which helps to isolate discrete sources amongst more general, spread-out radio emission. The radio data have $\approx$ 3 arcsec resolution. The results of these searches are in a number of catalogues, available with this manuscript, and described in Appendices~\ref{a:c-radio} and \ref{a:c-xray}. Maps showing a blow-up of each galaxy and their discrete sources are presented in Appendix~\ref{a:figs_r}; images of the entire field can also be downloaded as described in the appendix.

In the radio, we find 2507 sources at L-band over all galaxies in a field size of 16.7 arcmin $\times$ 16.7 arcmin and 1413 sources at C-band in a field size of 12.0 arcmin $\times$ 12.0 arcmin. The median spectral power is $L_{L-Band}\,=\,7.56\,\times\,10^{18}$ W/Hz
(log$L_{L-band}\,=\,18.9$), and
$L_{C-Band}\,=\,1.61\,\times\,10^{18}$ W/Hz
(log$L_{C-band}\,=\,18.2$), both largely consistent with SNRs. The median spectral index is $\alpha_m\,=\,-0.669\,\pm\,0.18$, clearly non-thermal and also consistent with SNRs.

We have cross-correlated the radio data between L-band and C-band over the smaller 12 arcmin field of C-band. 
The cross-matched L/C sources are given in Appendix~\ref{a:cross_L_C}. 
We also cross-correlated the (cross-correlated) radio sources with the X-ray sources.  The results are presented   Table~\ref{t:cross_LCX} and are available for download as described in the footnotes to that table.
Of these sources, we find a number of previously known X-ray sources, but many were not previously known to have radio counterparts. We present only one in Fig.~\ref{f:N4438_source}, but there are many others.

{Three new nuclear sources have been identified from the radio cross-correlation: NGC~3877, NGC~4192, and NGC~5792 (Table~\ref{t:new_AGNs}).  NGC~3877 and NGC~4192 also have X-ray sources right at the nucleus. When compared to previous observations, we find that NGC~3877 is also variable.
As suggested in Sect.~\ref{ss:cross_L_C}, follow-up data would help to confirm the nature of these apparent LLAGNs. It is interesting that, between the 18 galaxies identified to have AGNs in Table~10 of \citet{irw19b} and three more candidates from this paper, a total of 21 out of 35 CHANG-ES galaxies (60\%) appear to have AGNs. Note that the sample was {\it not} chosen for its AGN population, but was chosen for angular size, edge-on orientation, and observability \citep{Irwin_2012a}. } 

In addition, we find new nuclear sources in two companion galaxies: NGC~4298 (companion to NGC~4302) and NGC~4435 (companion to NGC~4438). These nuclear sources are also listed in Table~\ref{t:cross_LCX}.

An unexpected result is what appears to be a foreground double-star (Fig.~\ref{f:doublestar}).  Both stars have associated X-ray emission, and radio emission at both L-band and C-bands is seen in between the stars.

\section*{Acknowledgements}{The first author wishes to thank the Natural Sciences and Engineering Research Council of Canada for a Discovery Grant. {We also thank an anonymous referee for valuable comments.}This research has
made use of the NASA/IPAC Extragalactic Database (NED), which is funding by the 
National Aeronautics and Space Administration and operated by the California Institute of Technology. This research has used the Karl G. Jansky Very Large Array operated by the National Radio Astronomy Observatory (NRAO). NRAO is a facility of the National Science Foundation operated under cooperative agreement by Associated Universities, Inc. This research has made use of the "Aladin Sky Atlas" developed at CDS, Strasbourg Observatory, France, and the software provided by the Chandra X-ray centre (CXC) in the application packages CIAO and Sherpa, as well as data obtained from the Chandra Data Archive and from the European Space Agency (ESA) mission
{\it Gaia} (\url{https://www.cosmos.esa.int/gaia}), processed by the {\it Gaia}
Data Processing and Analysis Consortium (DPAC,
\url{https://www.cosmos.esa.int/web/gaia/dpac/consortium}). Funding for the DPAC
has been provided by national institutions, in particular the institutions
participating in the {\it Gaia} Multilateral Agreement.}

Facility keywords: VLA and CXO; Dataset identifiers (Chandra OBSID): 15333,  14376 ,2038 ,1972, 11310,  2883, 797, 20994,  15587, 19297, 20947, 767, 19390, 8042, 4018, 20995, 1633,  4613, 7091,  768, 4738,  3950,  12173, 18352 , 794, 19360, 952, 942,  404 , 19370, 4010,  11311, 2025 , 4013,  19392, 1586, 2940, 1636, 2039,  4739, 19397,  407, 12987, 3956, 395 , 19345, 12291, 9532, 14391, 19307, 1971, 7103, 1619, 9533, 20830

\section*{Data Availability}
The data underlying this research are available at
https://queensu.ca/changes and in the online supplementary material accompanying this article.



\bibliographystyle{mnras}
\bibliography{Compact_sources_PAPER} 



\newpage
\onecolumn

\appendix

\clearpage

\section{PyBDSF setup and testing}\label{a:param}

A qualitative introduction of PyBDSF was given in Sect.~\ref{ss:ana_radio} and further details of our tests are given below. For a recent example that makes use of PyBDSF, see \cite{shi19}.

To run PyBDSF, we have adopted a set of parameters to optimize the measurements of source locations, intensities, etc.   These parameters are listed in Table~\ref{t:pybdsf} with a brief explanation as to what they control\footnote{See https://www.astron.nl/citt/pybdsf for documentation with complete details of the parameters.}. Parameters that are not listed were set at their defaults. Of those listed, two of the most important are the island threshold (thresh$\_$isl) and the source detection threshold (thresh$\_$pix), which control the number of sources that will be detected.  After some experimentation, we have adopted the default values for these thresholds as well.

\begin{figure}
    \centering
    \includegraphics[width=0.45\textwidth]{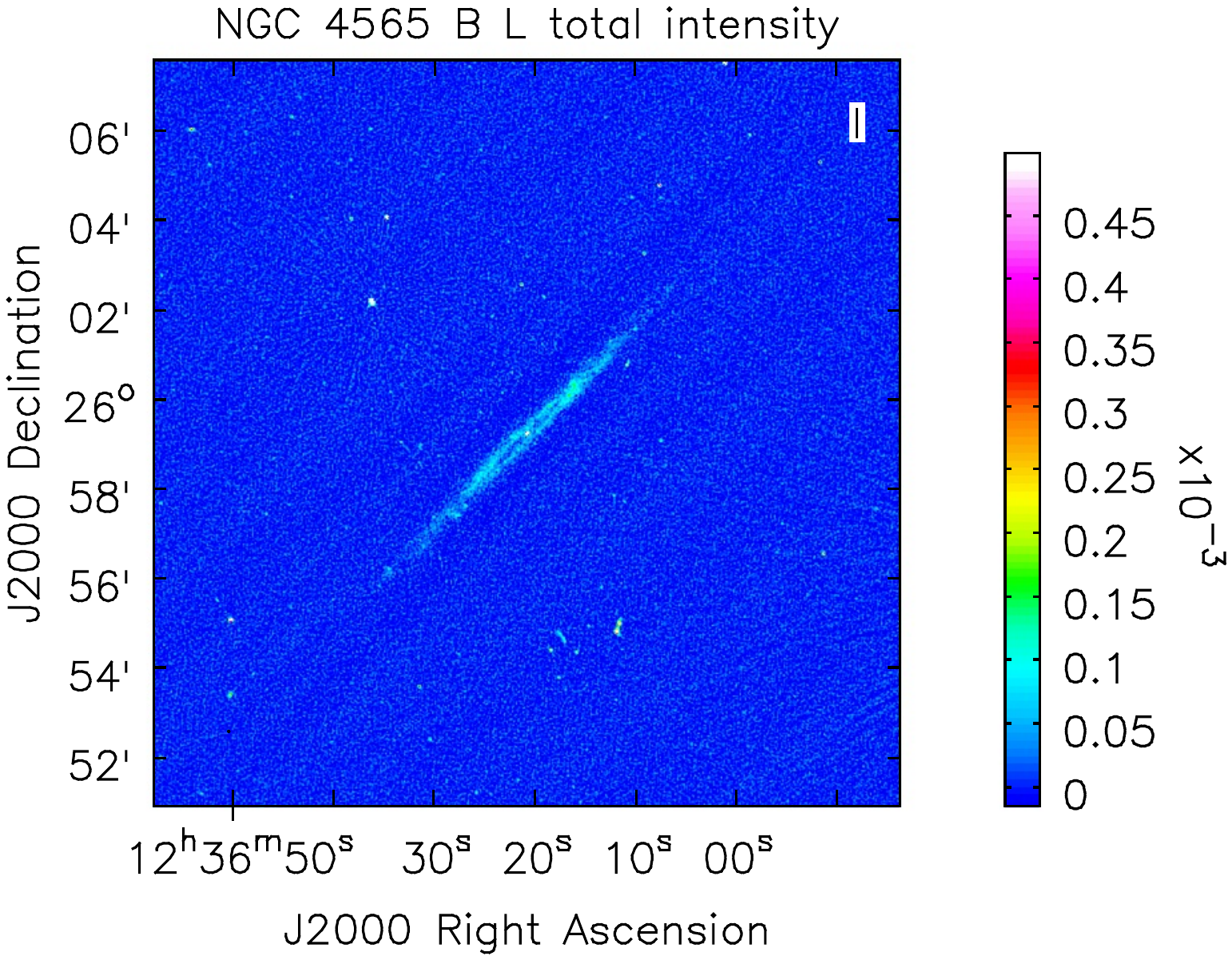}
    \includegraphics[width=0.45\textwidth]{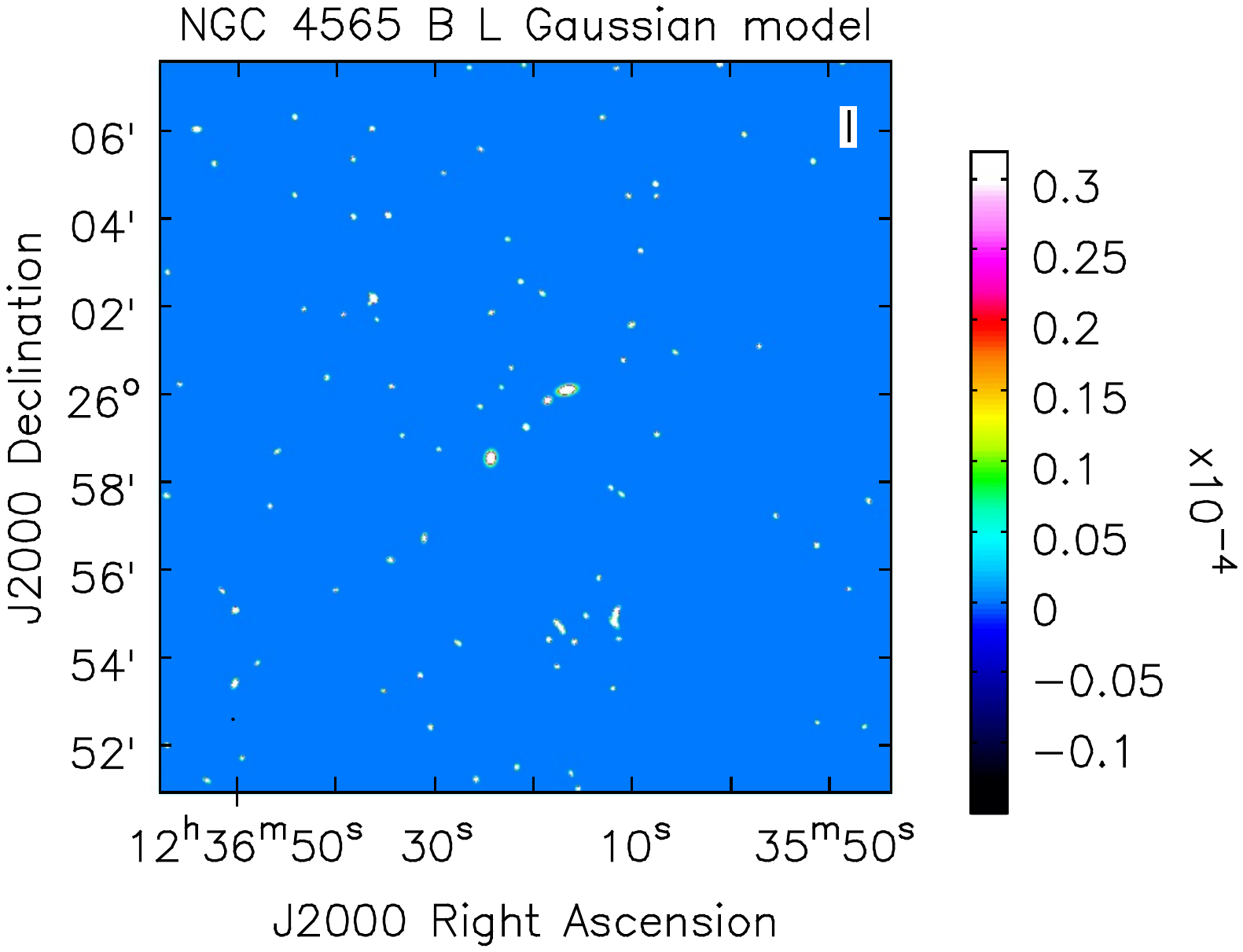}\\
    \includegraphics[width=0.45\textwidth]{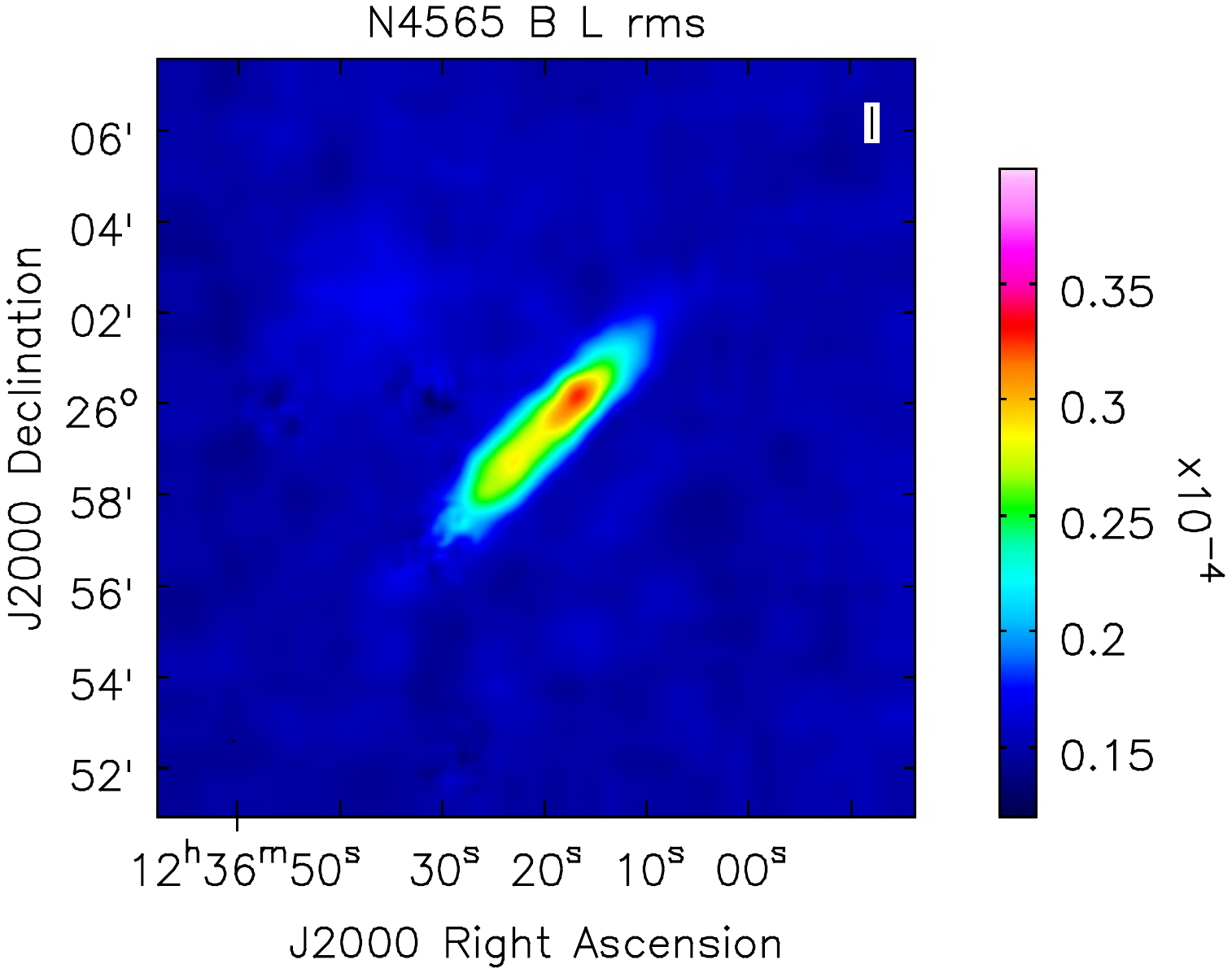}
    \includegraphics[width=0.45\textwidth]{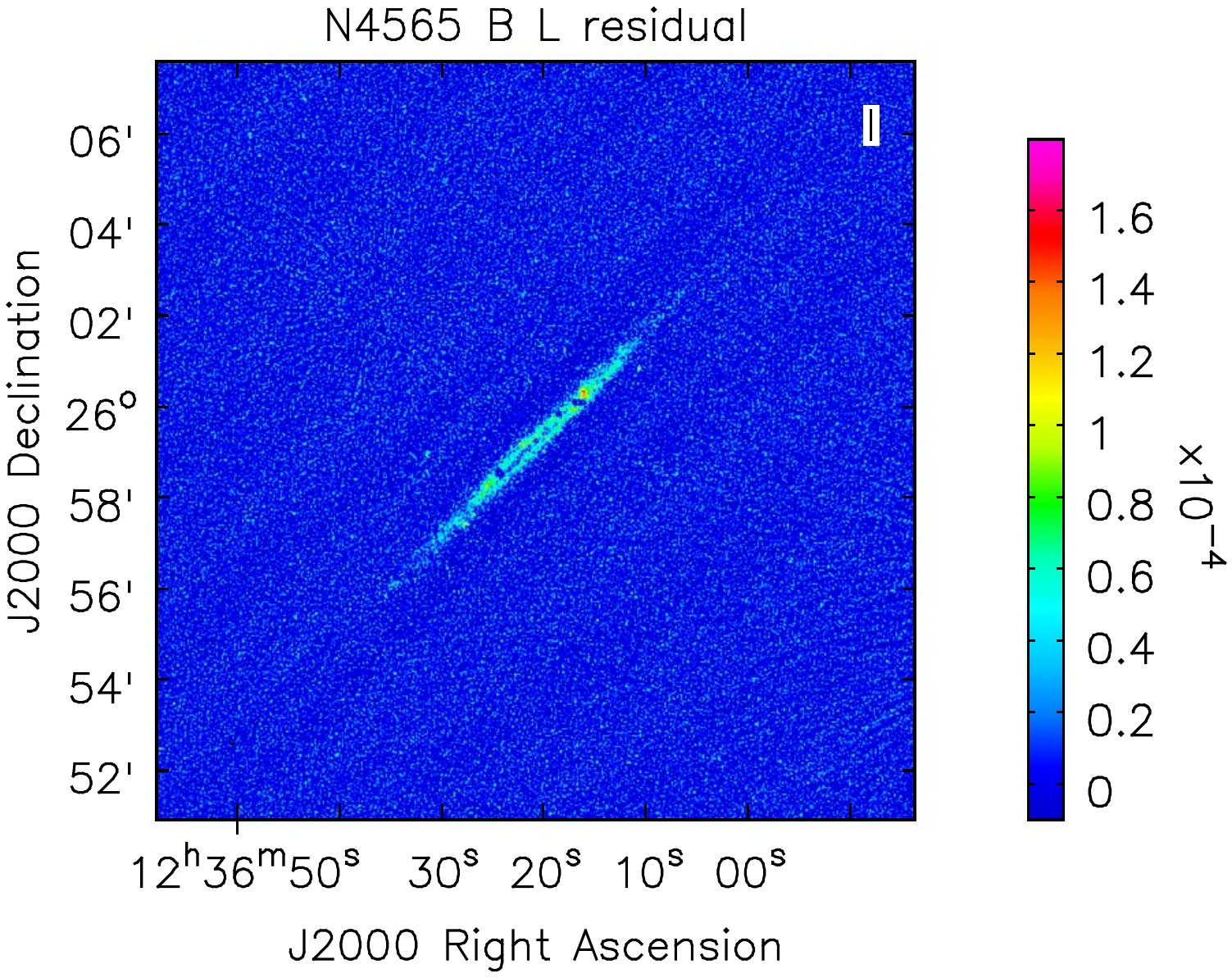}\\
    \caption{The NGC~4565 field.  The field is 2000 x 2000 pixels in size (16.7 arcmin) {for each panel}. A black dot at lower left denotes the synthesized beam. { This dot is very tiny in the images because it is only $\sim$ 3 arcsec in size. {The colour bars are in units of Jy beam$^{-1}$.} The `I' at top right indicates `intensity image'.} {\bf Top left:} Original total intensity image. {\bf Top right:} The Gaussian model. {\bf Bottom left:} The rms image. {\bf Bottom right:} The residual image.  }
    \label{f:appendix_tests}
\end{figure}

\begin{figure}
   \centering
    \includegraphics[width=1.0\textwidth]{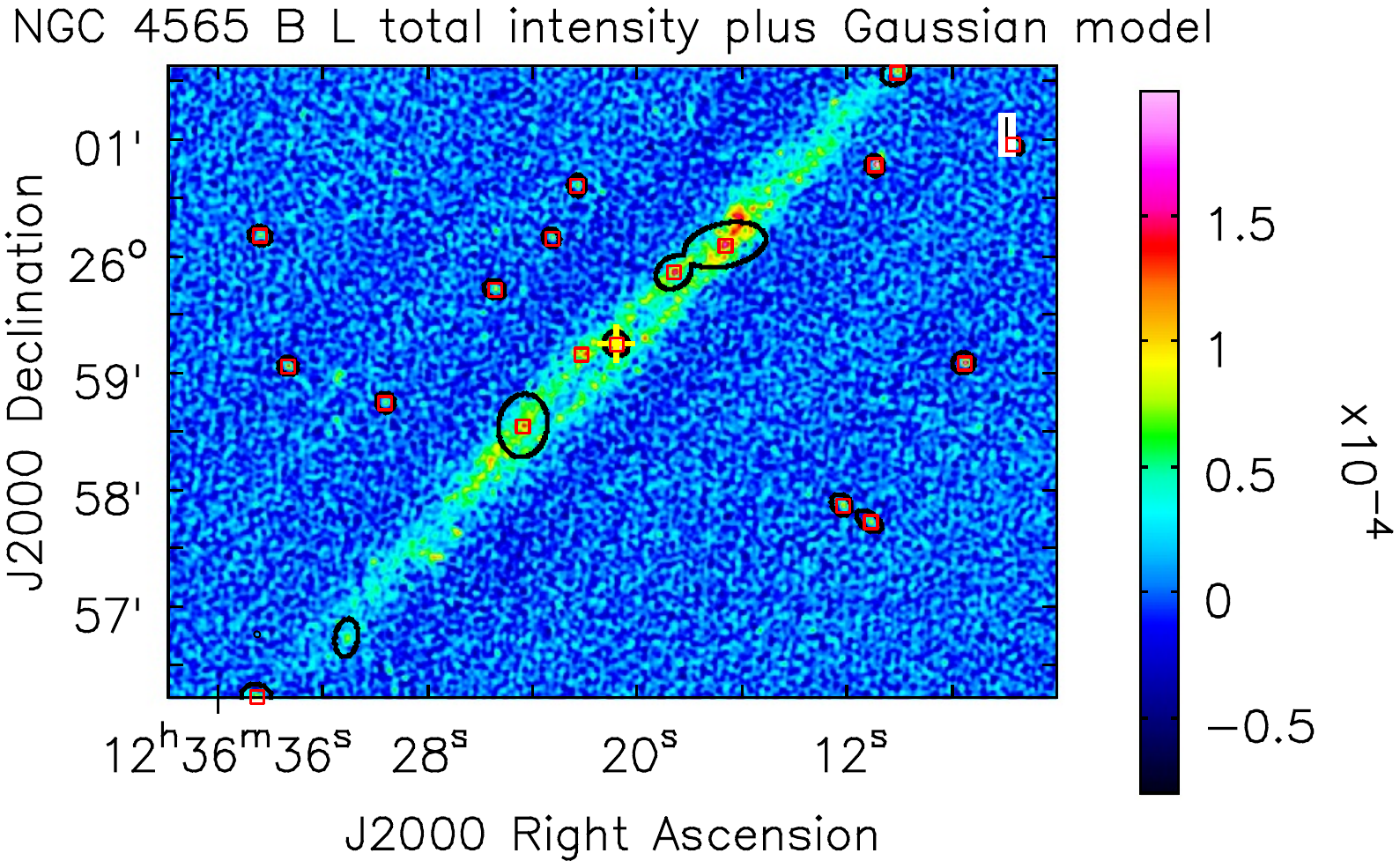}
    \caption{Blowup of the NGC~4565 field showing the total intensity image in colour superimposed with the 0.2 $\mu$Jy beam$^{-1}$ contour of the Gaussian model in black. A yellow cross denotes the galaxy centre and the output sources are shown as red squares, as in Appendix~\ref{a:figs_r}. {The colour bar is in units of Jy beam$^{-1}$. The `I' at top right means `intensity image'.}}
    \label{f:appendix_blowup}
\end{figure}

In Fig.~\ref{f:appendix_tests} we show the entire fitted field (2000 x 2000 pixels) of the B-configuration L-band image of NGC~4565, a galaxy of large angular size (Table~\ref{t:g_param}). Missing short uv spacings 
(Sect.~\ref{ss:Obs_radio}) results in a very slight negative depression parallel and close (about 0.5 arcmin) to the major axis at a level of $\approx\,5\,\mu$Jy beam$^{-1}$. Cleaning residuals then result in very faint linear artifacts parallel to the galaxy's major axis at a level of 1 - 2 $\mu$Jy beam$^{-1}$, decreasing with distance from the galaxy. The spacing of these artifacts is $\approx\,1.3$ arcmin. The total intensity image (Top Left, and Fig.~\ref{f:appendix_blowup}) reveals these
weak residual `wave-like' or `ripple-like' features in the noise of the image. 

To smooth over these minor artifacts, as recommended by the PyBDSF documentation, we have adopted a sliding box size and step (1.3 arcmin, 0.4 arcmin) (Table~\ref{t:pybdsf}) within which the rms and mean are calculated. The rms image (Bottom Left) shows the result.  Note the lower scale of this image (see colourbar).  Over the entire field, the rms values range from 12.8 to 32.7 $\mu$Jy beam$^{-1}$ and the mean is 15.4 $\mu$Jy beam$^{-1}$. Outside of the brighter central region containing the galaxy emission, the mean is 15.1 $\mu$Jy beam$^{-1}$ which agrees with the 15 $\mu$Jy beam$^{-1}$ rms that was measured manually in the original image (Table~\ref{t:g_param}). The rms map is used to set the threshold above which sources are identified.
In the event that there are strong sources which have strong localized residuals, a finer adjustable box size and step are also set, again, as recommended by the PyBDSF documentation.  For our maps, this is 30 arcsec, and 7.5 arcsec, respectively and the decision to switch to this mode is determined by an adaptive threshold that is automatically set within the software.
The mean image is also computed using the same sliding box sizes indicated above. When the dispersion in the rms or mean is very small over the map, PyBDSF adopts a constant value for the entire map.  This was the case for the mean of the NGC~4565 field whose value was calculated at -0.3 $\mu$Jy beam$^{-1}$.

{The PyBDSF algorithm fits Gaussians to islands simultaneously using the Levenberg-Marqhardt algorithm and computes errors following \cite{Condon}. Note that the errors include uncertainties in the fitting routines and map quality (rms, etc), and therefore describe random errors.  They do not include calibration errors in the primary flux calibrator which are of order 1\% \citep{per13}.}


The fitted Gaussian model for the NGC~4565 L-band image (Top Right) identifies the detected sources as given in Table~N4565\_B\_L.csv.  As indicated in Table~\ref{t:pybdsf}, an island is identified if contiguous emission is greater than 3 times the rms above the mean and a Gaussian source is accepted if the emission is greater than 5 times the rms above the mean.  Specifics of the identification of sources are given in Appendix~\ref {a:c-radio}.
A residual image (Bottom Right) is also formed when the Gaussian model (Top Right) is subtracted from the original image (Top Left).


\begin{table}
\caption{PyBDSF Input Parameters$^a$\label{t:pybdsf}}
\begin{center}
\begin{tabular}{lc}
\hline
Parameter & Value\\ \hline\hline
thresh\_isl$^{\rm a}$ & 3.0\\
thresh\_pix$^{\rm b}$ & 5.0\\
rms\_box, step $^{\rm c}$ & 160, 50 \\
adaptive\_rms\_box$^{\rm d}$ & True\\
~~~~~~rms\_box\_bright, step & 60, 15 \\
group\_by\_island$^{\rm e}$ & False\\
group\_tol$^{\rm f}$ & 10\\
flagging\_opts$^{\rm g}$ & True\\
\hline
\end{tabular}
\end{center}
$^{\rm a}$ The threshold for identifying an island is that the intensity be thresh\_isl $\times$ rms + mean, where rms is the local rms value and mean is the local mean value.\\
$^{\rm b}$ The threshold for identifying a source is thresh\_pix * rms + mean.\\
$^{\rm c}$ This is the sliding box size in pixels within which the rms and mean are calculated.   The box size is rms\_box pixels (1.3 arcmin in our maps) and it is computed in sliding units of step pixels (25 arcsec in our maps). One pixel is 0.5 arcsec.\\
$^{\rm d}$ Setting adaptive\_rms\_box to True allows for finer box and step sizes (rms\_box\_bright, and step, respectively) to be automatically calculated near bright sources. This ensures that the box within which the rms and mean are calculated minimizes the effect of possible residual sidelobes from bright sources.\\
$^{\rm e}$ A True setting for this parameter would group all sources within an island together, which is not desired in this analysis.\\
$^{\rm f}$ A parameter setting the criteria for grouping sources within an island. Gaussians are grouped if they are close together {\it and} the minimum value on a line separating them is close to the peak of at least one of the Gaussians.  Group\_tol controls both criteria, such that a larger value results in larger sources in an island.\\
$^{\rm g}$ This sets a variety of flagging options, such as flagging a Gaussian if its centre is outside of the map border, flagging a Gaussian if its size is more than 25 times the beam size, and other issues that result in questionable results.\\
\\
\end{table}

In Fig.~\ref{f:appendix_blowup}, we show a magnified image of the total intensity map (Top Left of Fig.~\ref{f:appendix_tests}) with contours showing the centring of the Gaussian fits.  It is clear that background sources off of the galaxy have been detected quite well. Sources that are superimposed on the galaxy disc have also been recovered, especially the source right at the core (RA = 12$^{\rm h}$ 36$^{\rm m}$ 20$\rasec$78, Dec = +25$^\circ$ 59$^\prime$ 15$\decsec$6) whose value is quite strong at 1.3 mJy beam$^{-1}$. Two Gaussians, slightly blended, can be seen to the north-west along the major axis, the smaller one at 30 arcsec from the centre and the larger at 57 arcsec. Both have point-like sources at their centres. The irregular linear feature that extends to the north-west in the larger ellipse has not been detected as a source.  

Two sources are potentially problematic, as seen in the blowup.  
The first is a source at RA = 12$^{\rm h}$ 36$^{\rm m}$ 22$\rasec$12, Dec = +25$^\circ$ 59$^\prime$ 10$\decsec$5 and can be seen as a red box with no Gaussian circle surrounding it. In other words, it has shown up as a source in the output list (which is generated from the Gaussian fits) but has not shown up in the Gaussian map itself. This source is indeed real and has indeed been fit with a Gaussian and detected as a source; the sizes and flux densities in the output list of sources are consistent with direct map measurements.  The fact that no Gaussian is seen in the model appears to be peculiar to the {plotting} program that generates the output image of the Gaussian model map. In other words, PyBDSF is {\it not} generating a spurious source at this position {in the tables}.

The second is the contour centred at the location, RA = 12$^{\rm h}$ 36$^{\rm m}$ 31$\rasec$11, Dec = +25$^\circ$ 56$^\prime$ 44$\decsec$2, which has been fit with a Gaussian (black ellipse) but for which no output source  has been generated. That is, PyBDSF has fit a Gaussian at this location, but has not accepted the result as representing a real source. At this location, the peak map pixel has a value of 5.3 times the rms plus the mean at the corresponding pixel.  However, the average map value within the displayed contour is only 1.4 times the rms plus the mean over the same region.  Consequently, the source satisfied the selection criterion for a Gaussian fit, but not, evidently, for source extraction.

PyBDSF flags Gaussians for a variety of reasons, including if the Gaussian centre is outside of the island or if the size of the Gaussian is exceptionally large.  Inspection of the flagged islands indicate that most have been flagged because they exceed 25 times the beam size. These have been excluded from this analysis.

The final C-band (magenta circles), L-band (red squares), and X-ray (blue crosses) sources have been overlaid in the figures of  Appendix~\ref{a:figs_r}. {The tables of Appendix~\ref{a:c-radio} contain a complete set of all sources that meet the criteria specified above. Broad scale emission tends to be removed, either because large sources are flagged, or because there is insufficient variation to the brightness distribution for them to be identified as `sources'.  Any larger sources that remain in the tables can easily be identified by examining the major and minor axis sizes, as described in Appendix~\ref{a:c-radio}.}

\newpage
\clearpage
\newpage
\mbox{~}
\vspace{-0.5truein}
\section{Description of Radio Source Catalogues}\label{a:c-radio}

Table~\ref{t:s_n4565_lb} shows the first two rows, representing the first two sources, detected in L-band for NGC~4565, with values rounded for brevity.
The C-band catalogues have the same format.
 The full catalogues for all galaxies are only published electronically online as ascii text in `comma-separated-variable' (csv) format. 
\smallskip

\noindent Example filenames are:\\
\indent\indent N4565\_B\_L.csv\\
\indent\indent N4565\_C\_C.csv\\
\noindent The first letter and number give the name of the galaxy.  Following that are the VLA array configuration and frequency band (e.g. B\_L for B-configuration, L-band) of the image from which PyBDSF measurements were made.  These files are available for download in the tar (Tape ARchive) files called
Appendix\_B\_tables\_Lband.tar and Appendix\_B\_tables\_Cband.tar for L-band and C-band files, respectively. 

We describe the entries of the catalogues here. {The reader should note that PyBDSF output gives many more figures than are justified by the quoted uncertainties and appropriate adjustments to the significant figures should be made.}
\smallskip

\noindent The first six rows contain limited header information. The galaxy centre from Table~\ref{t:g_param}, in degrees, is provided in Row 5.\\

\noindent{\bf Column 1:} Unique source identification number {\bf (Source\_id)}, starting at zero.\\

\noindent{\bf Column 2:} Unique Island identification number {\bf (Isl\_id)}, starting at zero.\\

\noindent {\bf Columns 3, 4, 5 and 6:} Source Right Ascension {\bf (RA)}, the error in RA {\bf (E\_RA)}, Declination {\bf (Dec)}, and error in Dec {\bf (E\_Dec)}. Quoted errors are 1-$\sigma$ values.\\ 

\noindent {\bf Columns 7 and 8:} The integrated Stokes I flux of the source {\bf (Total\_flux)} and its error {\bf (E\_Total\_flux)}. These values are {\it uncorrected} for the primary beam (PB). {Note that these uncertainties are propagated by PyBDSF as outlined in Appendix~\ref{a:param}. }\\

\noindent {\bf Columns 9 and 10:} Value of the flux at the location of the source maximum {\bf (Peak\_flux)} and its error {\bf (E\_Peak\_flux)}, uncorrected for the PB.\\

\noindent {\bf Columns 11, 12, 13 and 14:} The RA {\bf (RA\_max)}, its error {\bf (E\_RA\_max)}, the Dec {\bf (Dec\_max)} and its error {\bf E\_Dec\_max)}, associated with the peak flux given in Column 9.\\

\noindent {\bf Columns 15, 16, 17, 18, 19, and 20:} FWHM of the source major axis {\bf (Maj)} with its error {\bf (E\_Maj)}, FWHM of the source minor axis {\bf (Min)} with its error {\bf (E\_Min)}, and major axis position angle {\bf (PA)} with its error {\bf (E\_PA)}  measured North through East.\\

\noindent {\bf Columns 21, 22, 23, 24, 25, and 26 :} {\bf DC\_Maj, E\_DC\_Maj, DC\_Min, E\_DC\_Min, DC\_PA, and E\_DC\_PA} are the
same as Columns 15 through 20, but refer to the source after the synthesized beam has been deconvolved.  A zero means that the source is unresolved.\\

\noindent{\bf Columns 27, 28, 29, 30, 31, and 32:}
Integrated total flux of the island within which the source is located {\bf (Isl\_Total\_flux)} and its error {\bf (E\_Isl\_Total\_flux)}. The original background rms value  {\bf (Isl\_rms)} and original background mean {\bf (Isl\_mean)}, as computed using the sliding box (Appendix~\ref{a:param}). Average residual background rms over the island {\bf (Resid\_Isl\_rms)} and residual mean {\bf (Resid\_Isl\_mean)} after subtracting all sources in the island.
\\

\noindent{\bf Columns 33 and 34:}
Angular distance of the source from the galaxy centre, computed from RA and Dec using a standard spherical coordinate system {\bf (Dist\_to\_center)}.  Corresponding angle {\bf (Angle\_from\_x=0)}, measuring counterclockwise from x=0 (due west).  A source at 90 degrees would be due north from the galaxy centre and a source at 270 degrees would be due south. The two sources that are displayed in Table~\ref{t:s_n4565_lb} are both in the 3rd quadrant. These columns are introduced to help guide the eye when comparing sources in the catalogues to the maps shown in Appendix~\ref{a:figs_r}.
\\

\noindent{\bf Column 35 (PB\_corr\_factor):} Primary beam (PB) attenuation at the position of the centre of the source.  All listed fluxes should be divided by this factor  to correct for PB attenuation. The PB weighting is given in \cite{per16} who indicate that a typical error in this factor is 0.5\% in both bands.  This error does not take into account systematic errors that may be present due to departures from the assumption that the PB is spherically symmetric.  According to \cite{bha13}, the mapping parameters
 used for CHANG-ES (MTMFS+SI in their Fig. 4) could result in larger PB errors that increase with distance from the map centre.  These start to depart from the 0.5\% quoted above once PB\_corr\_factor falls below 0.85.  In the range,
0.85 $\ge$ PB\_corr\_factor $\ge$ 0.60, the error steadily increases from 0.5\%  to 1.5\%, and in the range, 0.60 $\ge$ PB\_corr\_factor $\ge$ 0.11, the error increases to 6.3\%.  Once PB\_corr\_factor falls below 0.11, the PB correction is unreliable.

In C-band, eight galaxies were observed in two different pointings for which the primary beams overlapped \citep{wie15}.  For these galaxies (NGC~891, NGC~3628, NGC~4244, NGC~4565, NGC~4594, NGC~4631, NGC~5084, and NGC~5907), the PB\_corr\_factor column contains the value determined from the combined, weighted PB map applicable to the galaxy in question.  An empty cell means that the position falls outside of the 20\% level of the combined beam.
\\


 \noindent {\bf Column 36 (S\_Code):}
Structure code of the source. This element has as output either ``S", ``C", or ``M", where ``S" indicates an isolated source that is fit with a single Gaussian; “C” represents sources that are fit by a single Gaussian but are within an island of emission that also contains other sources; and “M” is used for sources that are extended and have been fitted with multiple Gaussians.
 
\clearpage
\begin{center}
\begin{table*}
\caption{First two sources from L-band catalogue NGC~4565 (filename: N4565\_B\_L.csv)}\label{t:s_n4565_lb}
    \begin{tabular}[b]{| c | c | c | c | c | c | c| c|}
    \hline\hline
         1 & 2 & 3 & 4 & 5 & 6 & 7 & 8\\ \hline
         Source\_id & Isl\_id & RA & E\_RA& Dec & E\_Dec &Total\_flux & E\_Total\_flux\\
          & & (deg) & (deg) & (deg) & (deg)& ($\mu$Jy)& ($\mu$Jy)\\
         \hline\hline
         0 & 0 & $189.23815$ & $1.4\times 10^{-5}$ & $25.86675$ & $1.2\times 10^{-5}$ & 439.7 & 25.6
\\
 \hline
            1 & 1 & $189.23833$ & $7.6\times 10^{-5}$ & $25.96159$ & $5.1\times 10^{-5}$ & 182.9 & 36.8
\\
         \hline
    \end{tabular}
    \end{table*}

\begin{table*}
\begin{tabular}{| c | c | c | c | c | c | c | c |}
    \hline\hline
        9 & 10 & 11 & 12 & 13 & 14 & 15 & 16\\ \hline
          Peak\_flux & E\_Peak\_flux & RA\_max & E\_RA\_max & Dec\_max & E\_Dec\_max & Maj & E\_Maj  \\
          ($\mu$Jy beam$^{-1}$) & ($\mu$Jy beam$^{-1}$) & (deg) & (deg) & (deg) & (deg) & (arcsec) & (arcsec)\\
         \hline\hline
       $426.0$ & $14.4$ &	$189.23815$ &	$1.4\,\times\,10^{-5}$ &
       $25.86675$ & $1.2\,\times\,10^{-5}$
       & $3.46$ & $0.13$
\\
         \hline
       $118.1$ & $15.6$ &	$189.23833$ &	$7.6\,\times\,10^{-5}$ &
       $25.96159$ & $5.1\,\times\,10^{-5}$
       & $4.49$ & $0.66$
\\
         \hline
    \end{tabular}
    \end{table*}
    
    \begin{table*}
\begin{tabular}{| c | c | c | c | c | c | c | c | c | c |}
    \hline\hline
        17 & 18 & 19 & 20 & 21 & 22 & 23 & 24 & 25 & 26\\ \hline
          Min & E\_Min & PA & E\_PA & DC\_Maj & E\_DC\_Maj & 
          DC\_Min & E\_DC\_Min & DC\_PA & E\_DC\_PA \\
          (arcsec) & (arcsec) & (deg) & (deg) & (arcsec) & (arcsec) & (arcsec) & (arcsec)
          & (deg) & (deg)\\
         \hline\hline
       $2.97$ & $0.09$ &	$55.2$ & $9.8$ &
       $0$ & $0.13$
       & $0$ & $0.09$ & $0$ & $9.8$
\\
         \hline
       $3.43$ & $0.41$ & $73.8$ &	$23.3$ &
       $3.11$ & $0.66$
       & $1.49$ & $0.41$ & $80.0$ & $23.3$
\\
         \hline
    \end{tabular}
    \end{table*}
    
       \begin{table*}
\begin{tabular}{| c | c | c | c | c | c |}
    \hline\hline
        27 & 28 & 29 & 30 & 31 & 32 \\ \hline
          Isl\_Total\_flux & E\_Isl\_Total\_flux & Isl\_rms & Isl\_mean & Resid\_Isl\_rms & Resid\_Isl\_mean  
\\
          ($\mu$Jy) & ($\mu$Jy) & ($\mu$Jy beam$^{-1}$) & ($\mu$Jy beam$^{-1}$) & ($\mu$Jy beam$^{-1}$) & ($\mu$Jy beam$^{-1}$)
          \\
         \hline\hline
       $424.1$ & $18.6$ &	$14.3$ & $-0.300$ &
       $6.30$ & $0.053$
\\
         \hline
       $138.0$ & $16.5$ & $14.7$ &	$-0.300$ &
       $6.21$ & $-0.13$
\\
         \hline
    \end{tabular}
    \end{table*}
    
\begin{tabular}{| c | c | c | c |}
    \hline\hline
        33 & 34 & 35 & 36
        \\ \hline
          Dist\_to\_center & Angle\_from\_x=0 &  PB\_corr\_factor & S\_Code  
  \\
          (arcmin) & (deg) &  &   \\
         \hline\hline
       $10.93$ & $221.5$ & 0.64	& S
\\
         \hline
       $8.33$ & $190.8$ & 0.77 & S 
\\
         \hline
    \end{tabular}

\end{center}

\newpage
\clearpage
\newpage
\mbox{~}

\section{Description of Cross-Matched L-band and C-band Radio Catalogue}\label{a:cross_L_C}

Table~\ref{t:cross_L_C} shows the first two rows of the table that lists all sources that were cross-matched at the two radio bands.  All galaxies are in a downloadable single file in {csv} format, called Appendix\_C\_cross\_matched\_LC.csv. The first three rows contain table headings as shown in Table~\ref{t:cross_L_C} below.  The column descriptions are given below. {Again, as noted in Appendix~\ref{a:c-radio}, the reader should make appropriate adjustments to the significant figures according to the quoted errors.}\\

\noindent {\bf Column 1:} Name of the galaxy {\bf (Galaxy)}.\\

\noindent {\bf Column 2:} Unique L-band/C-band cross-match identification number {\bf (LC\_id)} starting at zero.\\

\noindent{\bf Columns 3 and 4:} Unique source identification number {\bf (Source\_id) } starting at zero, equivalent to Column 1 of Table~\ref{t:s_n4565_lb}, for L-band and C-band, respectively.\\

\noindent{\bf Columns 5, 6, 7, and 8:}  Flux density of the source {\bf (Flux\_cor)} and its error {\bf (E\_flux\_cor)} corrected for the primary beam (PB), for L-band and C-band, respectively. These values were computed using Columns 7 and 8 divided by Column 35 {\it of 
 Table~\ref{t:s_n4565_lb}},  for the respective bands.
 If the PB correction falls below 0.11, as outlined in Appendix~\ref{a:c-radio} (i.e. if PB\_corr\_factor $\le\,0.11)$, the flux density becomes unreliable, so the cell is left blank.  {The error in the flux density includes the error as computed by PyBDSF (Appendix~\ref{a:param}) plus the error in the primary beam (see comments for Column 35 of Appendix~\ref{a:c-radio}) plus 1\% to allow for a calibration error \citep{per13}, all added in quadrature. } \\

\noindent {\bf Columns 9 and 10:} Spectral index {\bf (Spectral index)} and its error {\bf (E\_SpIn)}, respectively.  The spectral index was computed from Eq.~\ref{eq:alpha}, where the flux densities refer to the primary beam corrected values (Columns 5 and 7 of this table). {As indicated above, if a flux density is unreliable, the spectral index cell is also left blank.  The error in the spectral index was calculated from
\begin{equation}
    \sigma_\alpha\,=\,
    \frac{1}
    {|ln(\nu_L/\nu_C)|}\,
    \sqrt{
    \left(\frac{\sigma_{S_L}}{S_L}\right)^2\,+\,
    \left(\frac{\sigma_{S_C}}{S_C}\right)^2
    }
    \end{equation}
    where $S$ is again taken from Columns 5 and 7 of this table, and $\sigma$ is taken from Columns 6 and 8 of this table, with subscripts indicating the band.}\\

\noindent{\bf Columns 11, 12, 13, and 14:} Distance to the centre of the map {\bf (Dist\_to\_center)} for L-band and then C-band, and angle from x=0 {\bf (Angle\_from\_x=0)} for L-band and then C-band.  These values are equivalent to those given in Columns 33 and 34 of Table~\ref{t:s_n4565_lb} for the respective bands.\\

\noindent{\bf Column 15:} Any source that is within 3 arcsec ($\approx$ the spatial resolution) from the centre of the galaxy, whose coordinates are given in Table~\ref{t:g_param}, have been identified in this column {\bf (Nuclear?) }. The cell is left blank if this condition is not met.\\

\newpage

\begin{table*}
\caption{First two rows from Cross-matched L-band and C-band Catalogues  (filename: Appendix\_C\_cross\_matched\_LC.csv)}\label{t:cross_L_C}
    \begin{tabular}[b]{| c | c | c | c | c | c | c| c|}
    \hline\hline
         1 & 2 & 3 & 4 & 5 & 6 & 7 & 8\\ \hline
         Galaxy & LC\_id & Source\_id& Source\_id& Flux\_cor & E\_flux\_{cor} &Flux\_{cor}  & E\_flux\_{cor}\\
          & & L-band & C-band & L-band & L-band& C-band& C-band\\
          & & &  &($\mu$Jy)  & ($\mu$Jy)& ($\mu$Jy)& ($\mu$Jy)\\
         \hline\hline
         NGC~660& 0& 4 & 0 & 238.8 & 55.6 & 455.7& 63.5
\\
 \hline
            NGC~660 & 1 & 7 & 1 & 156.1 & 52.2 & 63.7 & 21.4
\\
         \hline
    \end{tabular}
    \end{table*}
    
  \begin{table*}
\begin{tabular}{| c | c | c | c | c | c | c |}
    \hline\hline
        9 & 10 & 11 & 12 & 13 & 14 & 15 \\ \hline
          Spectral index  & E\_SpIn & Dist\_to\_center & Dist\_to\_center &Angle\_from\_x=0  &  Angle\_from\_x=0& Nuclear?  \\
            &  & L-band & C-band & L-band & C-band &\\
          &  & (arcmin) & (arcmin) & (deg) & (deg) &  \\
         \hline\hline
       +0.483& 0.205 &	5.5716 & 5.5705 &
        170.33 & 170.37   &  \\
         \hline
       -0.670& 0.358  &4.1041	 &4.1091	 &151.81
        & 151.87
       &  \\
         \hline
    \end{tabular}
    \end{table*}


\newpage
\clearpage
\newpage
\section{Description of X-ray source catalogue}\label{a:c-xray}

Table~\ref{t:660_xray} shows the first two rows of the X-ray sources detected in the field of NGC 660. The full catalogue of the 27 galaxies is only published electronically online. \\
\noindent The individual files are called:\\
NGC0660\_xray\_src.csv\\
NGC0891\_xray\_src.csv\\
etc., for all of the galaxies.  All files are available for download, along with a text file called, xray\_table\_ReadMe.txt which contains the descriptive information described below.  The tar file containing all files is called:\\
Appendix\_D\_tables\_Xray.tar\\

\noindent The first row contains column headings.  Units are shown in Table~\ref{t:660_xray}. The columns are:\\

\noindent {\bf Column 1 (ID):} Unique source identification number from the \textsc{wavedetect} output, starting at zero.\\

\noindent {\bf Columns 2-4:} Source Right Ascension ({\bf RA}, degrees), Declination ({\bf Dec}, degrees), and positional uncertainty ({\bf delta\_x}, arcsec). Each uncertainty is the addition in quadrature of the statistical and systematic contributions (see Section \ref{ss:Obs_xray}), and represents error radii in both RA and Dec.\\

\noindent{\bf Columns 5-16:} The total X-ray photon counts detected, the number of X-ray photon counts per second, and the count rate error, first for the soft band ({\bf Sft\_Cts, Sft\_Rate, E\_Sft\_Rate}) followed by the same quantities in the medium band 
({\bf Med\_Cts, Med\_Rate, E\_Med\_Rate}), followed by the same quantities in the hard band ({\bf Hrd\_Cts, Hrd\_Rate, E\_Hrd\_Rate}), followed finally by the same quantities in the broad band
({\bf Brd\_Cts, Brd\_Rate, E\_Brd\_Rate}). 
The energy bands are defined as: 0.3-1.2~keV (Soft, Sft), 1.2-2~keV (Medium, Med), 2-7~keV (Hard, Hrd) and 0.3-7~keV (Broad, Brd).\\

\noindent{\bf Columns 17-20 (HR1, E\_HR1, HR2, E\_HR2):}  Hardness ratios and their errors. Hardness ratios defined as HR1 = (Med\_Rate-Sft\_Rate)/(Med\_Rate+Sft\_Rate) and HR2 = (Hrd\_Rate-Med\_Rate)/(Hrd\_Rate+Med\_Rate). Hardness ratios are listed only for sources with broad band counts greater than 10.   \\

\newpage
\begin{table*}
\caption{First two rows from the NGC 660 Catalogue of X-ray Sources}\label{t:660_xray}
    \begin{tabular}[b]{| c | c | c | c|c | c | c |}
    \hline\hline
         1 & 2 & 3 & 4 & 5 & 6 & 7 \\ \hline
         ID & RA & Dec& delta\_x& Sft\_Cts& Sft\_Rate &E\_Sft\_Rate\\
          & (deg)& (deg) & (arcsec)& (counts) &($10^{-3}$ counts s$^{-1}$) & ($10^{-3}$ counts s$^{-1}$) \\
         \hline\hline
         0& 25.71908 &13.62213& 0.42 & 1.8 &  0.029&0.041 \\
 \hline
            1 & 25.75566 & 13.63984& 0.30 & 2.8 & 0.047&0.050 \\
         \hline
    \end{tabular}
    \end{table*}
    
    \begin{table*}
\begin{tabular}{| c | c | c | c  | c }
    \hline\hline
        8 & 9 & 10 & 11 & 12  \\ \hline
        Med\_Cts  & Med\_Rate & E\_Med\_Rate& Hrd\_Cts  &Hrd\_Rate \\
          (counts)  & ($10^{-3}$ counts s$^{-1}$) & 
          ($10^{-3}$ counts s$^{-1}$) &(counts) & ($10^{-3}$ counts s$^{-1}$)  \\
         \hline\hline
         2.8& 0.046&0.051 & 3.5& 0.063\\ 
        1.8&0.030&0.043&3.7&0.068\\
        \hline
        \hline
    \end{tabular}
    \end{table*}

 \begin{table*}
\begin{tabular}{| c | c | c | c }
    \hline\hline
        13 & 14 & 15 & 16  \\ \hline
        E\_Hrd\_Rate & Brd\_Cts & Brd\_Rate &  E\_Brd\_Rate\\
        ($10^{-3}$ counts s$^{-1}$) &  (counts)  & ($10^{-3}$ counts s$^{-1}$) & ($10^{-3}$ counts s$^{-1}$)\\
         \hline\hline
         0.062 & 8.1& 0.14 & 0.087 \\
         0.063 & 8.4& 0.15 & 0.087 \\
         \hline
         \hline
    \end{tabular}
    \end{table*}  
    
 \begin{table*}
\begin{tabular}{| c | c | c | c }
    \hline\hline
        17 & 18 & 19 & 20  \\ \hline
        HR1& E\_HR1& HR2 &  E\_HR2\\
         &    & &\\
         \hline\hline
         0& 0& 0& 0 \\
         \hline
         0& 0& 0& 0 \\
         \hline
         \hline
    \end{tabular}
    \end{table*}

\newpage

\clearpage
\newpage
\mbox{~}
\section{Images of CHANGES galaxies showing overlays of detected sources}\label{a:figs_r}

Here we present the images for all CHANGES galaxies, focussing on the region near the galaxy disc.  The entire 16.7 arcmin $\times$ 16.7 arcmin field (L-band field size, Table~\ref{t:fieldsize}) overlay images are available for download in the online file,  Appendix\_A\_figures\_pdf.tar with filenames designated, e.g.
N660\_16.7arcmin\_field.pdf. Information on these images is
as follows:\\

\noindent {\bf Greyscale:} H$\alpha$ images with arbitrary scaling. We do not show the entire field of view, but focus mainly on the galaxy disc and nearby environment. The images are from \cite{var19}, unless otherwise noted, freely available at queensu.ca/changes.  The remaining H$\alpha$ images are taken from the literature.  References for the literature images are as follows:

N891: \cite{pat12}    

N4217:   \cite{ran96}

N4244   \cite{dal09}

N4302: \cite{ran96}

N4438: \cite{ken08}

N4565: \cite{pat12}    

N4594: \cite{ken03} 

N4631: \cite{ken03} 

N5775:    \cite{col00}

N5907:   \cite{ran96}

The exception is NGC~5084 for which a good H$\alpha$ image was not readily available.  For this galaxy, we obtained a Digitized Sky Survey optical IIIaJ map from the Nasa Extragalactic Database and show it instead. \\

\noindent{\bf Yellow `Plus' symbol:} Position of the galaxy centre (Table~\ref{t:g_param}).\\

\noindent {\bf Green ellipse:} The projection of the optical galaxy disc at the 25 magnitude per square arcsec level, 
as specified in Table~\ref{t:g_param}. Note that the position angle has been measured from the K$_s$ passband as given in the Nasa Extragalactic Database (NED), with minor adjustments where necessary (all less than 5 degrees) so that the orientation more closely matched that of the C-band radio data \citep[see][]{kra20}. {In a few cases, the green ellipse is shown to extend beyond the map borders so that the image of the galaxy could be shown at sufficient resolution for clarity.}
\\

\noindent {\bf Magenta circles:} C-band sources detected by PyBDSF via Gaussian fitting, as described in Appendix~\ref{a:param}.  Note that the symbol sizes are selected for clarity and do not represent positional uncertainties, which are less than the 3 arcsec resolution. See the tables of Appendix~\ref{a:c-radio}for information on the sources, their fluxes and sizes.\\

\noindent {\bf Red squares:} L-band sources detected by PyBDSF, with symbol sizes adopted for clarity as was done for C-band.\\

\noindent {\bf Blue crosses:} X-ray sources. Note that some galaxies do not have X-ray data.  The 27 galaxies that do, are indicated in Table~\ref{t:obslog_xray}.

\clearpage
\newpage
\mbox{~}

\begin{figure}[!]
    \centering
    \includegraphics[width=0.8\textwidth]{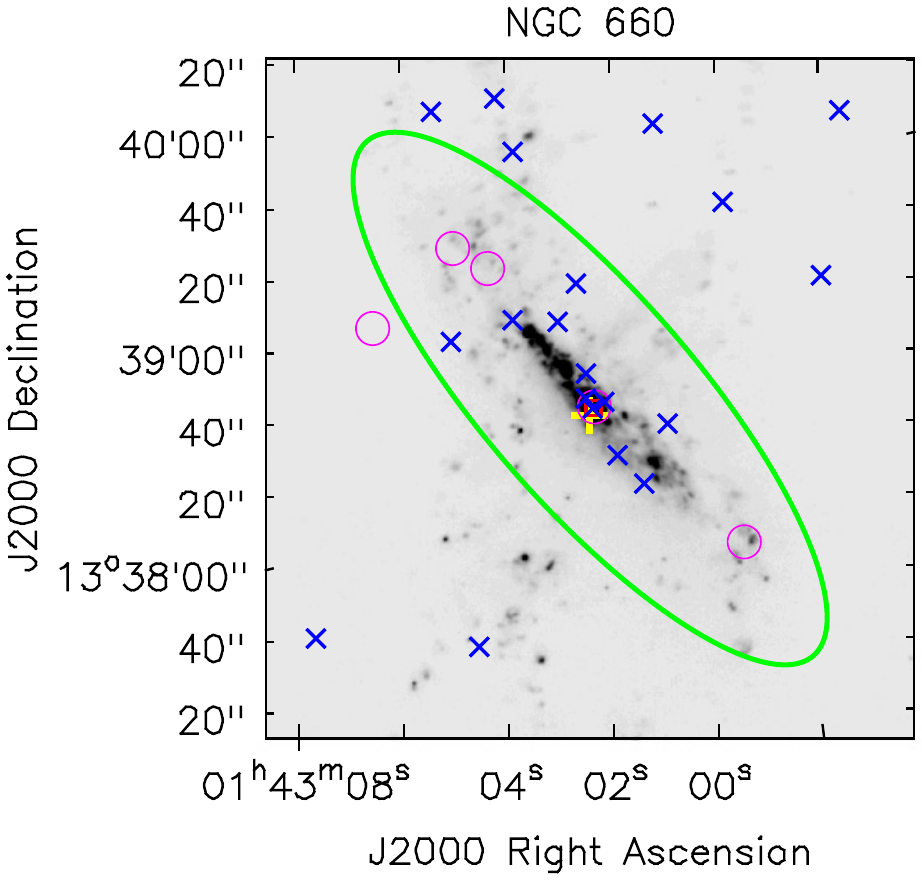}
    \label{f:NGC660}
    \caption{Overlays of the sources on H$\alpha$ greyscale images.  The green ellipse delineates the optical galaxy at the 25 magnitude per square arcsec level. The yellow `+' denotes the galaxy centre. Red squares and magenta circles show the positions of the L-band and C-band sources, respectively; positional uncertainties are much smaller than the symbol sizes. X-ray sources are marked with blue `x' symbols; only 27 galaxies have X-ray data. For more information, see details at the beginning of this appendix.}
\end{figure}

\begin{figure}[!]
    \centering
    \includegraphics[width=0.8\textwidth]{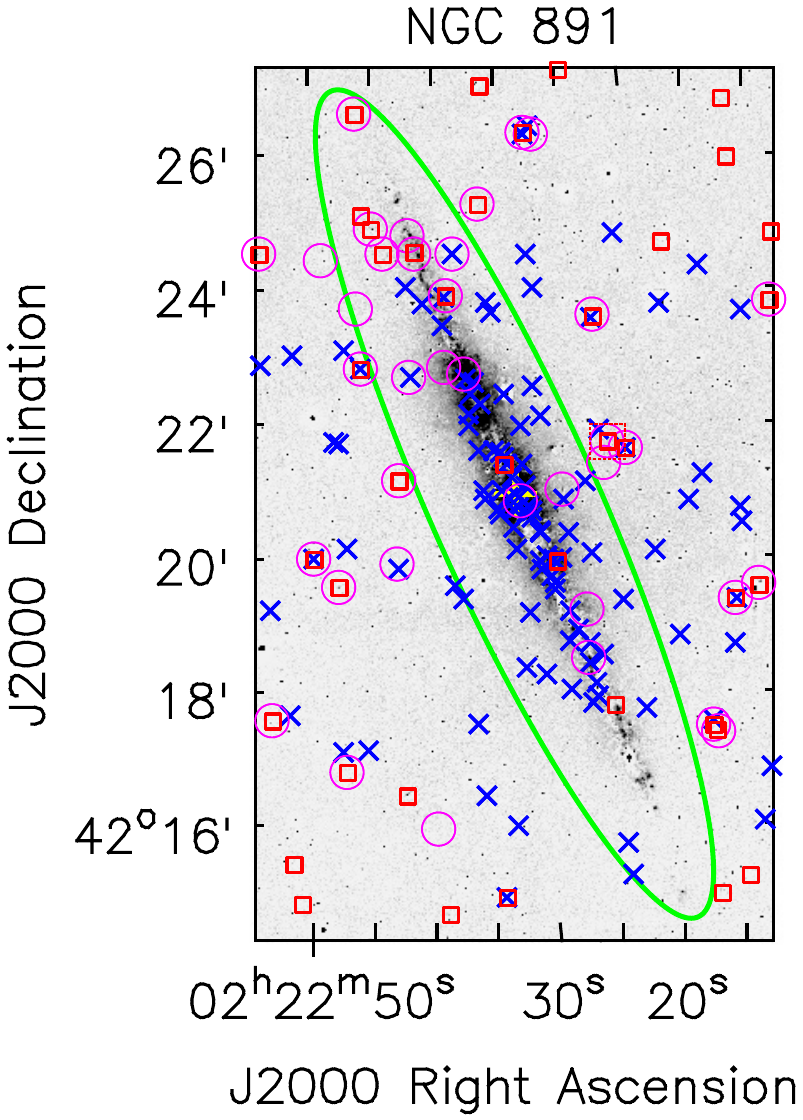}
    \label{f:NGC891}
\end{figure}

\begin{figure}[!]
    \centering
    \includegraphics[width=0.8\textwidth]{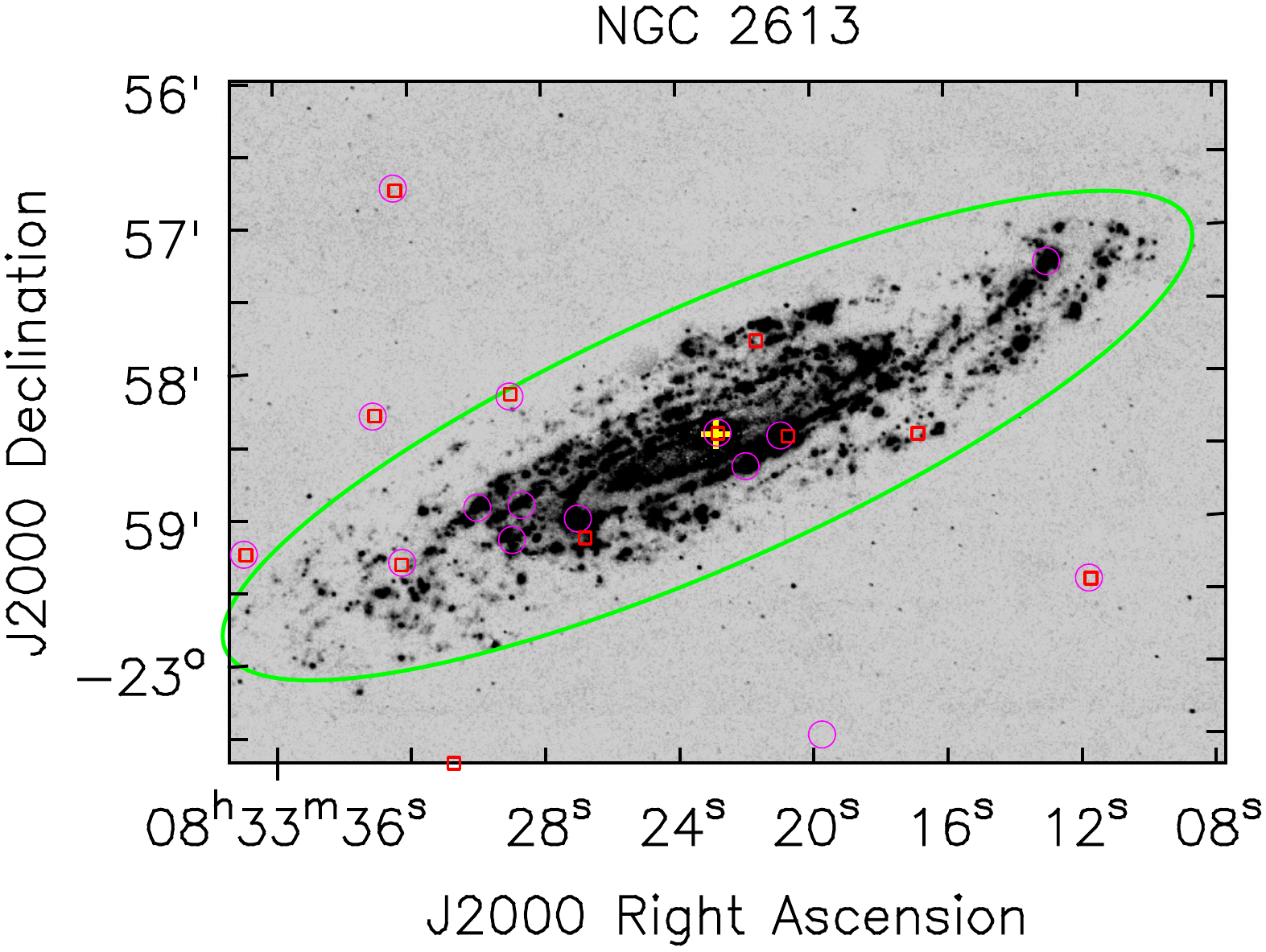}
   \label{f:NGC2613}
\end{figure}

\begin{figure}[!]
    \centering
    \includegraphics[width=0.8\textwidth]{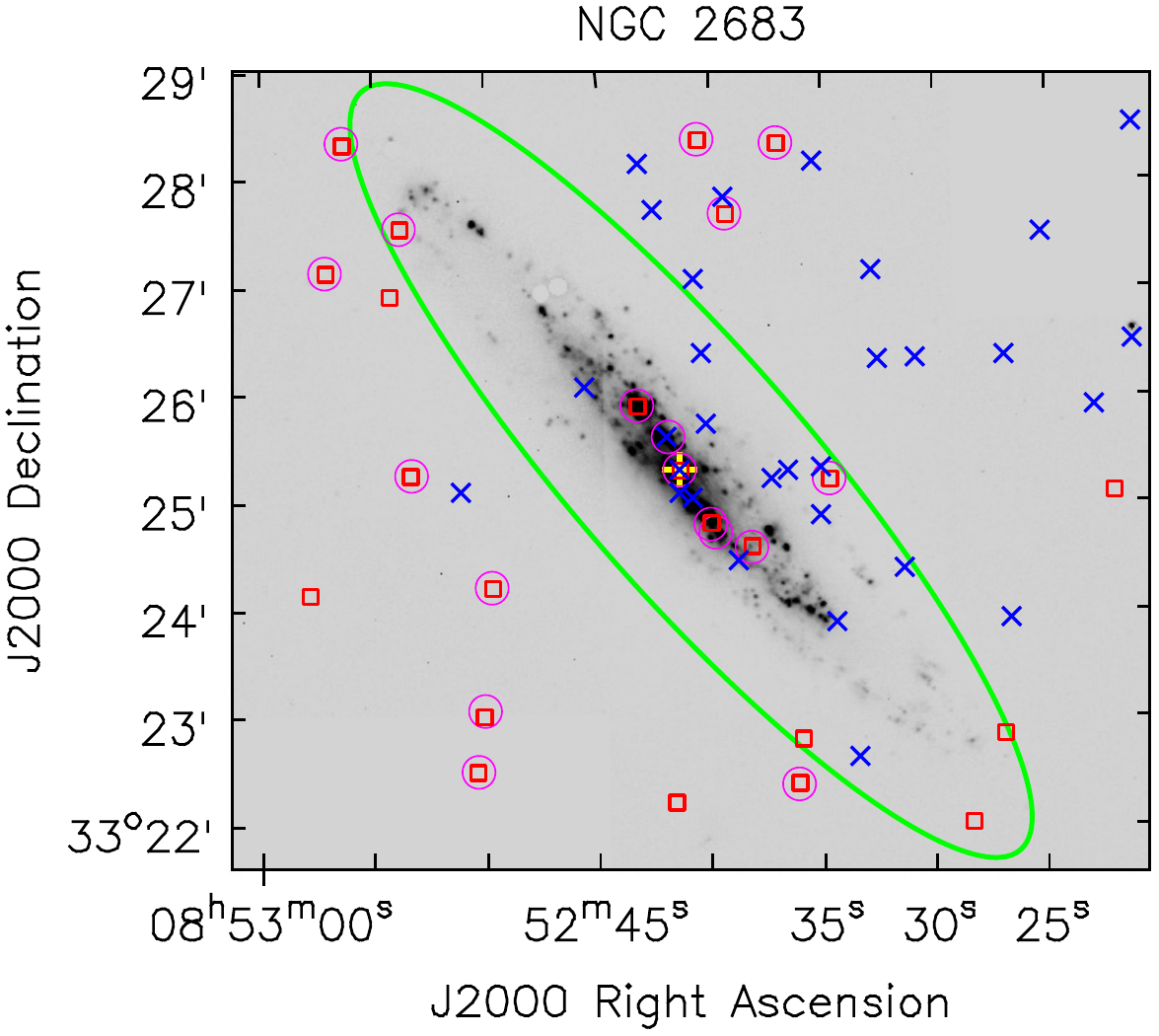}
    \label{f:NGC2683}
\end{figure}

\begin{figure}[!]
    \centering
    \includegraphics[width=0.8\textwidth]{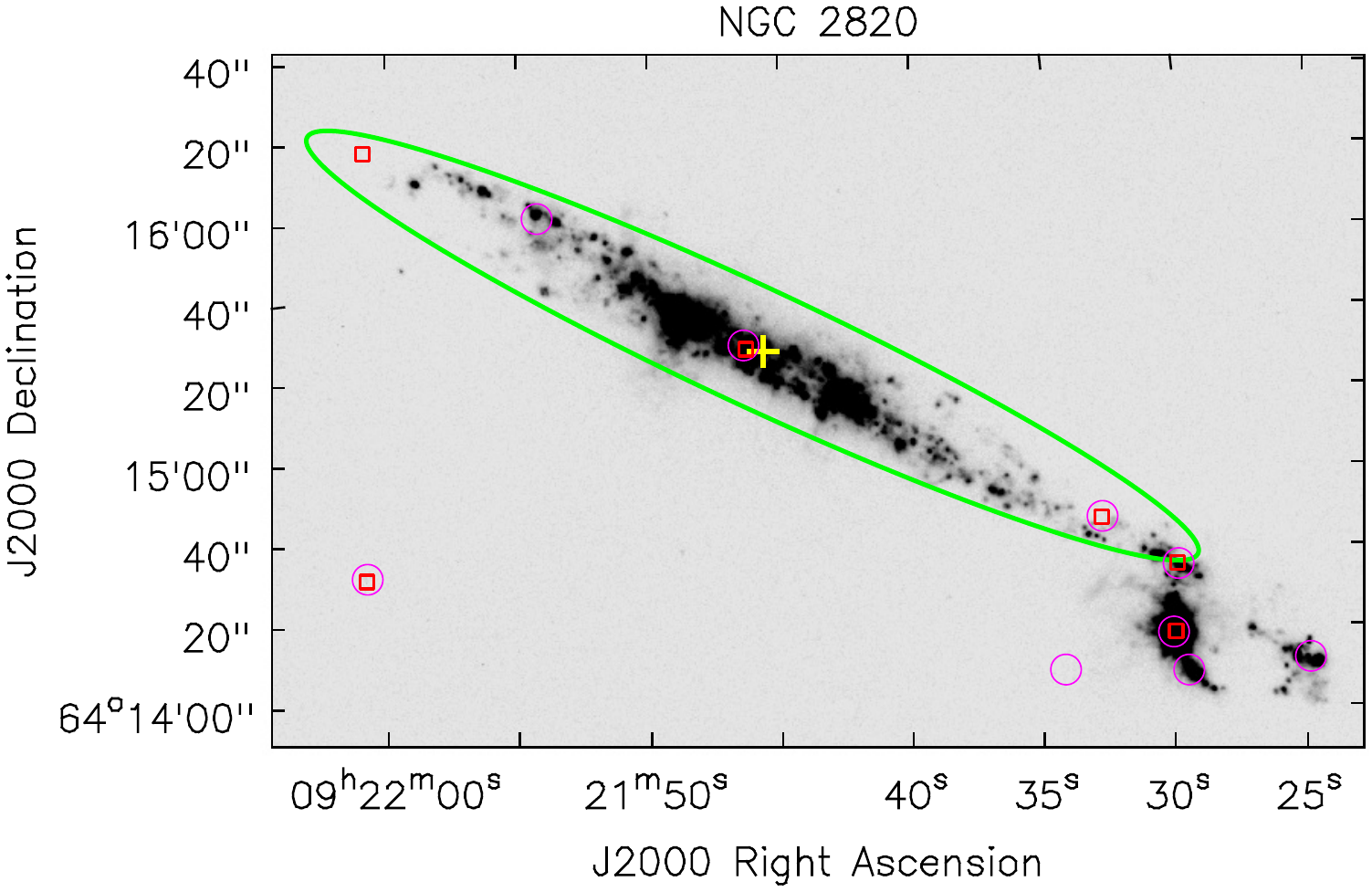}
    \label{f:NGC2820}
    \caption{No X-ray data were available for this galaxy.}
\end{figure}

\begin{figure}[!]
    \centering
    \includegraphics[width=0.8\textwidth]{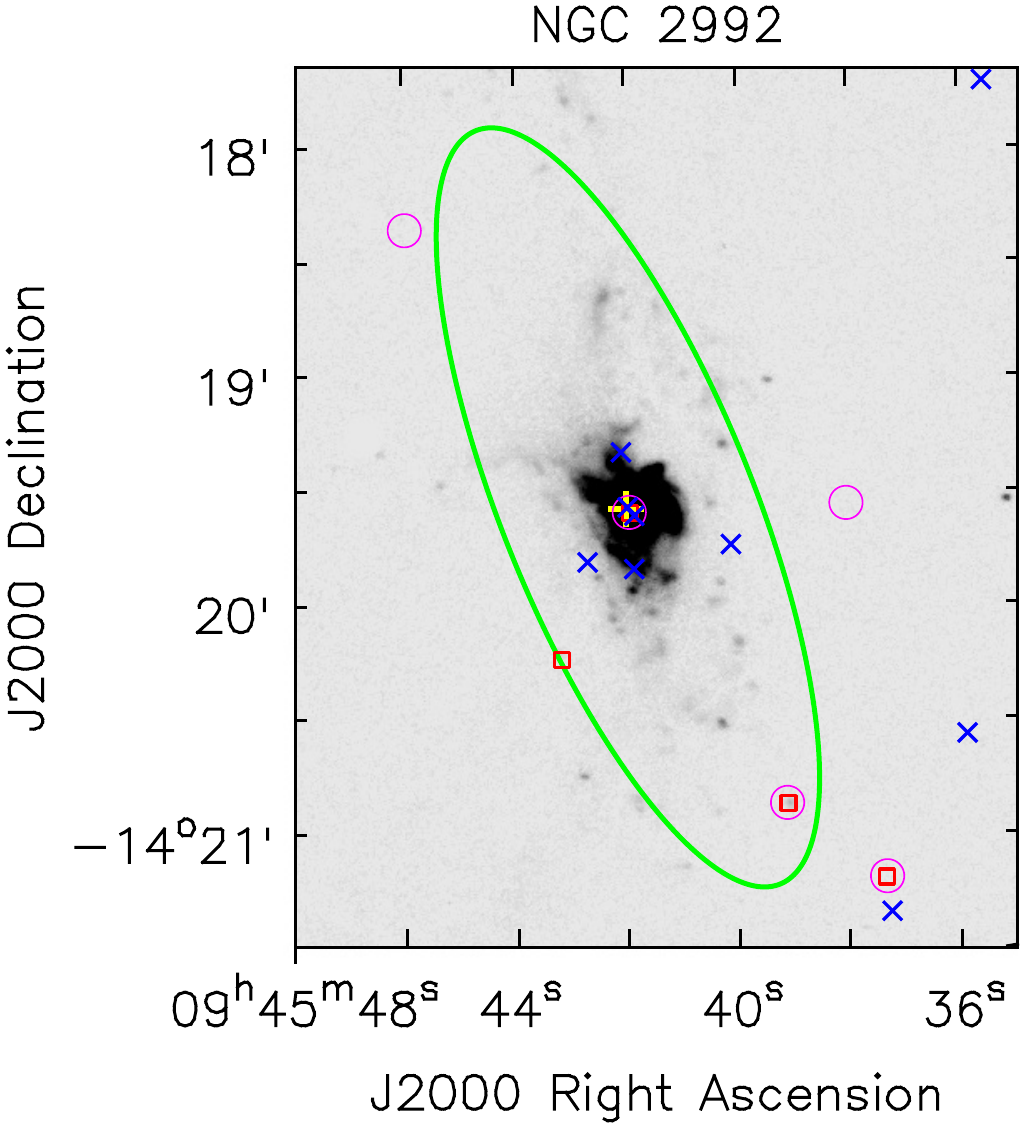}
    \label{f:NGC2992}
\end{figure}

\begin{figure}[!]
    \centering
    \includegraphics[width=0.8\textwidth]{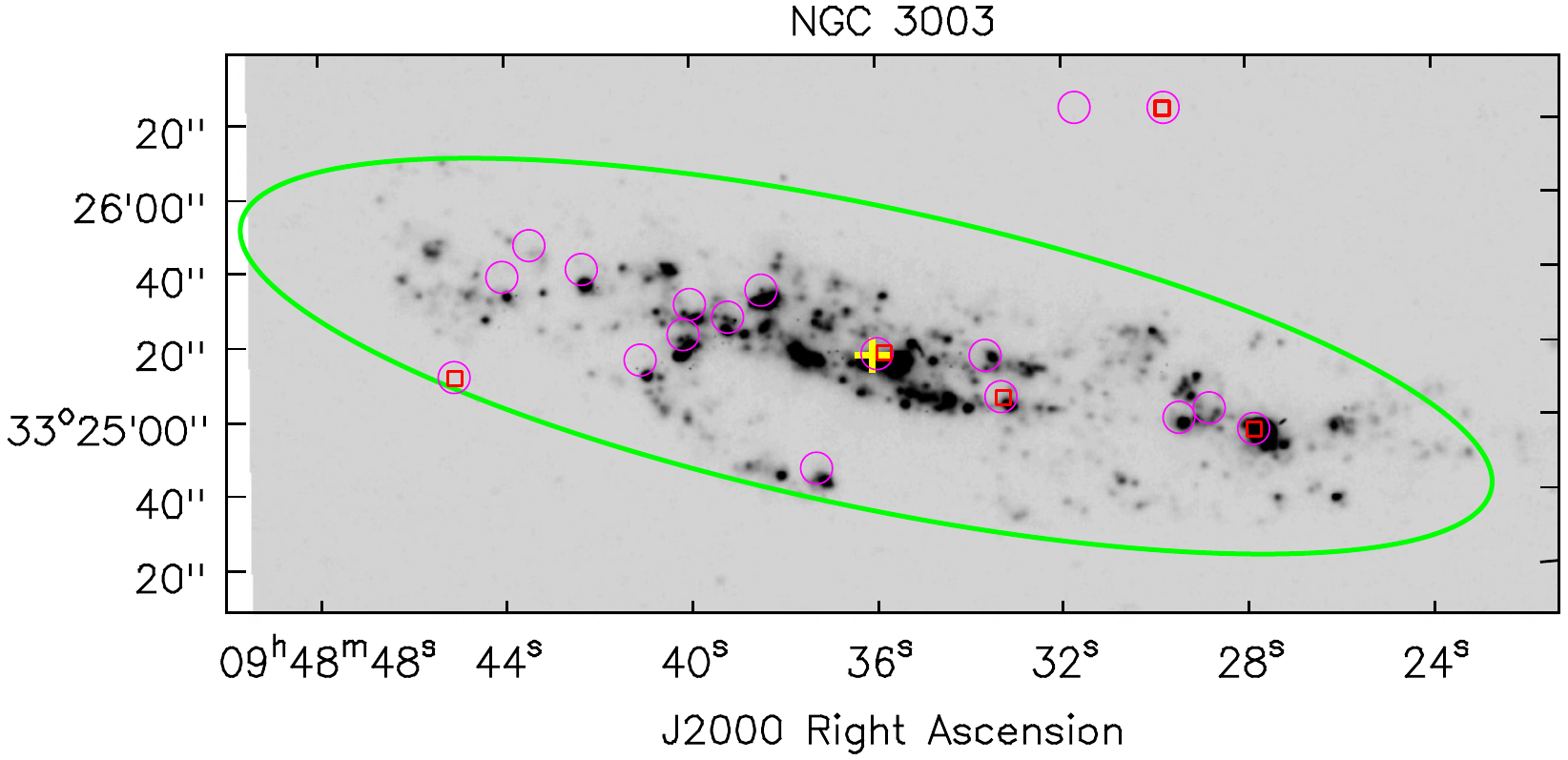}
    \label{f:NGC3003}
\end{figure}

\begin{figure}[!]
    \centering
    \includegraphics[width=0.8\textwidth]{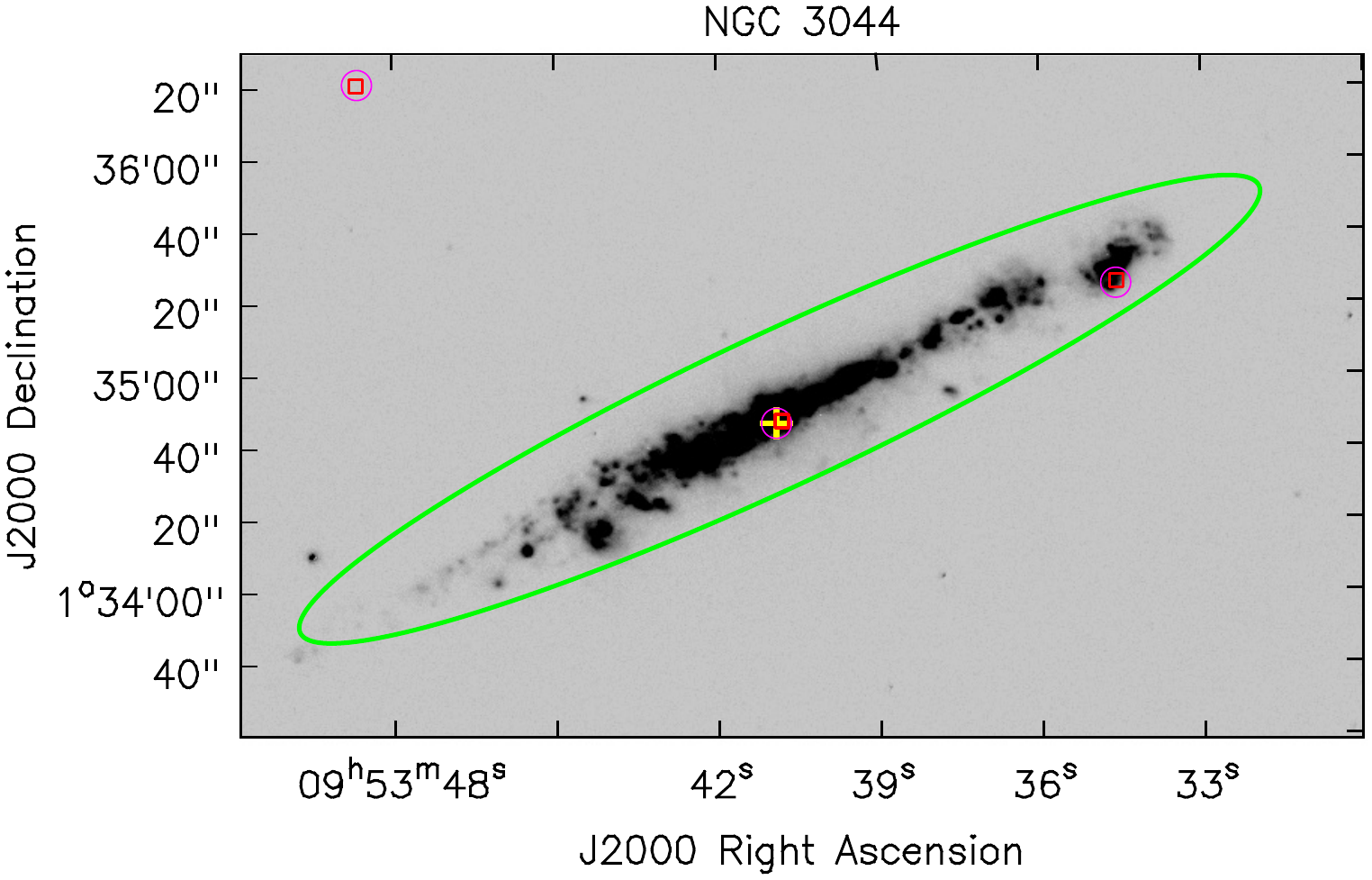}
    \label{f:NGC3044}
    \caption{No X-ray data are available.}
\end{figure}

\begin{figure}[!]
    \centering
    \includegraphics[width=0.8\textwidth]{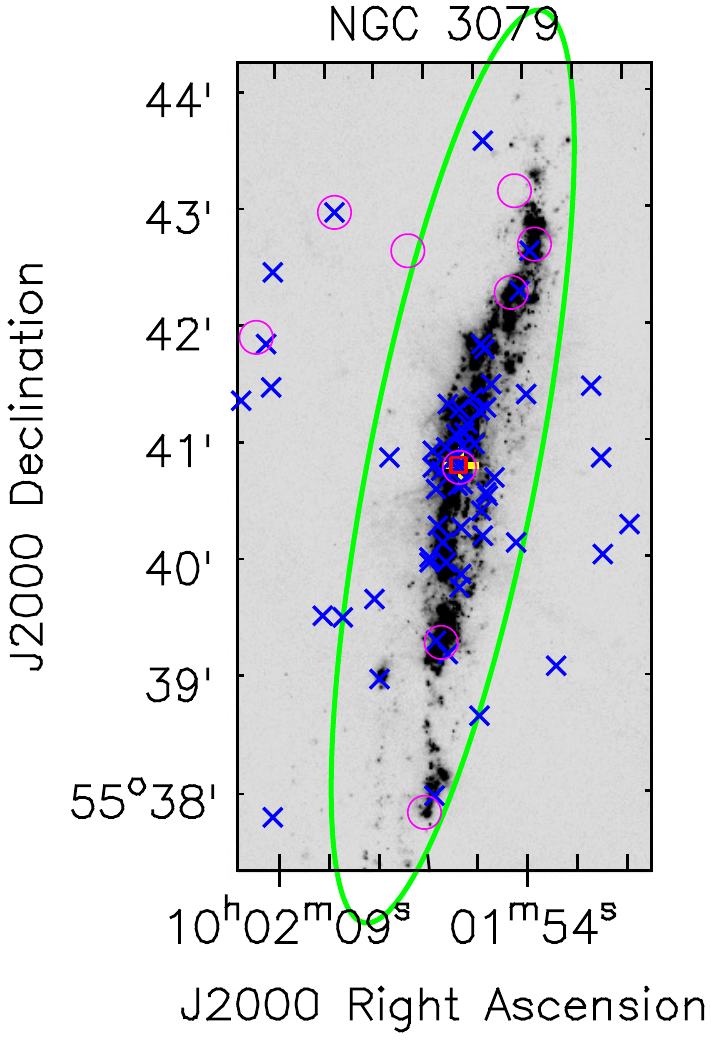}
    \label{f:NGC3079}
\end{figure}

\begin{figure}[!]
    \centering
    \includegraphics[width=0.8\textwidth]{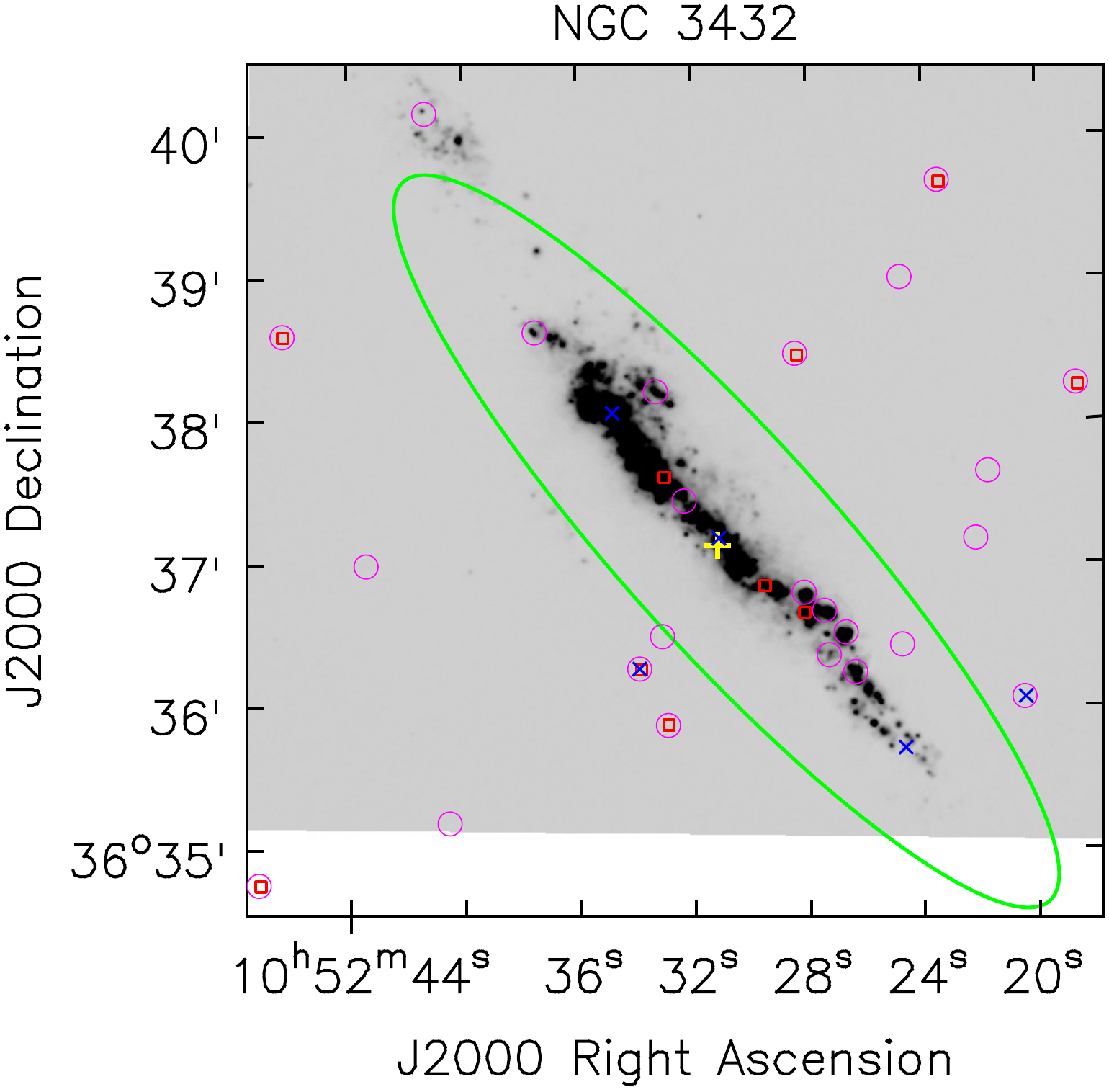}
    \label{f:NGC3432}
\end{figure}

\begin{figure}[!]
    \centering
    \includegraphics[width=0.8\textwidth]{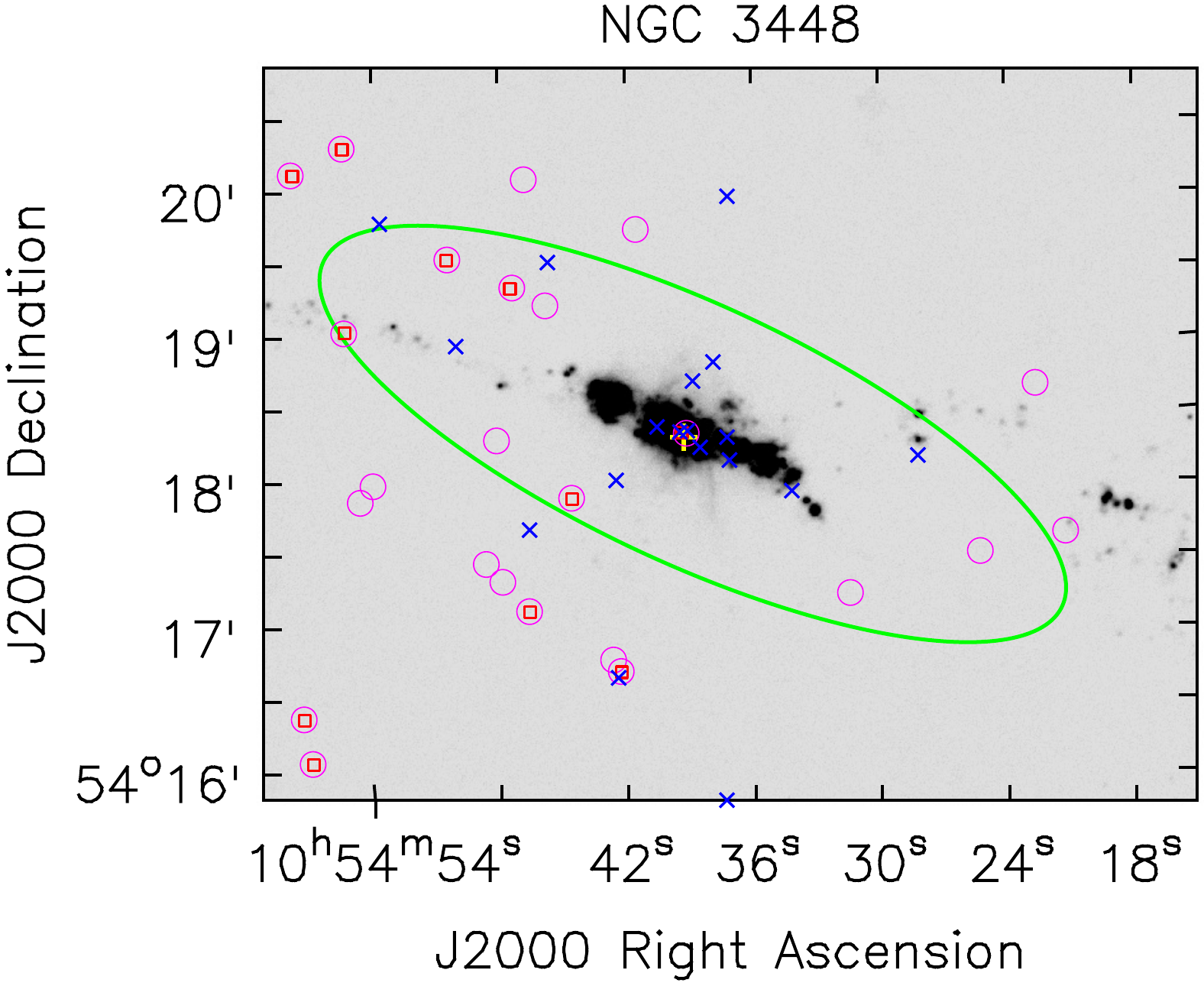}
    \label{f:NGC3448}
\end{figure}

\begin{figure}[!]
    \centering
    \includegraphics[width=0.9\textwidth]{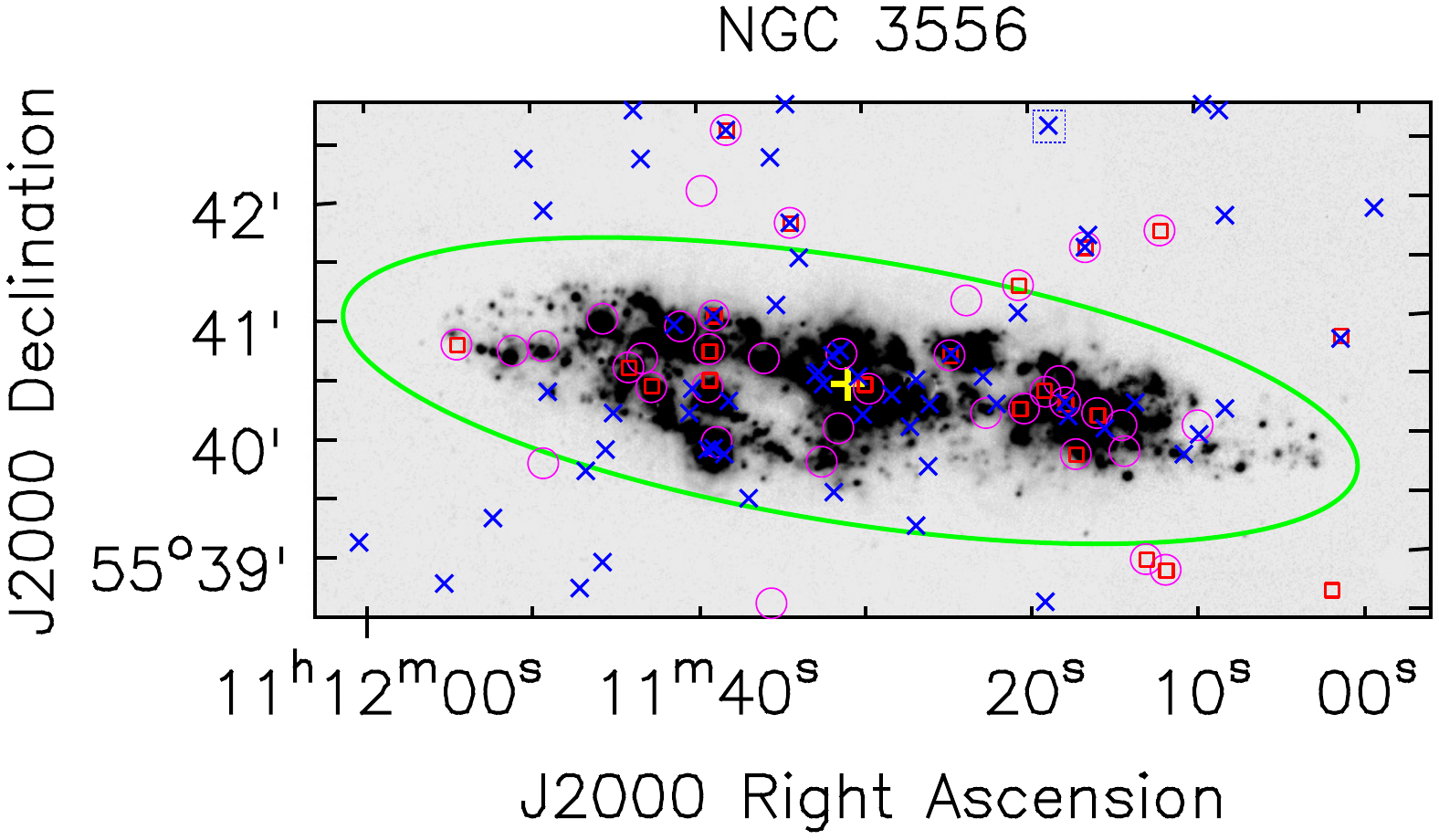}
    \label{f:NGC3556}
\end{figure}

\begin{figure}[!]
    \centering
    \includegraphics[width=0.8\textwidth]{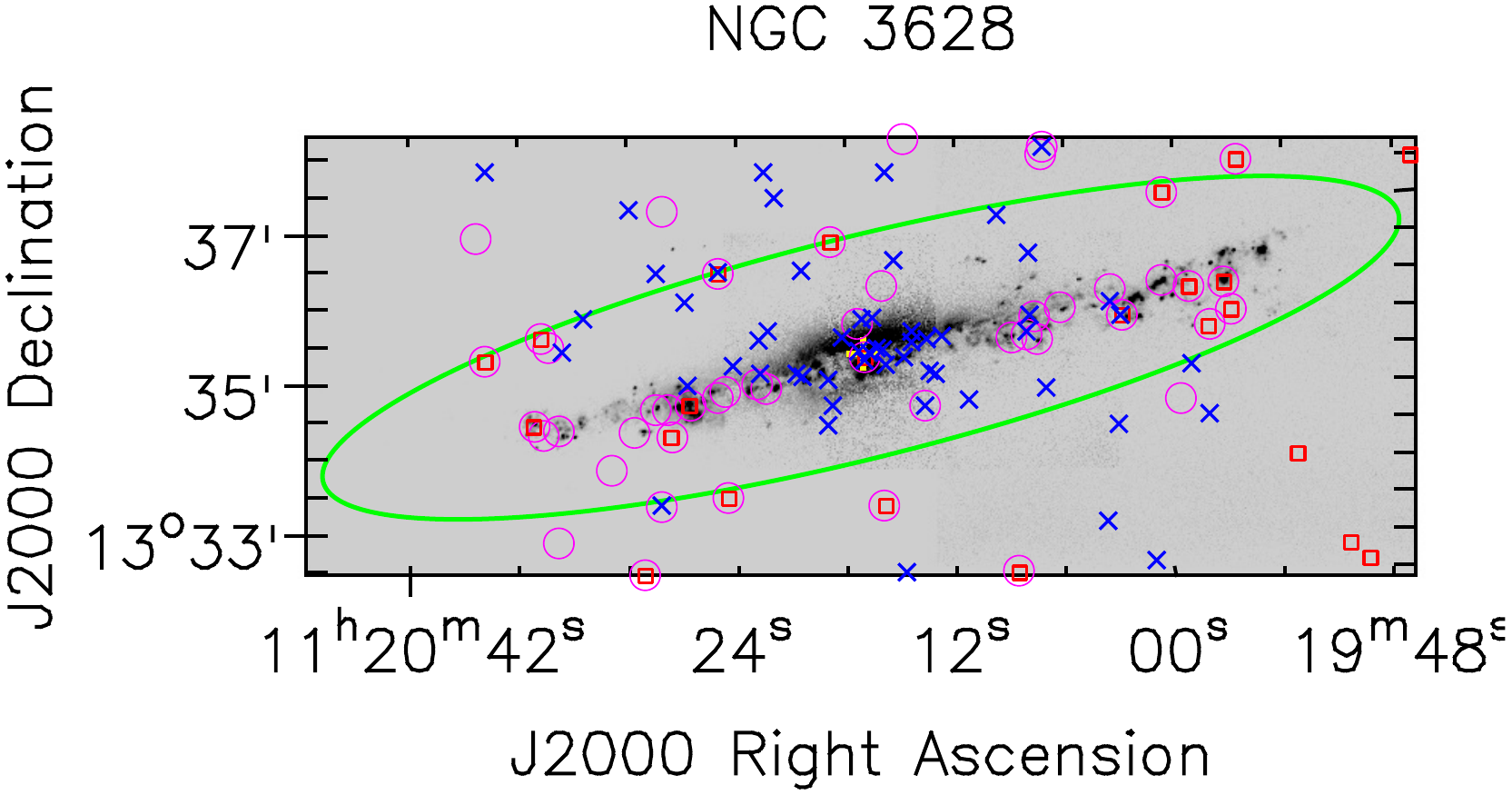}
    \label{f:NGC3628}
\end{figure}

\begin{figure}[!]
    \centering
    \includegraphics[width=0.8\textwidth]{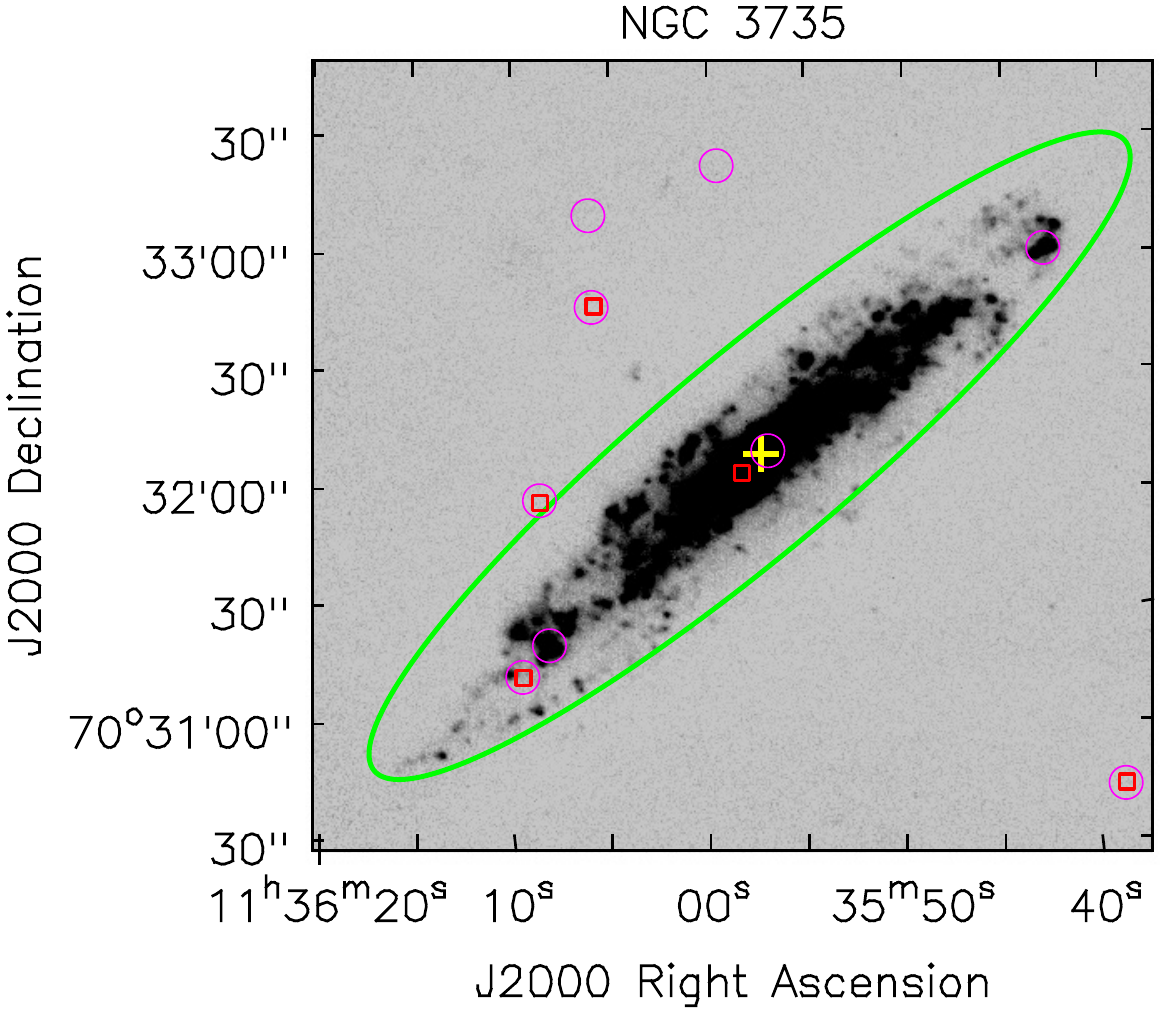}
    \label{f:NGC3735}
\end{figure}

\begin{figure}[!]
    \centering
    \includegraphics[width=0.8\textwidth]{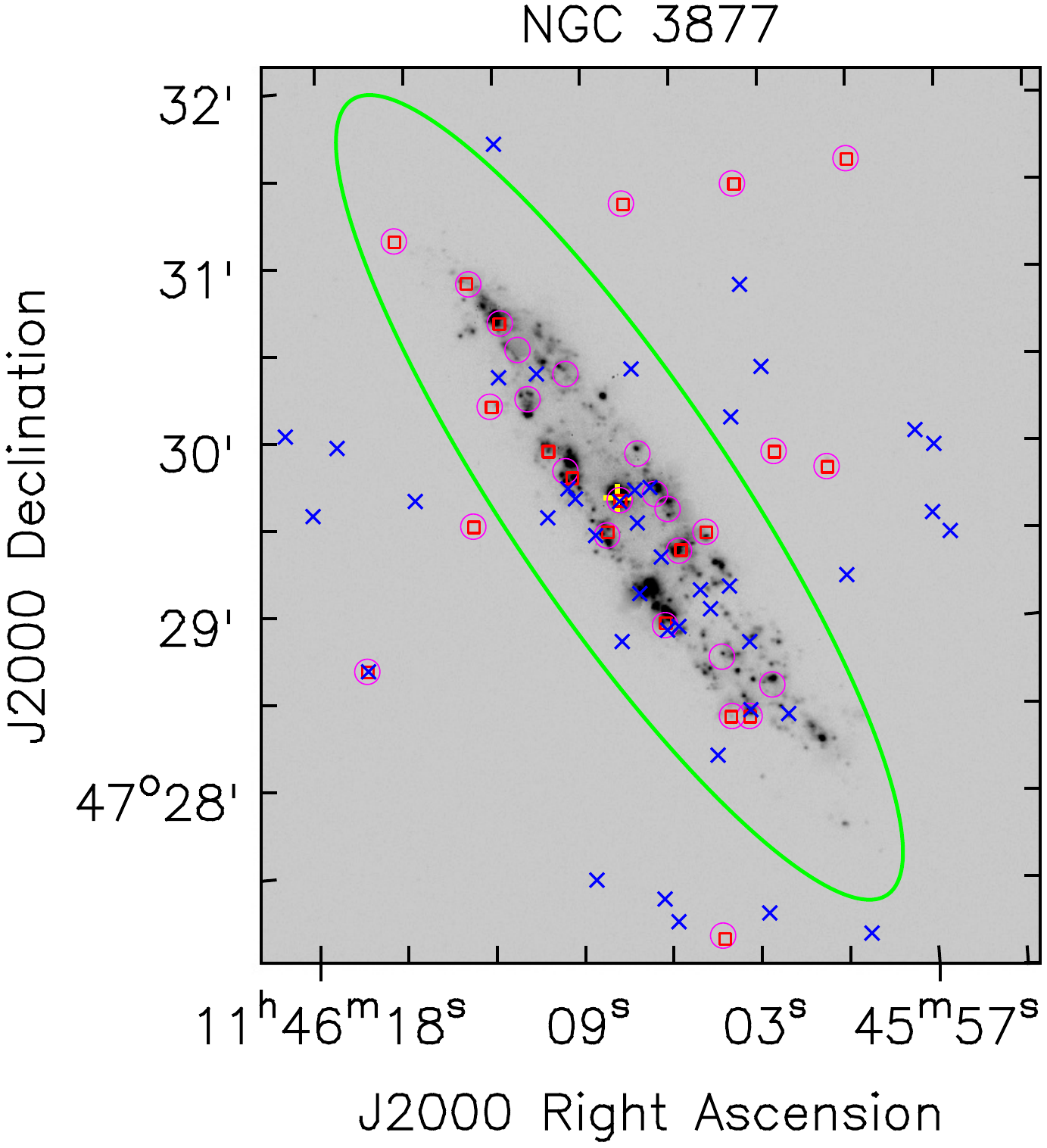}
    \label{f:NGC3877}
\end{figure}

\begin{figure}[!]
    \centering
    \includegraphics[width=0.8\textwidth]{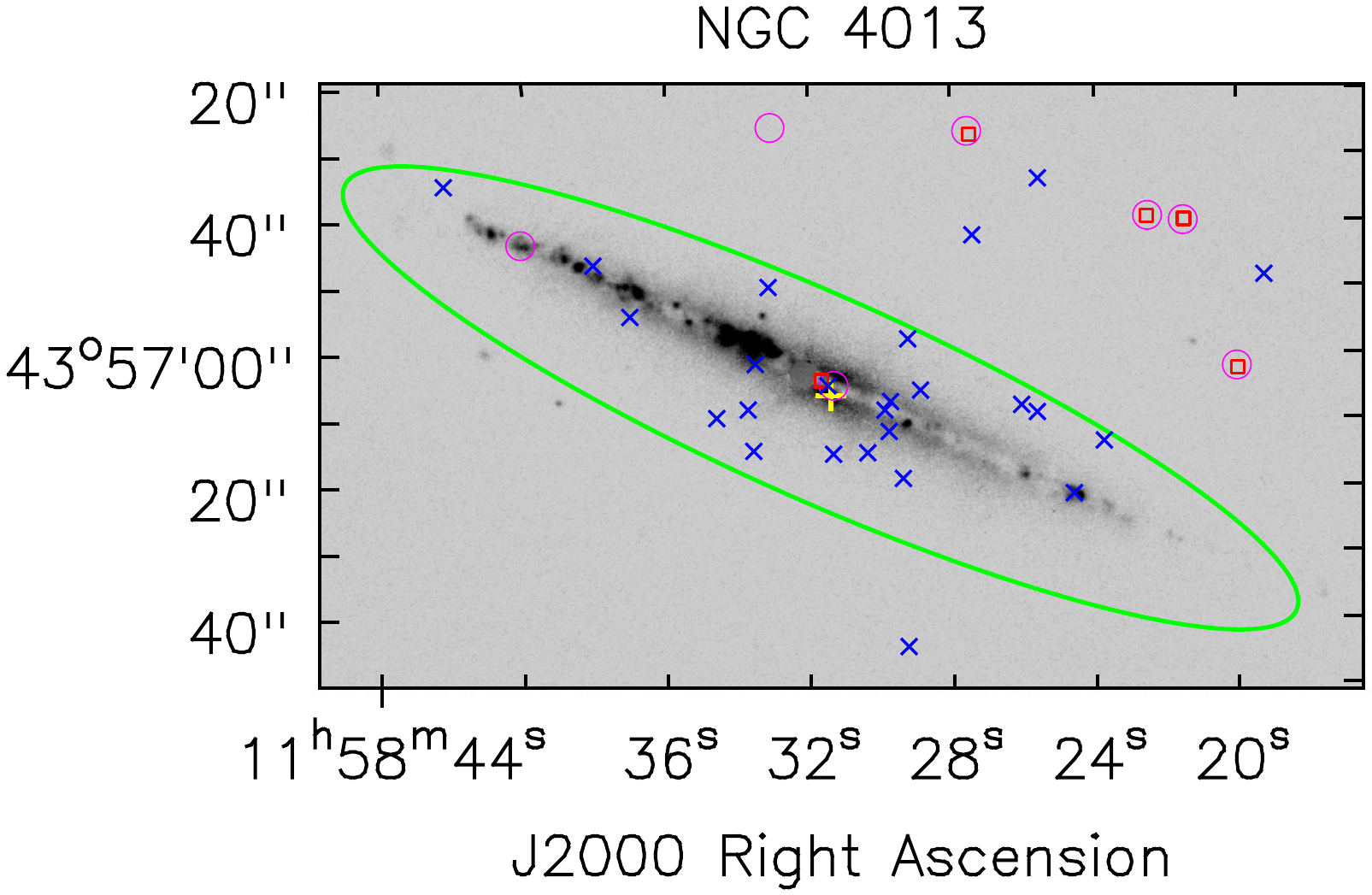}
    \label{f:NGC4013}
\end{figure}

\begin{figure}[!]
    \centering
    \includegraphics[width=0.8\textwidth]{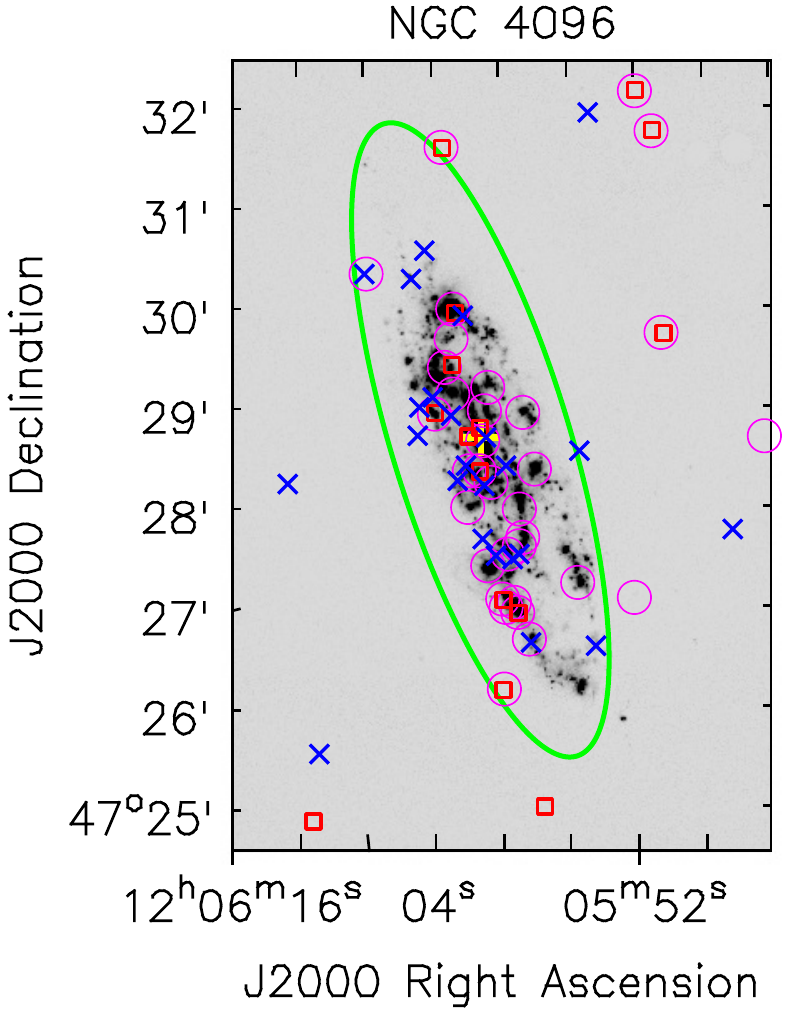}
    \label{f:NGC4096}
\end{figure}

\begin{figure}[!]
    \centering
    \includegraphics[width=0.8\textwidth]{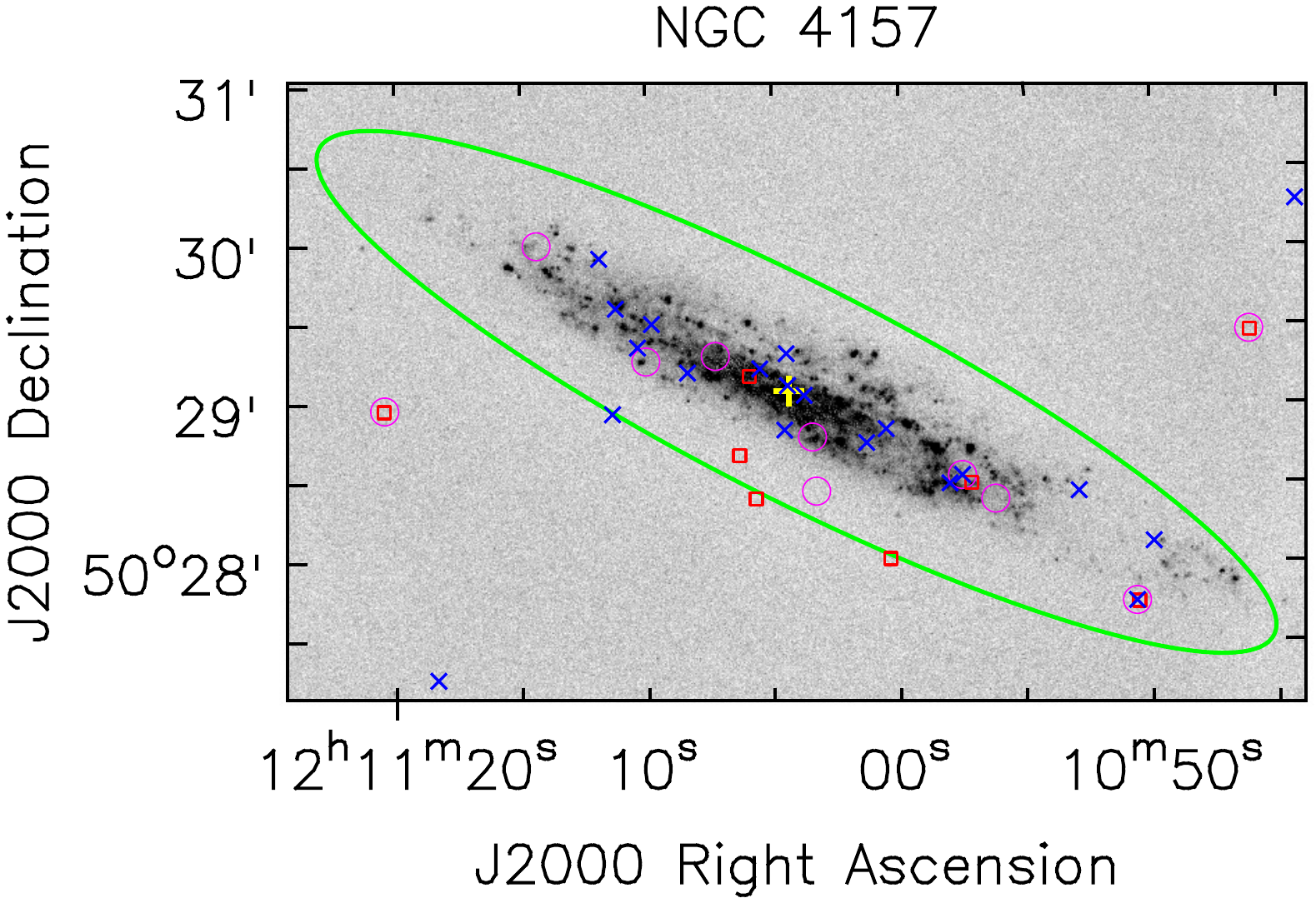}
    \label{f:NGC4157}
\end{figure}

\begin{figure}[!]
    \centering
    \includegraphics[width=0.8\textwidth]{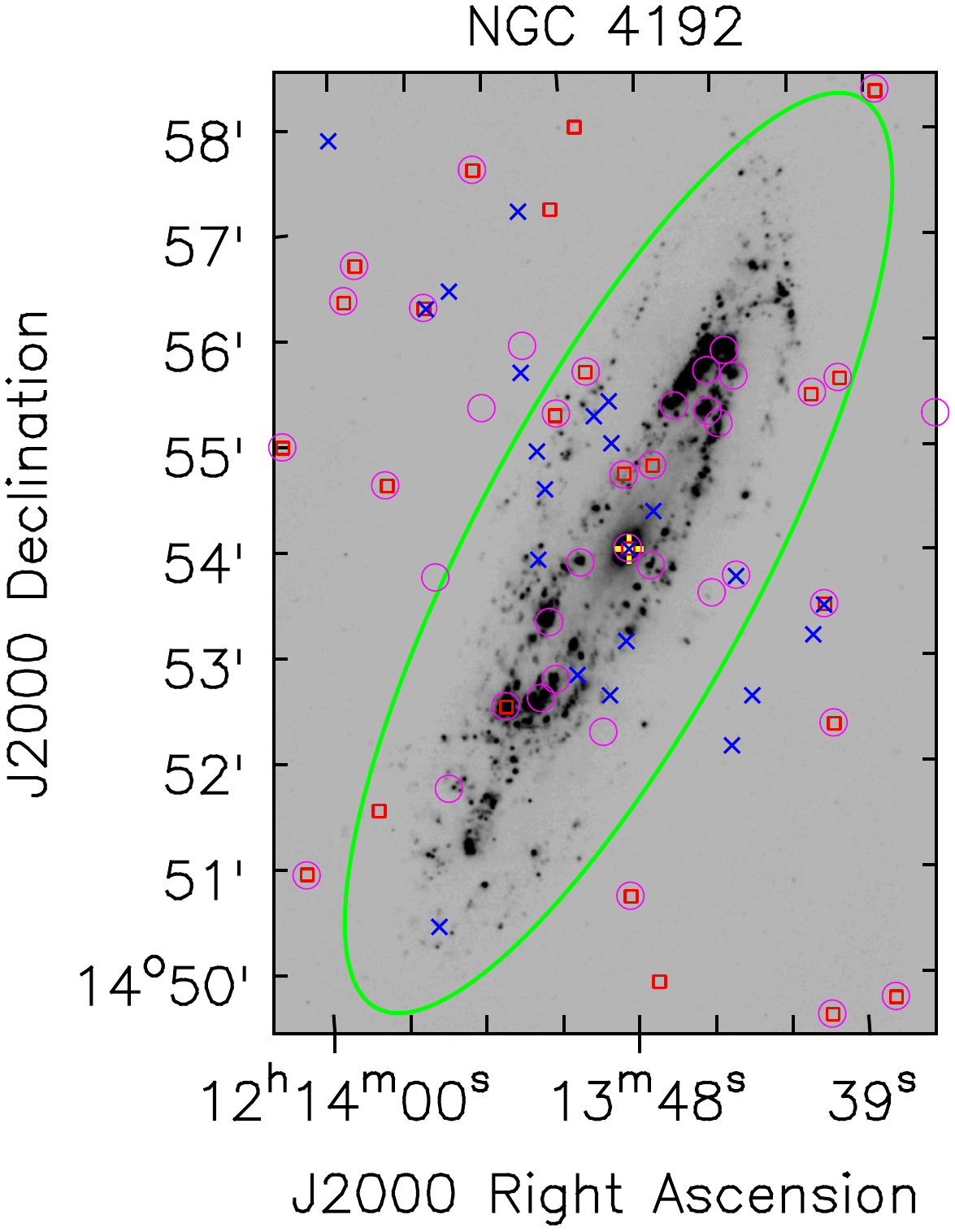}
    \label{f:NGC4192}
\end{figure}

\begin{figure}[!]
    \centering
    \includegraphics[width=0.8\textwidth]{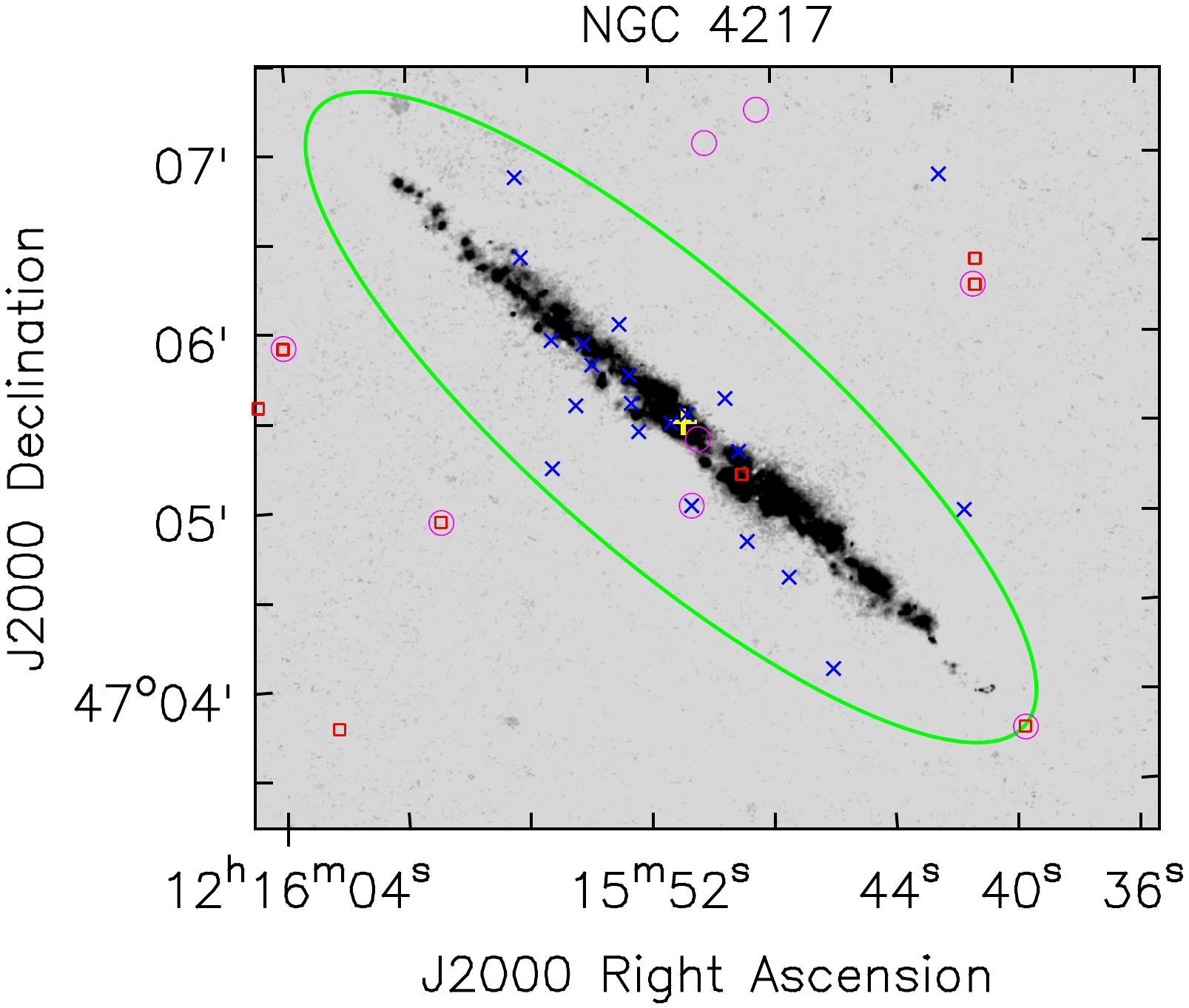}
    \label{f:NGC4217}
\end{figure}

\begin{figure}[!]
    \centering
    \includegraphics[width=0.8\textwidth]{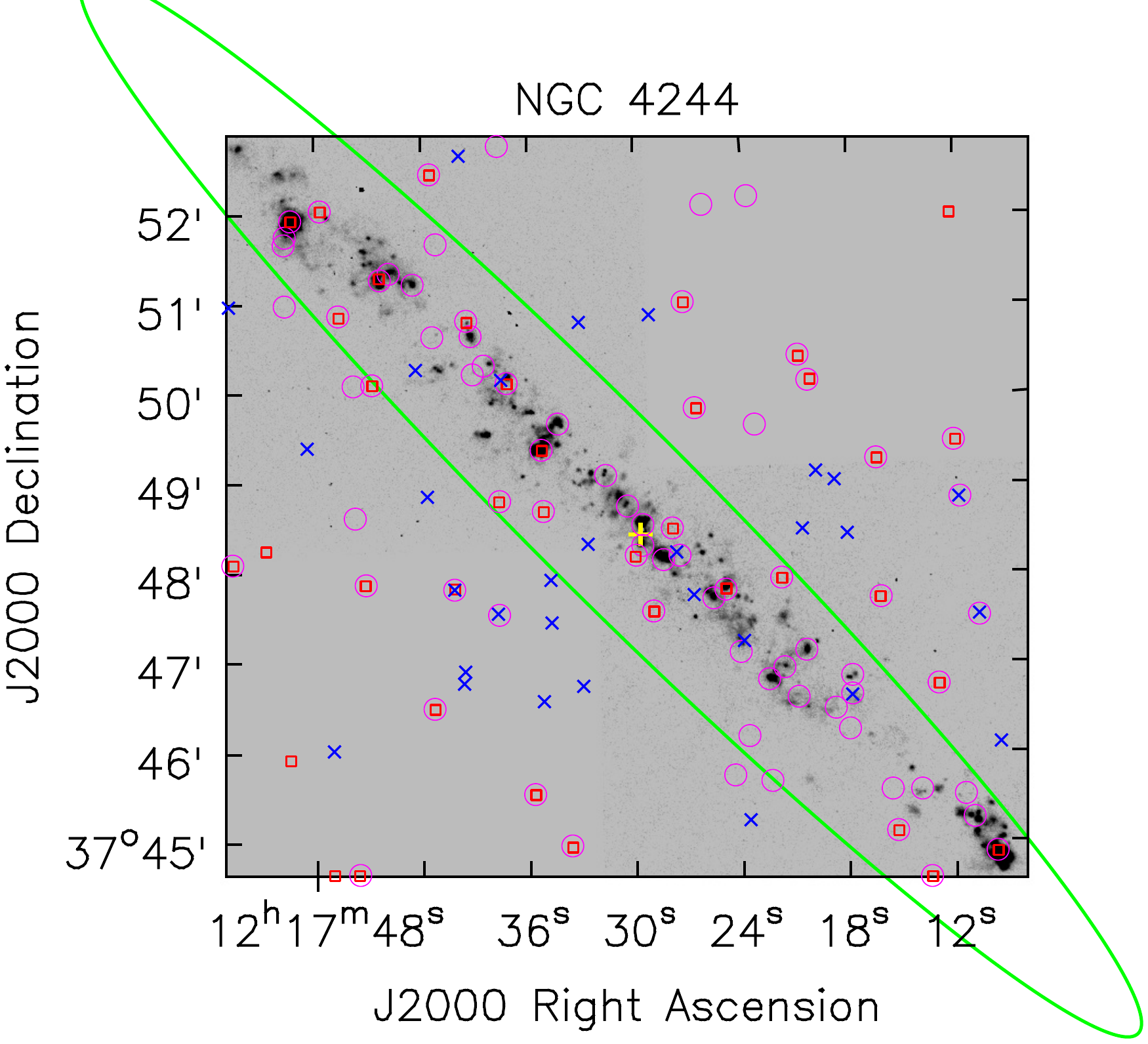}
    \label{f:NGC4244}
\end{figure}

\begin{figure}[!]
    \centering
    \includegraphics[width=0.8\textwidth]{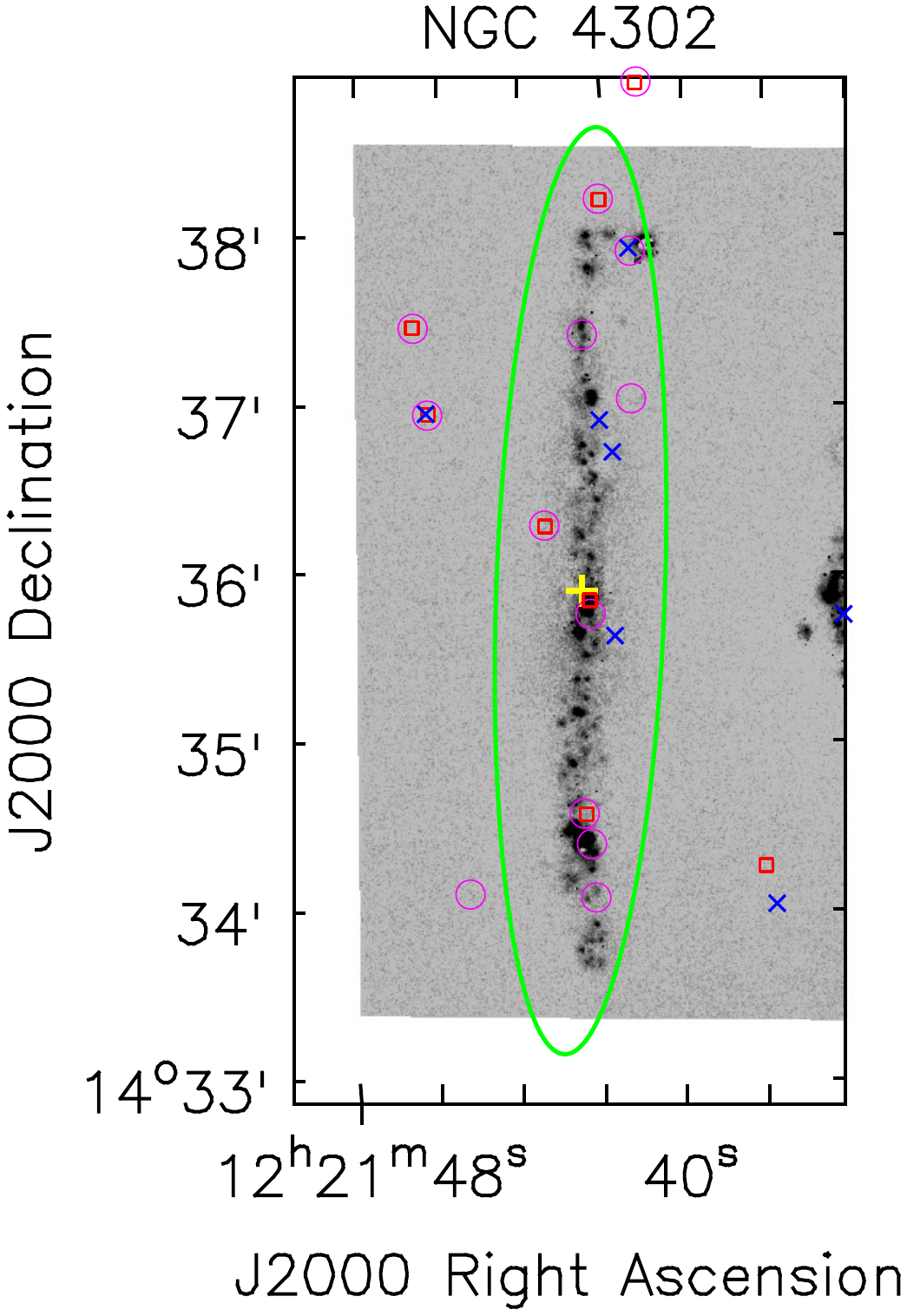}
    \label{f:NGC4302}
\end{figure}

\begin{figure}[!]
    \centering
    \includegraphics[width=0.8\textwidth]{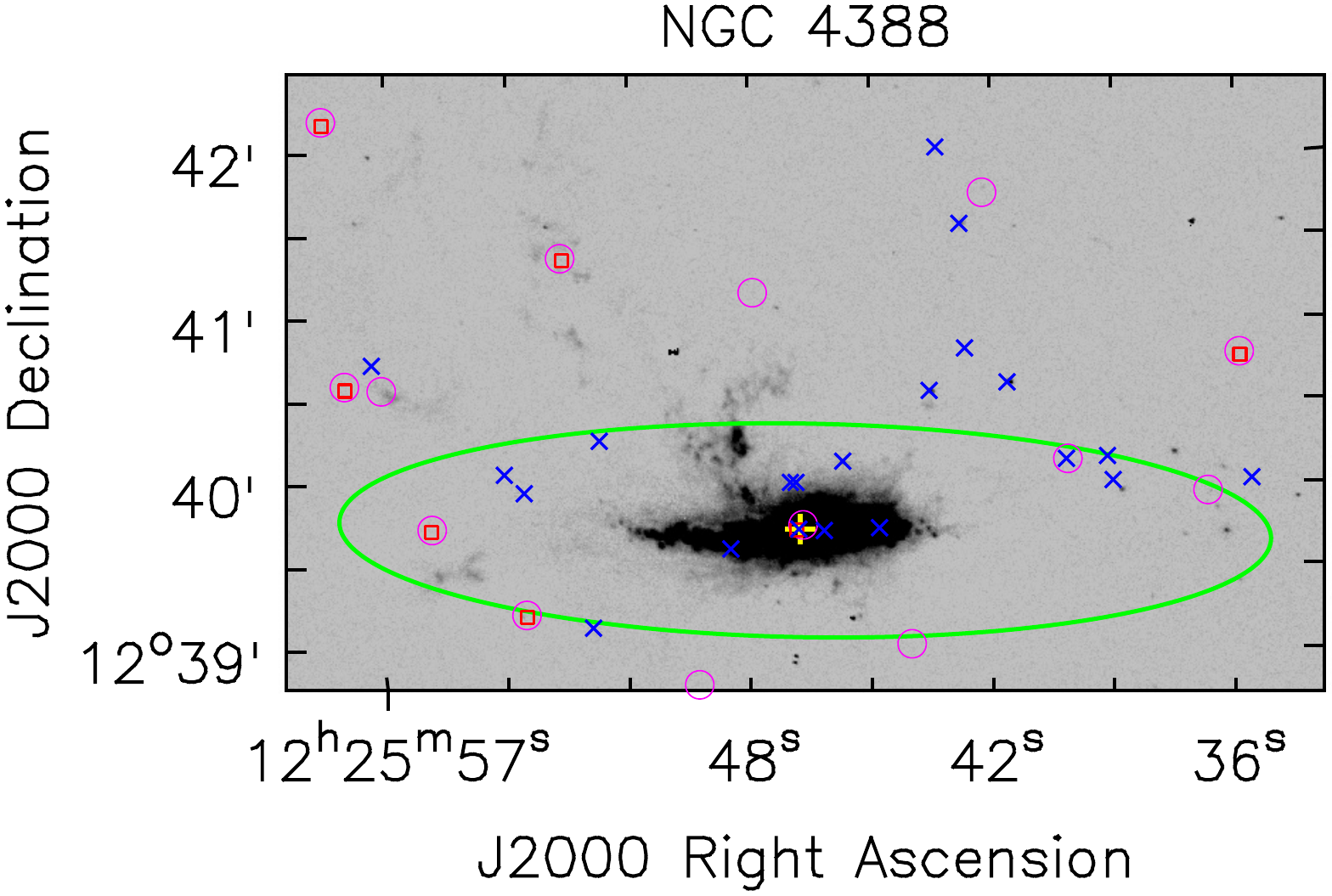}
    \label{f:NGC4388}
\end{figure}

\begin{figure}[!]
    \centering
    \includegraphics[width=0.8\textwidth]{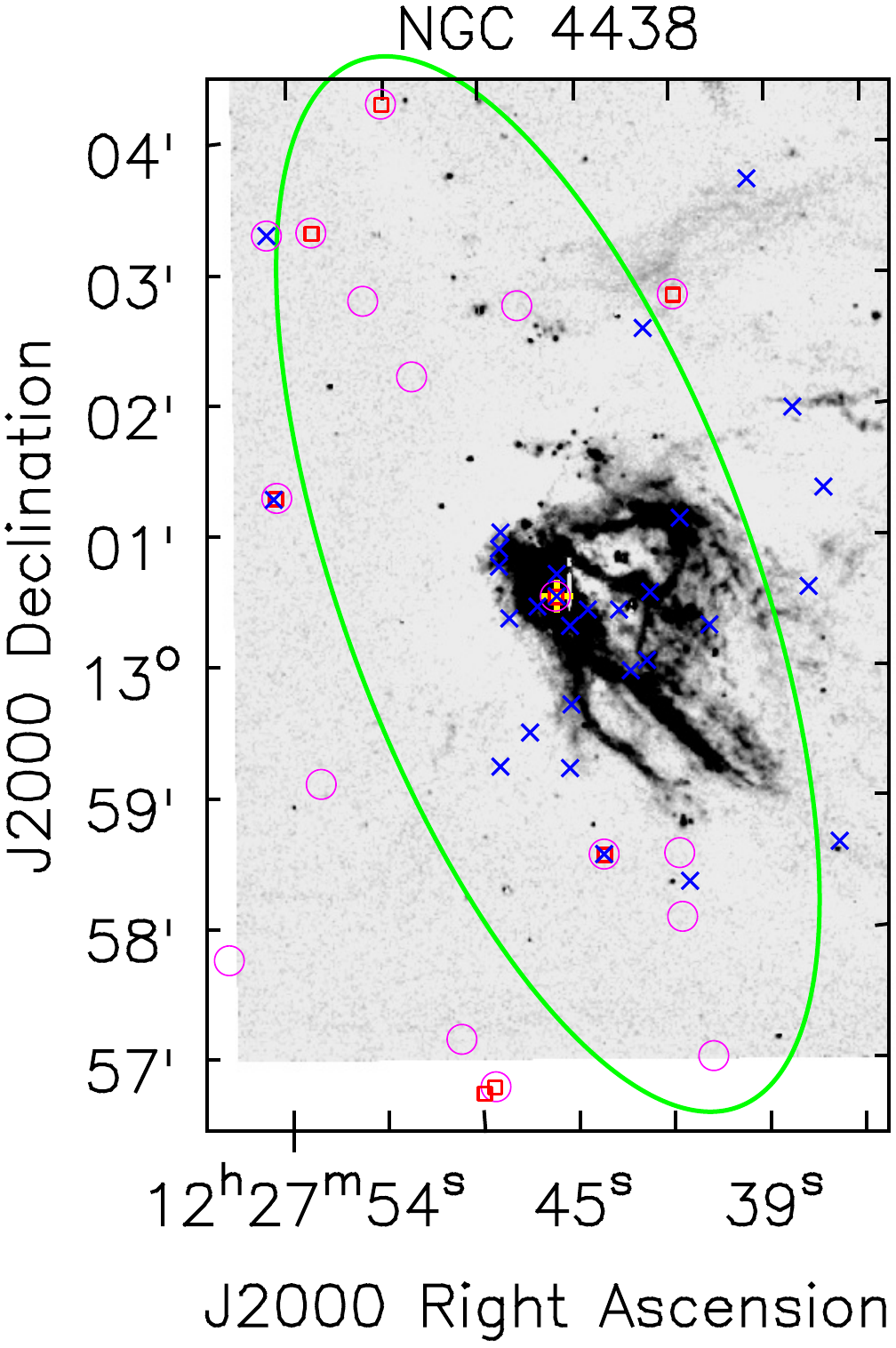}
    \label{f:NGC4438}
\end{figure}

\begin{figure}[!]
    \centering
    \includegraphics[width=0.8\textwidth]{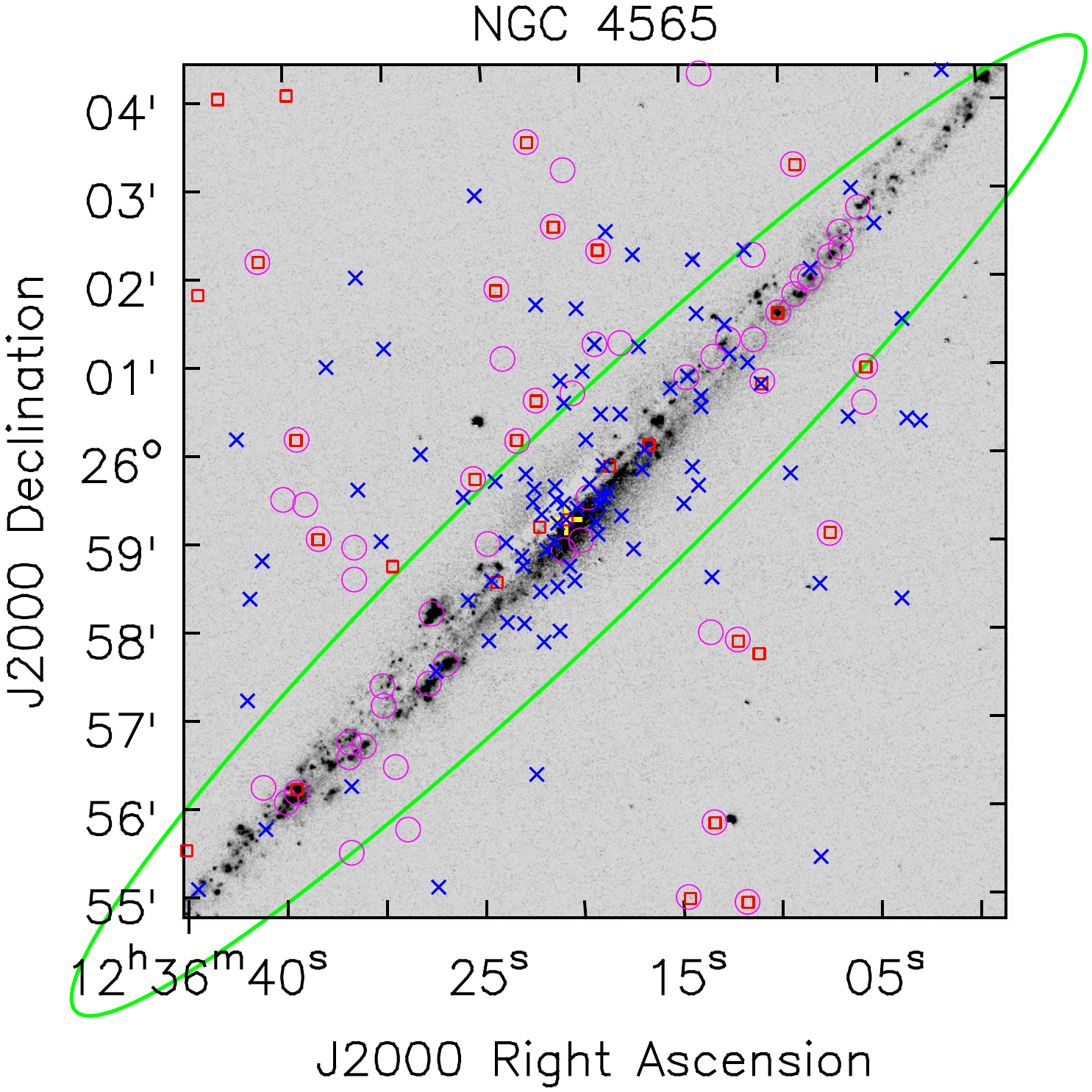}
    \label{f:NGC4565}
\end{figure}

\begin{figure}[!]
    \centering
    \includegraphics[width=0.8\textwidth]{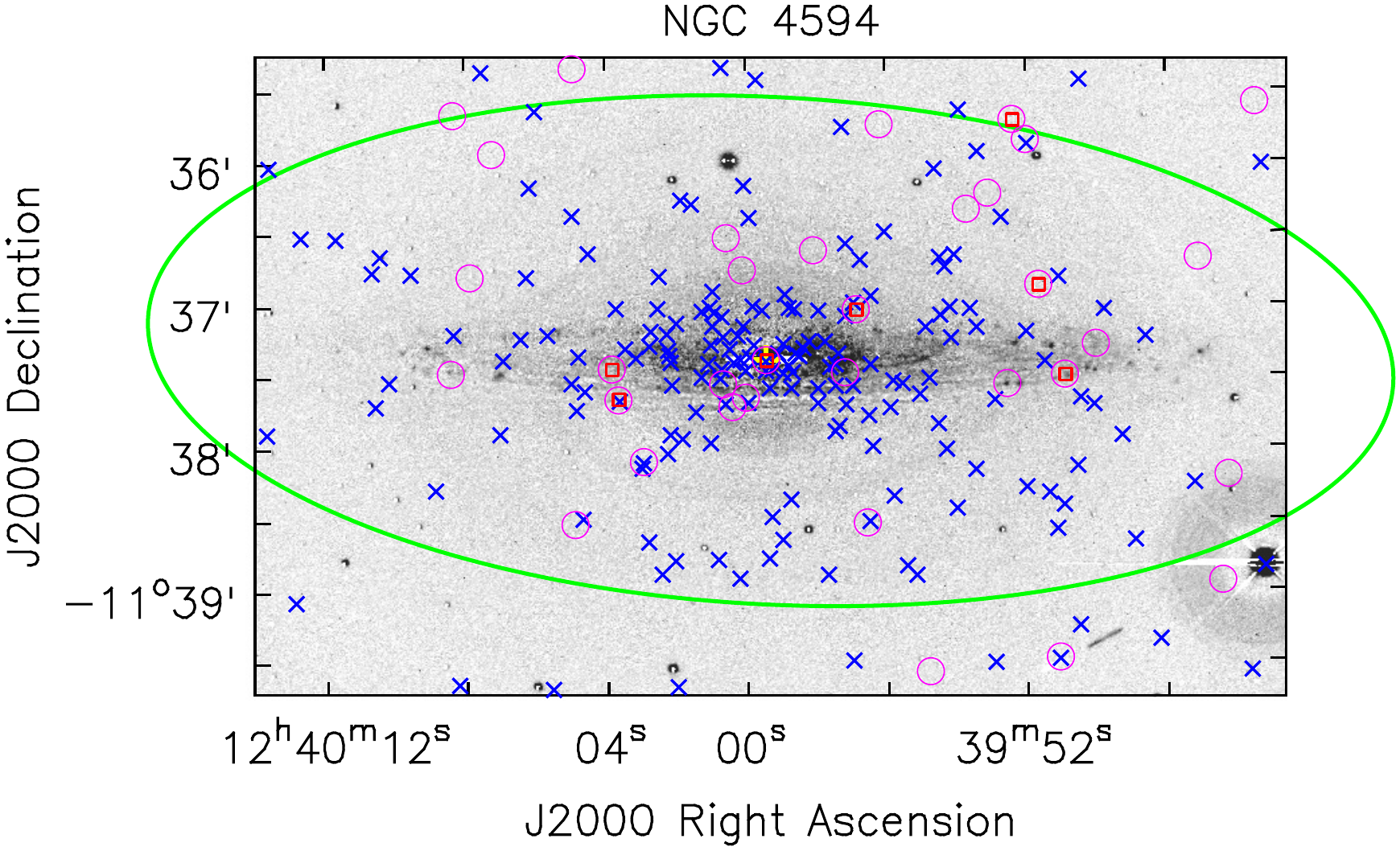}
    \label{f:NGC4594}
\end{figure}

\begin{figure}[!]
    \centering
    \includegraphics[width=\textwidth]{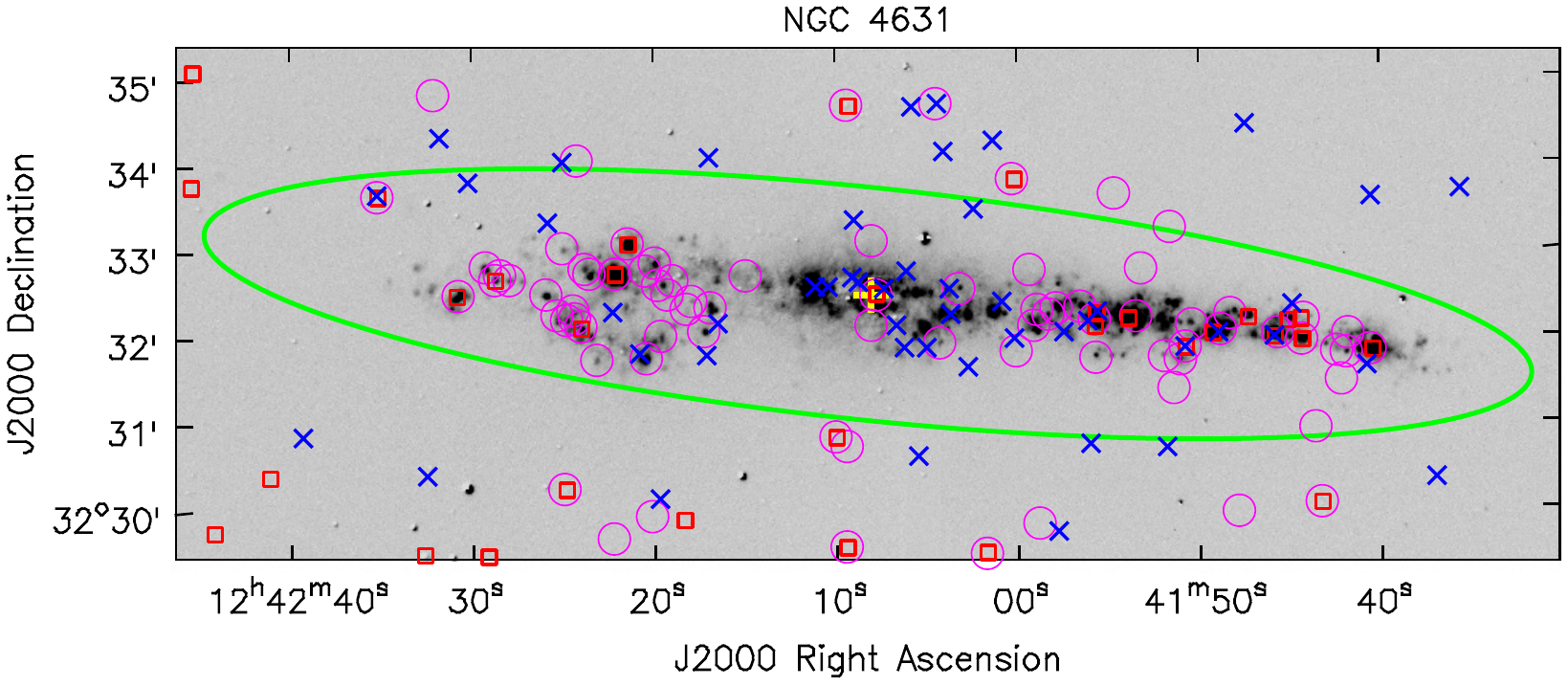}
    \label{f:NGC4631}
\end{figure}

\begin{figure}[!]
    \centering
    \includegraphics[width=0.8\textwidth]{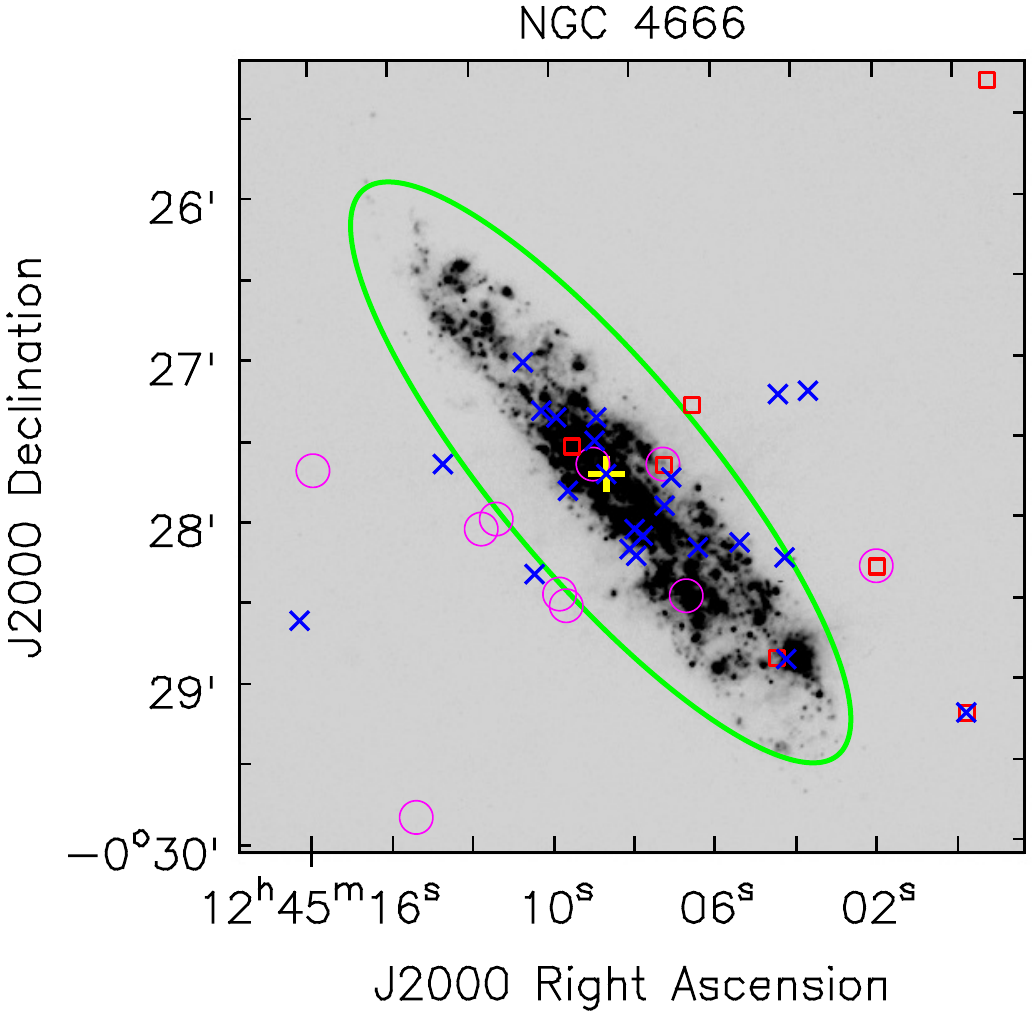}
    \label{f:NGC4666}
\end{figure}

\begin{figure}[!]
    \centering
    \includegraphics[width=0.8\textwidth]{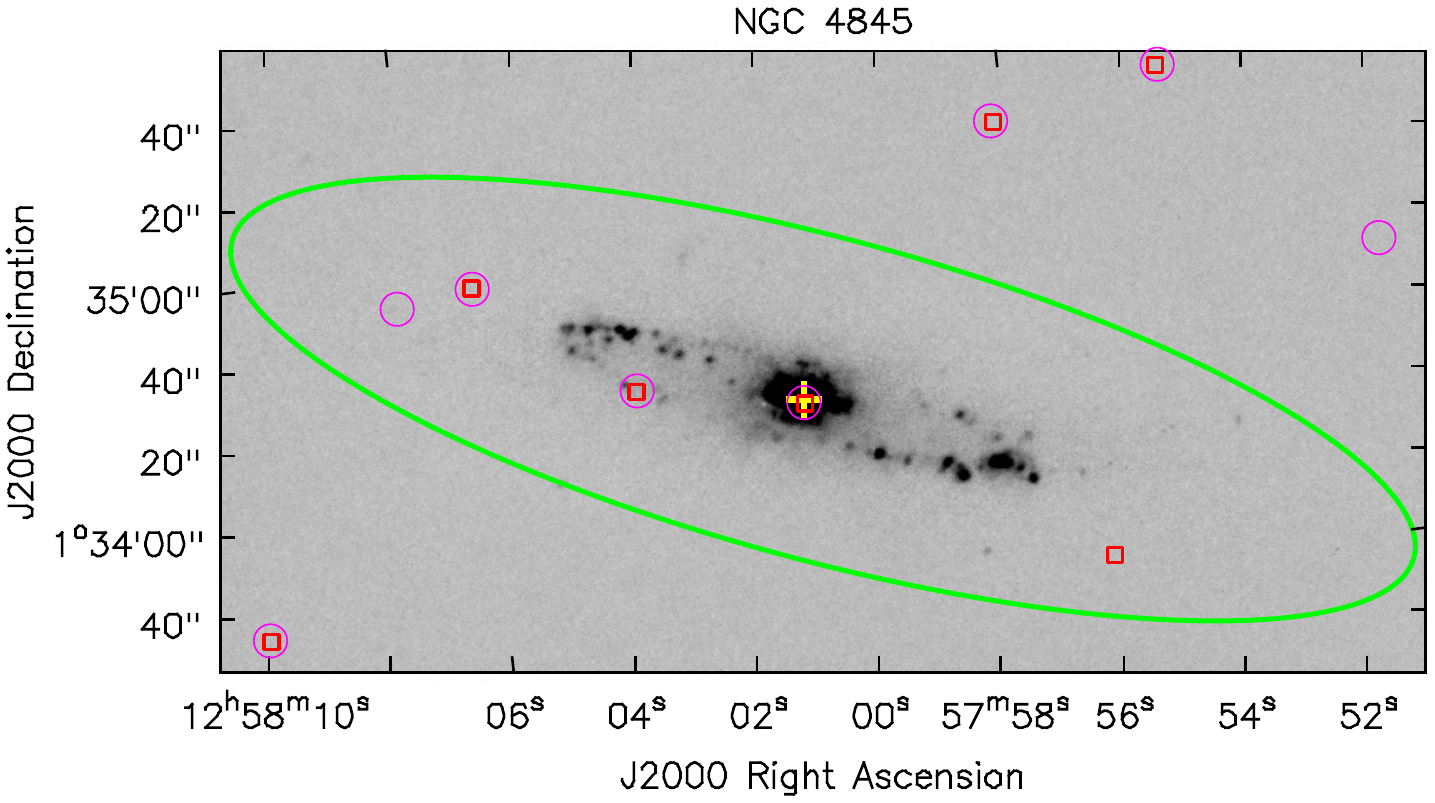}
    \label{f:NGC4845}
\end{figure}

\begin{figure}[!]
    \centering
    \includegraphics[width=0.8\textwidth]{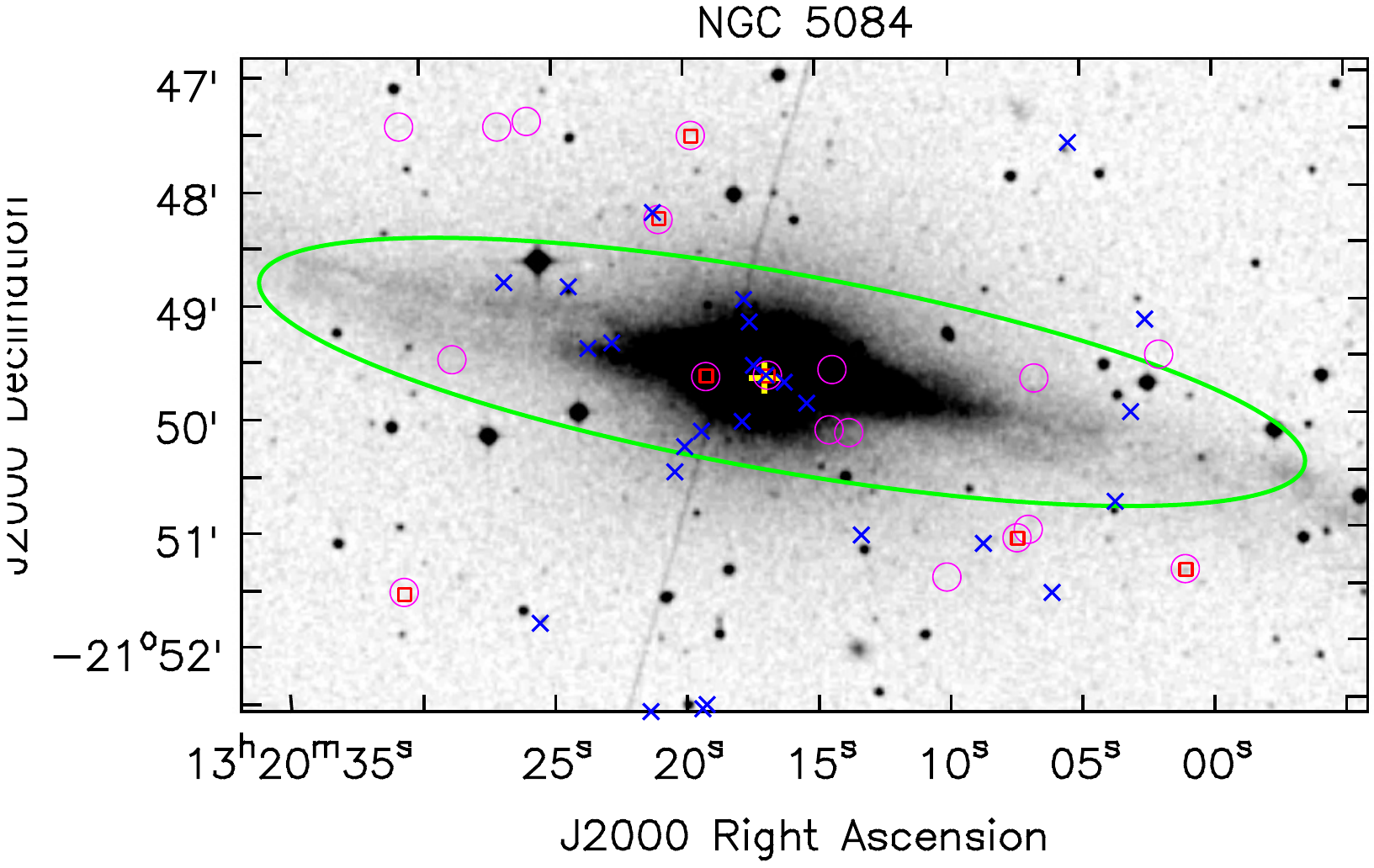}
    \label{f:NGC5084}
\end{figure}

\begin{figure}[!]
    \centering
    \includegraphics[width=0.8\textwidth]{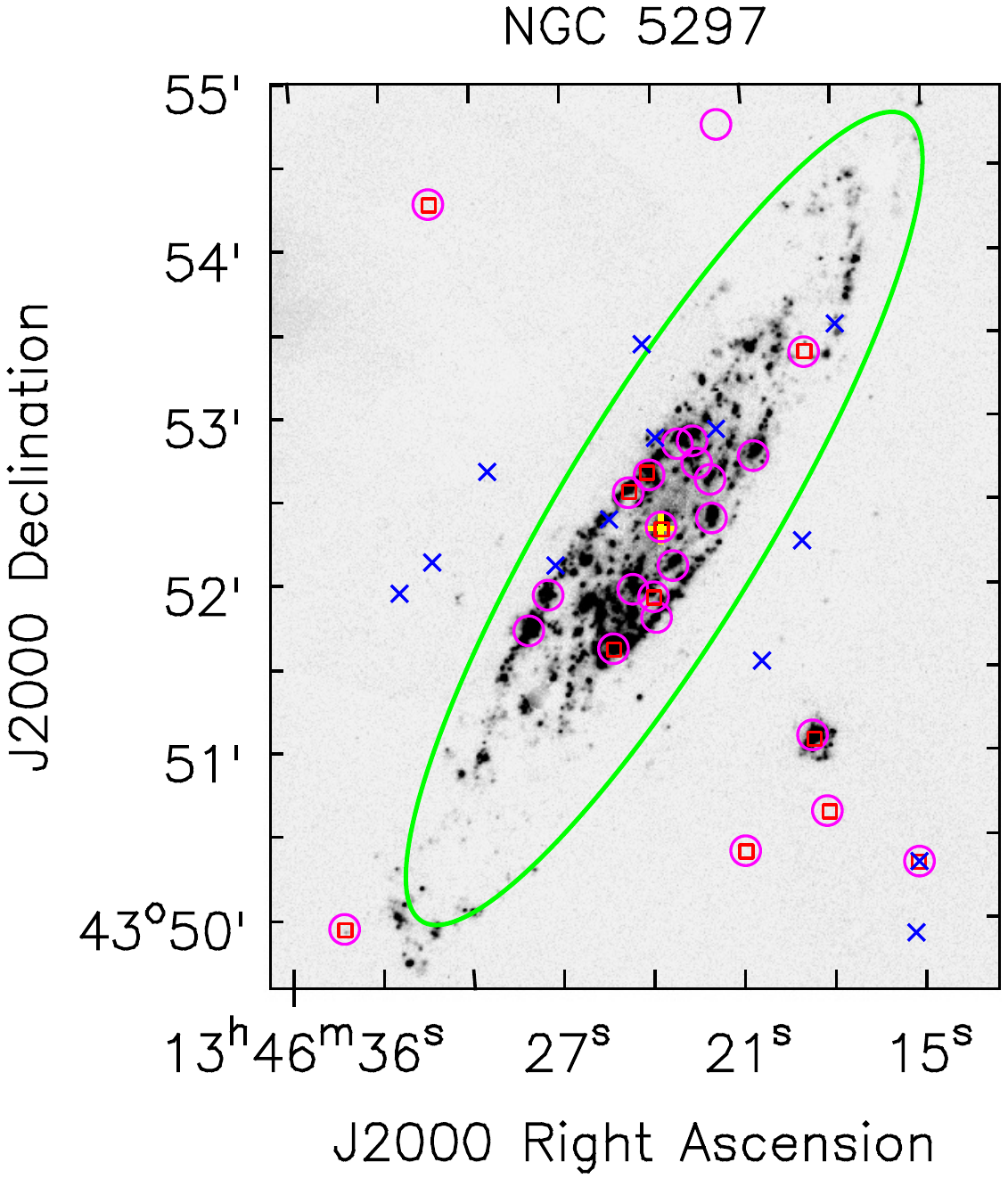}
    \label{f:NGC5297}
\end{figure}

\begin{figure}[!]
    \centering
    \includegraphics[width=0.8\textwidth]{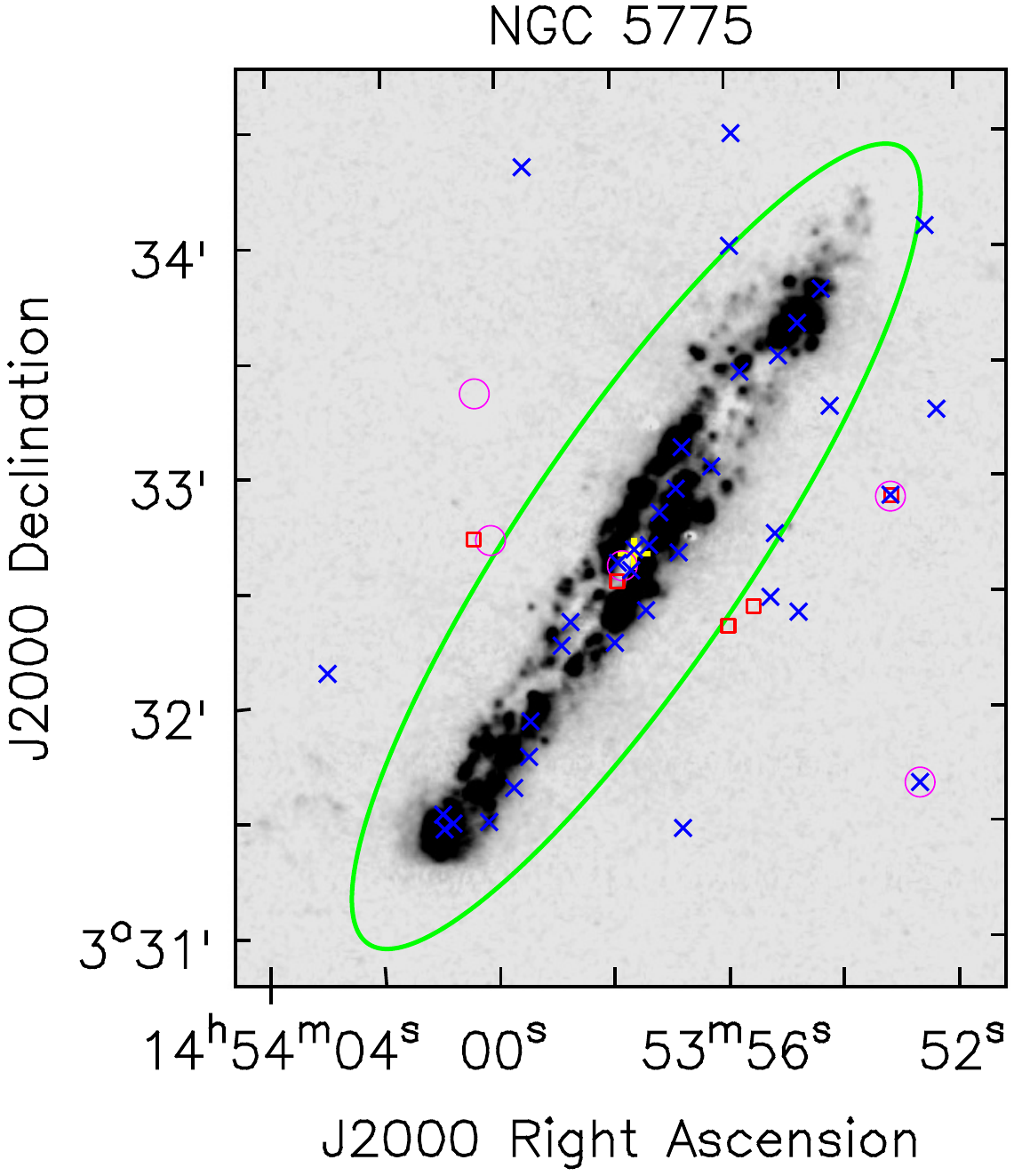}
    \label{f:NGC5775}
\end{figure}

\begin{figure}[!]
    \centering
    \includegraphics[width=0.8\textwidth]{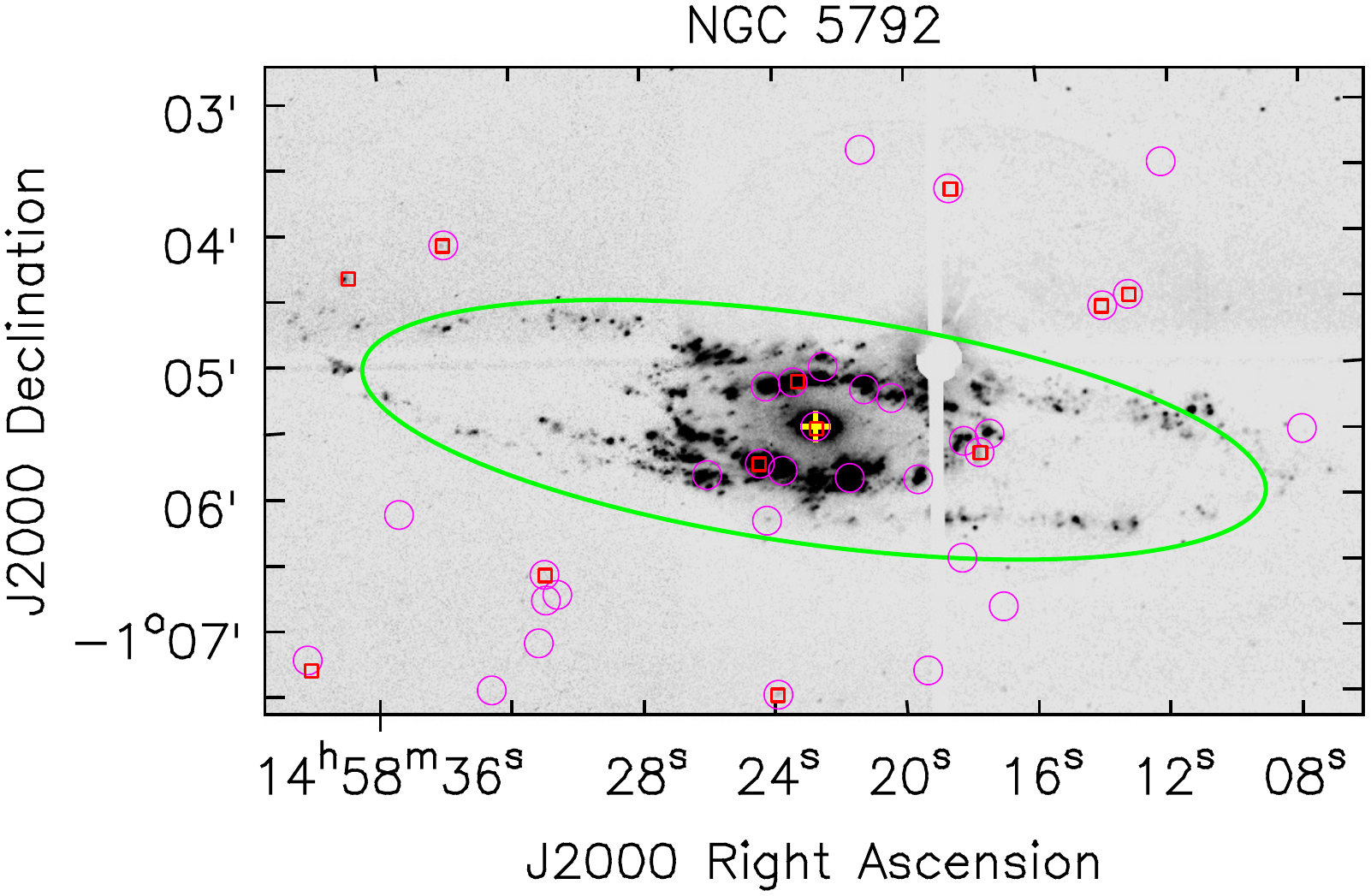}
    \label{f:NGC5792}
\end{figure}

\begin{figure}[!]
    \centering
    \includegraphics[width=0.8\textwidth]{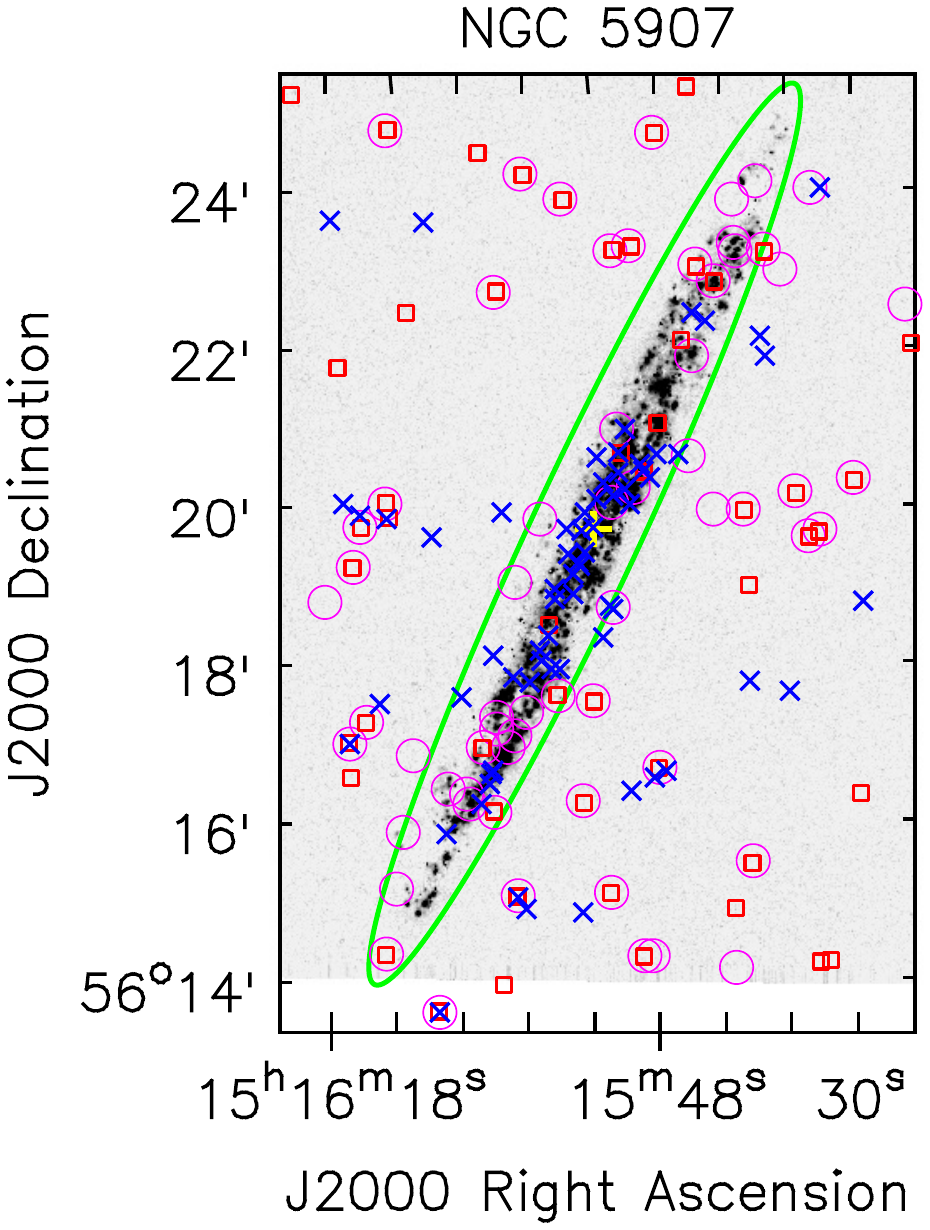}
    \label{f:NGC5907}
\end{figure}

\begin{figure}[!]
    \centering
    \includegraphics[width=0.8\textwidth]{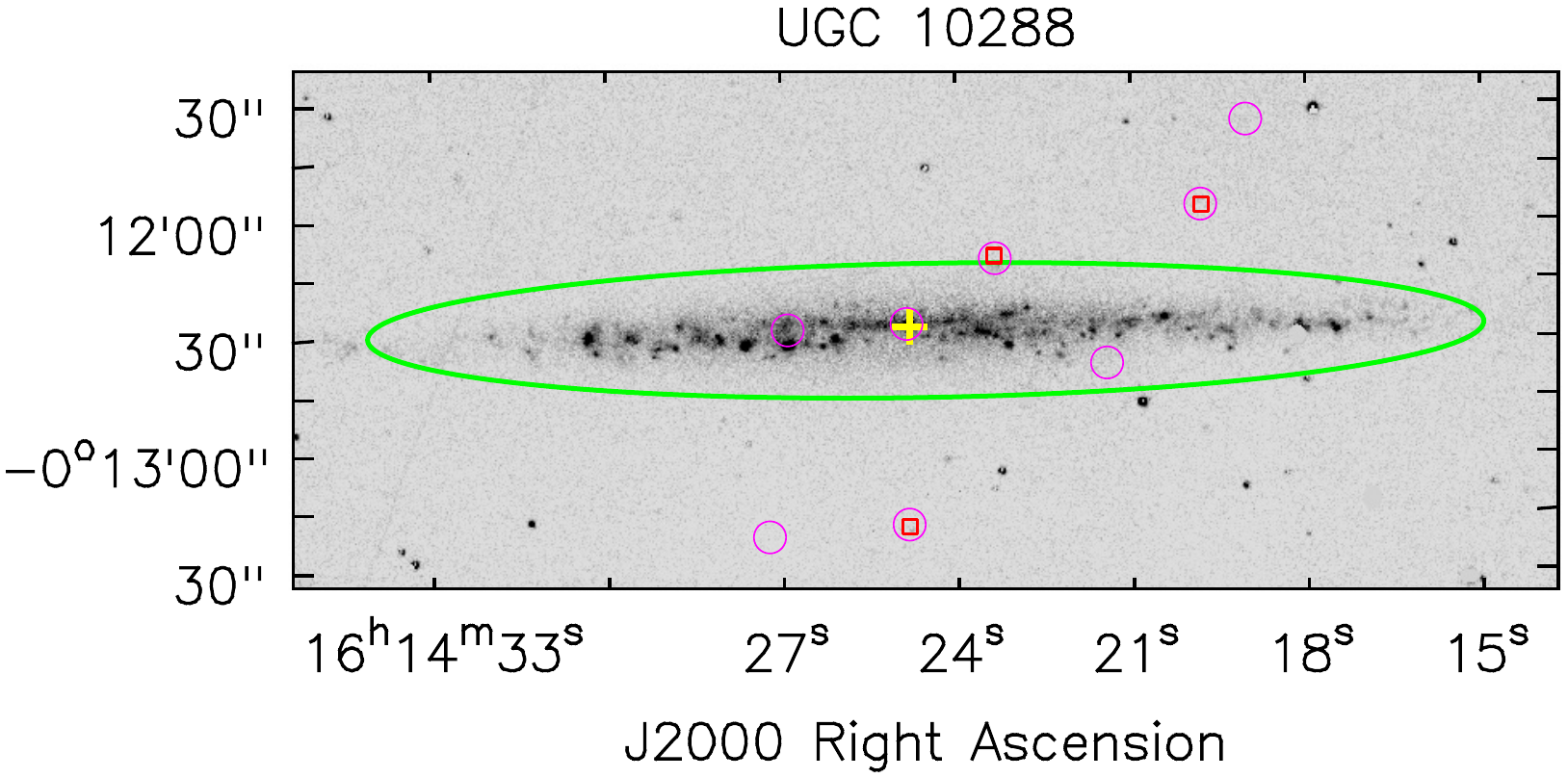}
    \label{f:UGC10288}
\end{figure}



\vfill
\bsp	
\label{lastpage}
\end{document}